\documentclass[aps,twocolumn,superscriptaddress,longbibliography,floatfix,notitlepage]{revtex4-2}
\usepackage{bbm}
\usepackage{physics}
\usepackage{amsmath}
\usepackage{epsfig}
\usepackage{subfigure,mathrsfs}
\usepackage{array}
\usepackage{amssymb}
\usepackage{braket}
\usepackage{lmodern,amssymb}
\usepackage{physics}
\usepackage[dvipsnames]{xcolor}
\usepackage{lipsum}

\newcommand{\beq}[0]{\begin{equation}}
\newcommand{\eeq}[0]{\end{equation}}

\def\be{\begin{equation}}
\def\ee{\end{equation}}
\def\bea{\begin{eqnarray}}
\def\eea{\end{eqnarray}}
\newcommand{\ba}{\begin{eqnarray}}
\newcommand{\ea}{\end{eqnarray}}
\usepackage{hyperref}

\usepackage{caption}
\usepackage[bb=boondox]{mathalfa}
\usepackage{sidecap}

\begin{document}

\title{Quantum Thermal Logic Gates}

\author{Shuvadip Ghosh}
\affiliation{Indian Institute of Technology Kanpur, 
	Kanpur, Uttar Pradesh 208016, India}
    \author{Papiya Maity}
    \affiliation{Department of Physics, Indian Institute of Technology Indore, Simrol, Indore, Madhya Pradesh 453552, India}
    \author{Bivas Dutta}
\thanks{bivas@iiti.ac.in}
\affiliation{Department of Physics, Indian Institute of Technology Indore, Simrol, Indore, Madhya Pradesh 453552, India}
\author{Arnab Ghosh}
\thanks{arnab@iitk.ac.in}
\affiliation{Indian Institute of Technology Kanpur, 
	Kanpur, Uttar Pradesh 208016, India}

\date{\today}

\begin{abstract}
We propose a new concept for quantum thermal logic gates -- analogous to classical electronic logic gates -- that exploit the heat current in a coupled quantum-dot system tunnel-coupled to metallic thermal reservoirs for logic operations in quantum circuits. We obtained a remarkable one-to-one correspondence with the structure of classical electronic logic gate circuits. An experimental setup is presented that demonstrates a realizable nano-electronic quantum circuit architecture for implementing such quantum thermal logic operations.
\end{abstract}

\maketitle


\textit{Introduction:} In the emerging era of quantum computation, achieving error-free and thermally efficient quantum circuits is the ultimate goal~\cite{maurand2022transistor,aamir2025thermally,blok2025quantum,Karimi2024bolometric,simbierowicz2024inherent,champain2025heatresilientholespinqubit}. Toward this effort, recent advancements in building compact quantum circuit architecture with unprecedented control over individual quantum systems are remarkable~\cite{rossnagel2016single,thierschmann2015three,jezouin2013quantum,banerjee2017observed,dutta2022isolated,myers2022quantum}. These include the development of quantum-advantageous devices on multiple platforms, e.g., nano-electronic quantum transistors~\cite{joulain2016quantum,dutta2019direct,perrin2015single,gupt2022PRE,wijesekara2021darlington}, diodes~\cite{perrin2016agate,shuvadip2022univarsal,khan2021efficient,upadhyay2024microwave}, capacitors~\cite{moskalets2008quantized}, batteries~\cite{Francesco2024colloquium,ferraro2026opportunities}, heat engines and refrigerators~\cite{kosloff2014quantum,arnab2017catalysis,ghosh2018two-level,binder2019thermodynamics}, transformers~\cite{maity2026quantum}, etc. In addition to developing innovative and compact designs of nano-electronic quantum circuits, managing thermal dissipation during their operation is another important issue~\cite{kurizki2022thermodynamics} for ensuring reliable and error-free device performance. To address these challenges, based on local thermal-measurement techniques~\cite{giazotto2006opportunities,pekola1994thermometry,nahum1993ultrasensitive,meschke2009calorimetric,halbertal2016nanoscale,karimi2020reaching}, various quantum devices with optimized thermal operations have been proposed and demonstrated~\cite{pekola2021colloquium,Majidi2024Heat}. Some noteworthy developments include strong electron interactions mediated enhanced heat-flow in single-electron junctions~\cite{dutta2017thermal,cui2017quantized,mosso2017heat}, near-optimal quantum heat engines with efficiencies close to the Carnot limit~\cite{josefsson2018a,volosheniuk2026asingle}, and quantum heat valves that regulate heat-flow using quantum levels~\cite{dutta2020single,maillet2020electric,ronzani2018tunable}, among others. These advances in controlling heat-flow and dissipation in quantum devices have already set the stage for further quantum thermodynamic explorations toward integration in quantum circuits~\cite{campbell2026roadmap,devvrat2025quantum}. However, establishing a quantum analog of classical electronic logic-gates for heat-based logic operations, while being a crucial element in integrated quantum circuits, remains largely unexplored.

In this letter, we introduce for the first time the concept of \textit{quantum thermal logic gates} (QTLG), which exploit temperature gradients and heat currents in a nano-electronic coupled quantum-dot (CQD) ~\cite{wang2022cycleflux,tesser2022heat,shuvadip2022univarsal,gupt2024graph,ghosh2026inverse} junction to perform logical operations --- analogous to how electronic logic gates process electrical signals. Such devices not only promise new modes of energy-efficient computation but also hold potential for intelligent thermal management in nano-electronic architectures. With the growing demand for sustainable energy solutions and waste-heat recovery in nanoscale systems, we focus on controlling thermal transport at the quantum level by exploring the fundamental principles and minimal models required to realize QTLG --- providing a step toward programmable, experimentally realizable heat-based computation and energy-efficient quantum technologies~\cite{wang2007thermal,paolucci2018phasetunable}.

\begin{figure}[t]
\centering    
\includegraphics[width=\columnwidth]{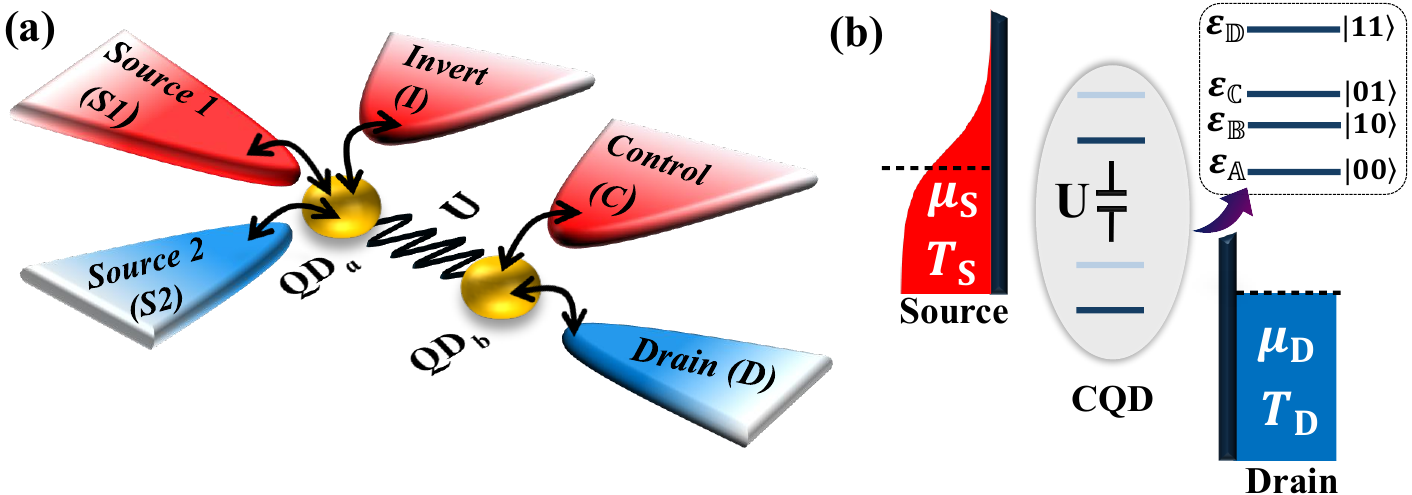}
\caption{(a) Schematic of the proposed QTLG: the CQD system is tunnel-coupled to S1 (hot/cold, here hot), S2 (hot/cold, here cold), and D (cold) leads; source temperatures define input thermal-logic (hot (red):~1, cold (blue):~0). I and C leads, coupled to $\rm{QD_a}$ and $\rm{QD_b}$, are kept at high temperature for NOT and AND gate operations, respectively. The thermal current measured at D (or at I lead for NOT gates) defines the output thermal-logic. (b) The basic building block of a QTLG -- a thermal-diode in the forward-bias configuration, where input logic-1 (broadened FD-distribution) allows heat-flow from S to D. Inset: CQD energy levels and its states. 
}
\label{General Model}
\end{figure}

\textit{General Model:} The general model of the proposed QTLG, based on a CQD system tunnel-coupled to reservoirs, is presented in Fig.~\ref{General Model} (a). The two quantum-dots $\rm{QD}_{\alpha}\;(\alpha=\rm {a, b})$ are capacitively coupled and interact strongly via Coulomb-interaction~\cite{ruokola2011single,koski2015onchip,thierschmann2015thermal,zhang2017three}. We consider each $\rm{QD}_\alpha$ to have a large level spacing, such that only a single energy state is thermally accessible from the reservoirs, with the inter-dot Coulomb-interaction energy $U$~\cite{whitney2018quantum} [Fig.~\ref{General Model} (b)]. The dots can thus be either empty ($|0\rangle$) or occupied ($|1\rangle$). Repulsive Coulomb-interaction blocks particle exchange between QDs but allows energy transfer via inter-dot coupling~\cite{sanchez2011optimal}. Each QD is tunnel-coupled with one or multiple Fermionic-reservoirs~\cite{gupt2024graph,ghosh2026inverse,pyurbeeva2026quantum}, allowing both particle and energy exchange between the reservoir and the QD through the tunnel barrier. The total Hamiltonian for the CQD, baths and their tunneling interactions is given by $H_{\rm{tot}}=H_{\rm s}+H_{\rm Bath}+H_{\rm T}$, where $H_{\rm{s}}=\varepsilon_{\rm a} \mathcal{N}_{\rm a}+\varepsilon_{\rm b} \mathcal{N}_{\rm b}+U\mathcal{N}_{\rm a}\mathcal{N}_{\rm b}$. Here, $\varepsilon_\alpha$ and $\mathcal{N}_{\rm{\alpha}}= d^\dagger_{\rm{\alpha}} d_{\rm{\alpha}}$ are the single-particle energy and the number operator of the $\alpha$-th QD; $d_{\rm{\alpha}}$($d^\dagger_{\rm{\alpha}}$) are annihilation (creation) operators, satisfying $\{d_{\rm{\alpha}},d^\dagger_{\rm{\alpha}}\}=1$~\cite{ghosh2012fermionic,nikhil2021statistical,damas2023cooling}. Thus, the CQD can be seen as a four-level system: $|\mathbb{A}\rangle=|00\rangle$, $|\mathbb{B}\rangle=|10\rangle$, $|\mathbb{C}\rangle=|01\rangle$, and $|\mathbb{D}\rangle=|11\rangle$ with increasing energies $\varepsilon_{\mathbb{A}}=0\;\mu$eV, $\varepsilon_{\mathbb{B}}=\varepsilon_{\rm {a}}=10\;\mu$eV, $\varepsilon_{\mathbb{C}}=\varepsilon_{\rm {b}}=15\;\mu$eV and $\varepsilon_{\mathbb{D}}=\varepsilon_{\rm{a}}+\varepsilon_{\rm b}+U=37\;\mu$eV, respectively [Fig.~\ref{General Model} (b): Inset]. The $H_{\rm Bath}=\sum_{\lambda} H^{\lambda}_{\rm B}$, where $H_{\rm{B}}^{\lambda}=\sum_{k} (\epsilon_{k}-\mu_{\lambda})c^{{\lambda}\dagger}_{k} c^{\lambda}_{k}$ and $\lambda=\{\rm{S1,S2,D,I,C}\}$; $\mu_{\lambda}$ is the chemical potential of the $\lambda$-th bath, $\epsilon_k$ being the free-electron energy, and $c^{{\lambda}\dagger}_{k}(c^{{\lambda}}_{k})$ are the creation (annihilation) operators of the $k$-th bath mode. The tunneling Hamiltonian~\cite{wang2022cycleflux,gupt2024graph} is $H_{\rm T}=\sum_{\lambda\alpha} H^{\lambda\alpha}_{\rm T}$, where, $H_{\rm{T}}^{{{\lambda{\alpha}}}}=\hbar\sum_k[t^{{{\lambda{\alpha}}}}_k c^{{\lambda}\dagger}_{k} d_{{\alpha}}+t^{{\lambda{\alpha}}*}_kd_{{\alpha}}^{\dagger}c^{{\lambda}}_{k}]$; with tunneling amplitudes $t_{k}^{{\lambda{\alpha}}}$.
The joint state of the CQD is described by the Lindblad master equation (LME) $\frac{d}{dt}\rho_{\rm{s}}(t)=\sum_{\lambda}\mathcal{L}_{\lambda}[\rho_{\rm{s}}(t)]$ \cite{breuer2002book,strasberg2022quantum}, where the Lindbladians $\mathcal{L}_{\lambda}[\rho_{\rm{s}}(t)]$ are written for Ohmic-bath~[See Appendix~\ref{Appendix-A}]. At steady state, the LME is a system of four equations with diagonal elements satisfying  $\dot{\rho}_{\mathbb{i}}=0$ ($\mathbb{i=A,B,C,D}$). The LME is used to calculate the steady state heat current ($J^{\lambda}_{\rm{Q}}$) at the output lead as a function of the temperatures ($T_{\lambda}$) of the input leads, thereby determining the logic operations of the QTLG-circuits. The heat current $J^{\lambda}_{\rm{Q}}=J^{\lambda}_{\rm{E}}-\mu_{\lambda}J^{\lambda}_{\rm{N}}$, where energy current $J^{\lambda}_{\rm{E}}=\sum_{\{\varepsilon_{\mathbb{ij}}\}}{\varepsilon_{\mathbb{ij}}}\Gamma_{\mathbb{ij}}^{\lambda}$ and particle current $J^{\lambda}_{\rm{N}}=\sum_{\{\varepsilon_{\mathbb{ij}}\}}\Gamma_{\mathbb{ij}}^{\lambda}$ are given in terms of the net-excitation rate $\Gamma_{\mathbb{ij}}^{\lambda}\; (\mathbb{j}>\mathbb{i})$ induced by $\lambda$-th bath, and the energy difference $\varepsilon_{\mathbb{ij}}=\varepsilon_{\mathbb{j}}-\varepsilon_{\mathbb{i}}$ between the CQD states $|\mathbb{i}\rangle$ and $|\mathbb{j}\rangle$. We note $\Gamma_{\mathbb{ij}}^{\lambda}=\gamma[f^{\lambda}_{{\varepsilon}_{\mathbb{ij}}}\rho_{\mathbb{i}}-(1-f^{\lambda}_{{\varepsilon}_{\mathbb{ij}}})\rho_{\mathbb{j}}]=-\Gamma_{\mathbb{ji}}^{\lambda}$, where, $\gamma \simeq3$ GHz~\cite{thierschmann2015three} is taken as the bare tunneling-rate for all gates and  $f^{\lambda}_{{\varepsilon}_{\mathbb{ij}}}=[e^{({{\varepsilon}_{\mathbb{ij}}}-\mu_\lambda)/k_{\rm B}T_{\lambda}}+1]^{-1}$ is the Fermi-Dirac (FD) distribution of the bath-$\lambda$ [Appendix~\ref{Appendix-B}].

The basic building-block of a QTLG is a CQD placed between a Source (S) and the Drain (D) lead [Fig.~\ref{General Model}(b)], forming a thermal-diode~\cite{shuvadip2022univarsal}. Here, temperatures play a role analogous to electrical voltages, while \textit{forward} or \textit{reverse bias} are determined by the position of the Fermi-levels relative to the QD-states. For example, in Fig.~\ref{General Model}(b), the diode is in forward thermal-bias, with $\mu_{\rm D}$ lying within the occupied and unoccupied states of $\rm{QD_b}$, while $\mu_{\rm S}$ lies below those of $\rm{QD_a}$. Hence, heat flows from the hot S to the cold D (See Appendix~\ref{Appendix-C1} for detailed operation). The opposite alignment corresponds to reverse-bias, where no heat flows from S to D~\cite{shuvadip2022univarsal}, irrespective of the source temperature $T_{\rm S}$ [Appendix~\ref{Appendix-C2}]. In QTLGs, $T_{\rm S}$ of the S-lead(s) defines input-thermal-logic: $T_{\rm S} \lesssim T^{0}_{\rm{S}}\;(\sim 50 \pm 10$ mK), corresponds to logic-0, while $T_{\rm S} \gtrsim T^{1}_{\rm{S}}\;(\sim 200 \pm 10$ mK) corresponds to logic-1. In addition to the S and D, the general QTLG setup includes the Invert (I) and Control (C) leads, coupled to $\rm{QD}_{\rm a}$ and $\rm{QD}_{\rm b}$, respectively. Both I and C leads are held at high temperature [red in Fig.~\ref{General Model}(a)], with $T_{\rm{I(C)}} \gtrsim 230$ mK, analogous to a constant +5V bias in electronic logic-gates~\cite{millman1972integrated}. The I-lead enables inverting gate operations (e.g., NOT, NOR, and NAND), whereas the C-lead is employed to obtain AND gate operations. The output thermal-logic is determined by $J_{\rm Q}^{\lambda}$ ($\lambda=\rm{D}$ or $\rm{I}$) measured at  D or I (specifically for inverting gates). If $|J_{\rm Q}^{\rm{D}}|,J_{\rm Q}^{\rm{I}} \lesssim J_{\rm Q}^{0} (\sim 65 \pm 10$ aW): output logic-0; and if $|J_{\rm Q}^{\rm{D}}|,J_{\rm Q}^{\rm{I}} \gtrsim J_{\rm Q}^{1} (\sim 100 \pm 10$ aW): output logic-1. Here, $J_{\rm Q}^{\rm D}$ is negative since it enters the D-lead, while $J_{\rm Q}^{\rm I}$ is positive since it leaves the I-lead. The logic thresholds ($J_{\rm Q}^{0}$ and $J_{\rm Q}^{1}$) are chosen based on realistic experimental parameters. A striking one-to-one correspondence between the proposed QTLG and electronic logic-gates becomes apparent as we proceed to discuss the individual gate operations.

\begin{figure}[t]
\centering    
\includegraphics[width=\columnwidth]{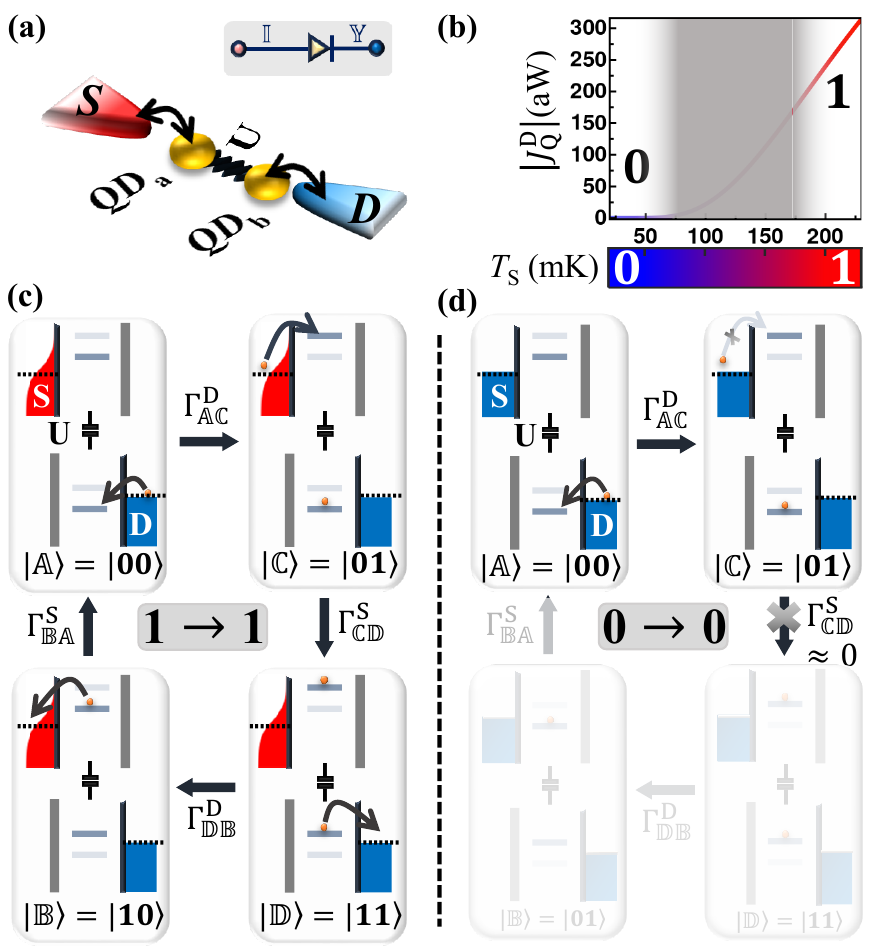}
\caption{(a) CQD setup for the QTBG, with its equivalent electrical-diode circuit (Inset).  (b) $|J_{\rm Q}^{\rm D}|$ w.r.t input-temperature $T_\mathrm{S}$, for forward-bias $\mu_{\rm{D}}=+21\;\mu\rm{eV}$ and $\mu_{\rm{S}}=-21\;\mu\rm{eV}$. (c) QTBG operational cycle for \textit{1}  $\rightarrow$ \textit{1} and (d) \textit{0}  $\rightarrow$ \textit{0}; The net transition-cycle runs in $\circlearrowright$ direction; while the anti-clockwise cycle $(\circlearrowleft)$ is suppressed, as no hot-lead is coupled to $\rm{QD_b}$~[See Appendix-~\ref{Appendix-C2} for details].}
    \label{Buffer Gate}
\end{figure}

\textit{Buffer Gate:} We begin with the simplest thermal logic gate, the \textit{quantum thermal buffer gate} (QTBG), in which the output replicates the input logic, and is equivalent to a thermal-diode configured in forward-bias condition. The thermal-circuit of QTBG, shown in Fig.~\ref{Buffer Gate}(a), consists of a single-input (S) coupled to $\rm{QD_a}$ and an output (D) coupled to $\rm{QD_b}$. The S can be hot (logic-1) or cold (logic-0), while the D remains cold at the minimum temperature $\sim 20$ mK (true for all QTLGs). For logic-1, a thermal-bias drives heat from S to D, enforcing unidirectional heat-flow [Fig.~\ref{Buffer Gate}(b)], analogous to current in an electronic-diode [Fig.~\ref{Buffer Gate}(a): Inset].

To understand the operation of the QTBG, we analyze the heat-flow cycle for input logic-1 [Fig.~\ref{Buffer Gate}(c)], where $T_{\rm S} \gtrsim 200$ mK. The hot-source is shown by a broadened-FD-distribution (red), while the cold-drain has a sharp FD-distribution (blue). With forward-bias, the cycle proceeds only in clock-wise direction ($\circlearrowright$: $\mathbb{A}$$\rightarrow$$\mathbb{C}$$\rightarrow$$\mathbb{D}$$\rightarrow$$\mathbb{B}$$\rightarrow$$\mathbb{A}$) and begins with the thermodynamically favorable tunneling of an electron from the D to $\rm{QD_b}~$~(Appendix~\ref{Appendix-C2}), transferring an amount of heat-current equal to $\varepsilon_{\rm b}\Gamma_{\mathbb{AC}}^{\rm D}$ and changing the CQD state from $|\mathbb{A}\rangle$ to $|\mathbb{C}\rangle$ (panel-1). Due to the Coulomb-interaction, the $\rm{QD_a}$-level shifts to its high-energy state $\varepsilon_{\rm a}+U$ (panel-2), requiring an additional amount of energy $U$ for tunneling. Hot electrons in S (at the tail of the FD) overcome this barrier and tunnel into $\rm{QD_a}$, driving the heat-flow from S and changing the state to $|\mathbb{D}\rangle$. This transfer of electrons into $\rm{QD_a}$, in turn, raises the $\rm{QD_b}$-level to $(\varepsilon_{\rm b}+U)$ (panel-3), allowing the electron in $\rm{QD_b}$ to tunnel back to D with an average amount of heat-current $-(\varepsilon_{\rm b}+U)\Gamma_{\mathbb {DB}}^{\rm D}$, changing the state to $|\mathbb{B}\rangle$. Finally, this transition again brings the $\rm{QD_a}$-level back to $\varepsilon_{\rm a}$ (panel-4) and subsequently the electron in $\rm{QD_a}$ tunnels back to S, restoring the initial-state $|\mathbb{A}\rangle$. The system thus completes a full-cycle, producing a heat current $J_{\rm Q}^{\rm D}=-U\Gamma_{\circlearrowright}^{\rm Buf}$ entering into D, with steady state net clock-wise cycle-rate $\Gamma_{\circlearrowright}^{\rm Buf} > 0$~(Appendix~\ref{Appendix-C2}). The output-current $|J_{\rm Q}^{\rm D}|$ is measurable and significantly higher than  $J_{\rm Q}^{\rm 1}$, corresponding to logic-1.

For input logic-0, $T_{\rm S} \lesssim 50$ mK, and the S is described by a sharp FD-distribution (blue), similar to the cold D. The cycle again begins with electron tunneling from D to $\rm{QD_b}$ (panel-1), changing the state from $|00\rangle$ to $|01\rangle$ and raising the $\rm{QD_a}$ level to $\varepsilon_{\rm a}+U$ (panel-2). However, electrons in the cold-source lack sufficient thermal energy to overcome the Coulomb-barrier and tunnel into $\rm{QD_a}$. Consequently, the cycle cannot proceed [Fig.~\ref{Buffer Gate}(d)], and the net heat-current at the drain remains negligibly small, yielding output logic-0. The calculated output current $|J_{\rm Q}^{\rm D}|$ as a function of $T_{\rm S}$ [Fig.~\ref{Buffer Gate}(b)] clearly displays the two regions of the QTBG truth-table: for $T_{\rm S} \lesssim 50$ mK, $|J_{\rm Q}^{\rm D}| \ll 20$ aW ($\textit{0} \rightarrow \textit{0}$); and $T_{\rm S} \gtrsim 200$ mK, $|J_{\rm Q}^{\rm D}| \gg 100$ aW  ($\textit{1} \rightarrow \textit{1}$).

\begin{figure}[t]
\centering    
\includegraphics[width=\columnwidth]{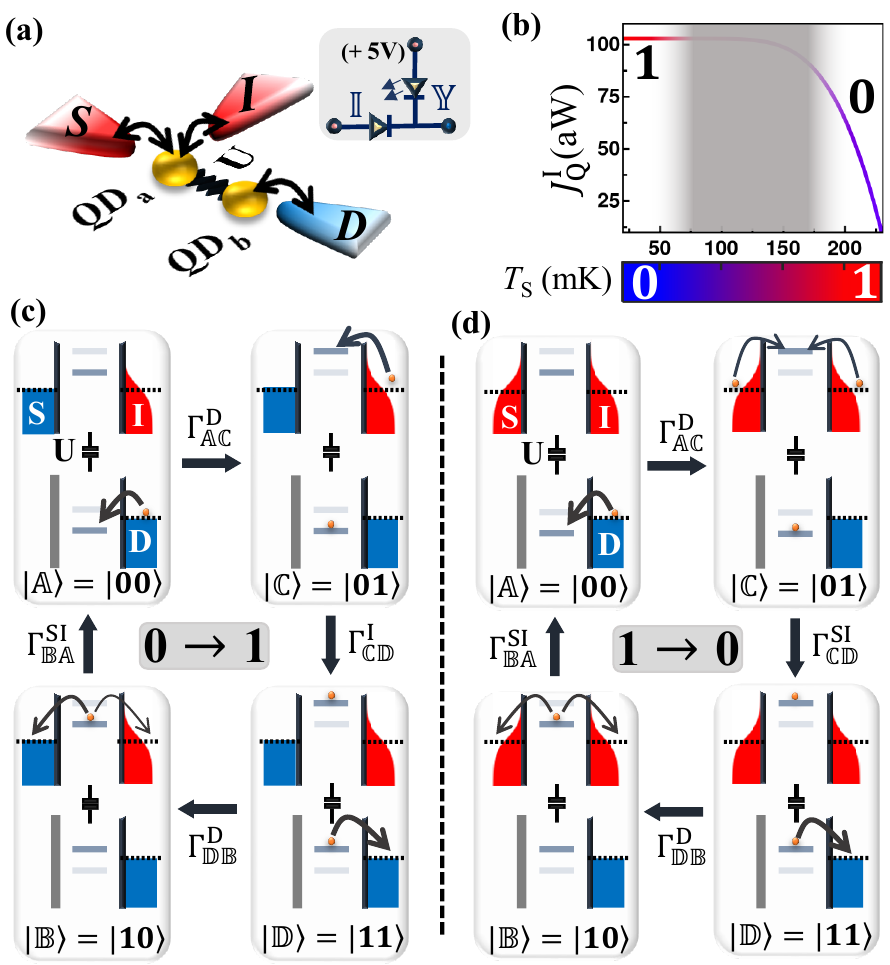}
\caption{(a) QTNG setup and its electrical circuit (Inset). (b) $J_{\rm Q}^{\rm I}$ w.r.t $T_\mathrm{S}$ for the same parameters as buffer gate, with additional parameter $\mu_{\rm{I}}=-21\;\mu\rm{eV}$, kept below the $\rm{QD_a}$ level, maintaining forward-bias. (c) QTNG operational cycle for \textit{0}  $\rightarrow$ \textit{1} and (d) \textit{1}  $\rightarrow$ \textit{0}. Like in QTBG, the net transition-cycle runs in $\circlearrowright$ direction; while the anti-clockwise cycle is suppressed~(Appendix~\ref{Appendix-C2}). Here, the widths of the arrows indicate the relative strength of the heat-flow.}
    \label{NOT Gate}
\end{figure}

\textit{NOT Gate:} A \textit{quantum thermal NOT gate} (QTNG) can be developed from the QTBG by adding an I-lead to $\rm{QD_a}$ [Fig.~\ref{NOT Gate}(a)], enabling logic inversion analogous to an electronic NOT gate (Inset). A thermal-bias is applied from S to D, while a constant high-temperature $T_{\rm I} \gtrsim 230$ mK is maintained at the I-lead (similar to constant +5V in electronic NOT gate~\cite{millman1972integrated}). The thermal current measured in the I-lead defines the output-logic. For a cold-input ($T_{\rm S} \lesssim 50$ mK, logic-0), negligible heat flows from S to D, but the strong thermal-bias at I-lead drives a large output heat-current $J_{\rm Q}^{\rm I}$ to both S and D. This current exceeds threshold, yielding output logic-1 [Fig.~\ref{NOT Gate}(b)]. For a hot input ($T_{\rm S}\gtrsim 200$ mK, logic-1), both I and S equally contribute to heat-flow to D, lowering $J_{\rm Q}^{\rm I}$ below $J_{\rm Q}^{\rm 0}\sim65$ aW and producing output logic-0.

The truth-table for QTNG is illustrated in Fig. \ref{NOT Gate}(c,d). For thermal input-logic-0 [Fig. \ref{NOT Gate}(c)], the heat-flow cycle begins with tunneling from D to the $\rm {QD_b}$ (panel-1), which shifts the $\rm{QD_a}$ level to its high-energy state $\varepsilon_{\rm {a}}+U$ (panel-2). The cold S-lead cannot support electron tunneling into the shifted-level $\varepsilon_{\rm a} + U$. However, hot electrons from the I-lead can tunnel into $\mathrm{QD}_{\rm a}$, causing a heat-flow and shifting the electron in $\rm{QD_b}$ level to the higher energy-state $\varepsilon_{\rm b}+U$ (panel-3). Subsequently, the electron in $\rm{QD_b}$ tunnels back to D, while the electron in $\rm{QD_a}$ mostly tunnels into the S-lead (panel-4), thereby completing the heat-flow cycle. This produces a heat current from the I-lead that exceeds $J_{\rm Q}^{\rm 1}$, implementing the $\textit{0} \rightarrow \textit{1}$ operation of the QTNG (See Appendix~\ref{Appendix-C3} for details).

For the $\textit{1} \rightarrow \textit{0}$  operation, the heat-flow cycle also completes. However, now as S is equally hot as I (input logic-1), both contribute to the heat-cycle and as a result the strength of heat-flow in and out of the I is comparable (indicated by the equal-width arrows to/from I) [Fig. \ref{NOT Gate}(d)], and hence the net output heat-current $J_{\rm Q}^{\rm I}$ become negligibly small ($<65$ aW), corresponding to output logic-0. The calculated $J_{\rm Q}^{\rm I}$ versus $T_{\rm S}$ is shown in Fig. \ref{NOT Gate}(b), from which the NOT gate truth-table follows.

\begin{figure}[t]
\centering    
\includegraphics[width=1\columnwidth,height=3.0in]{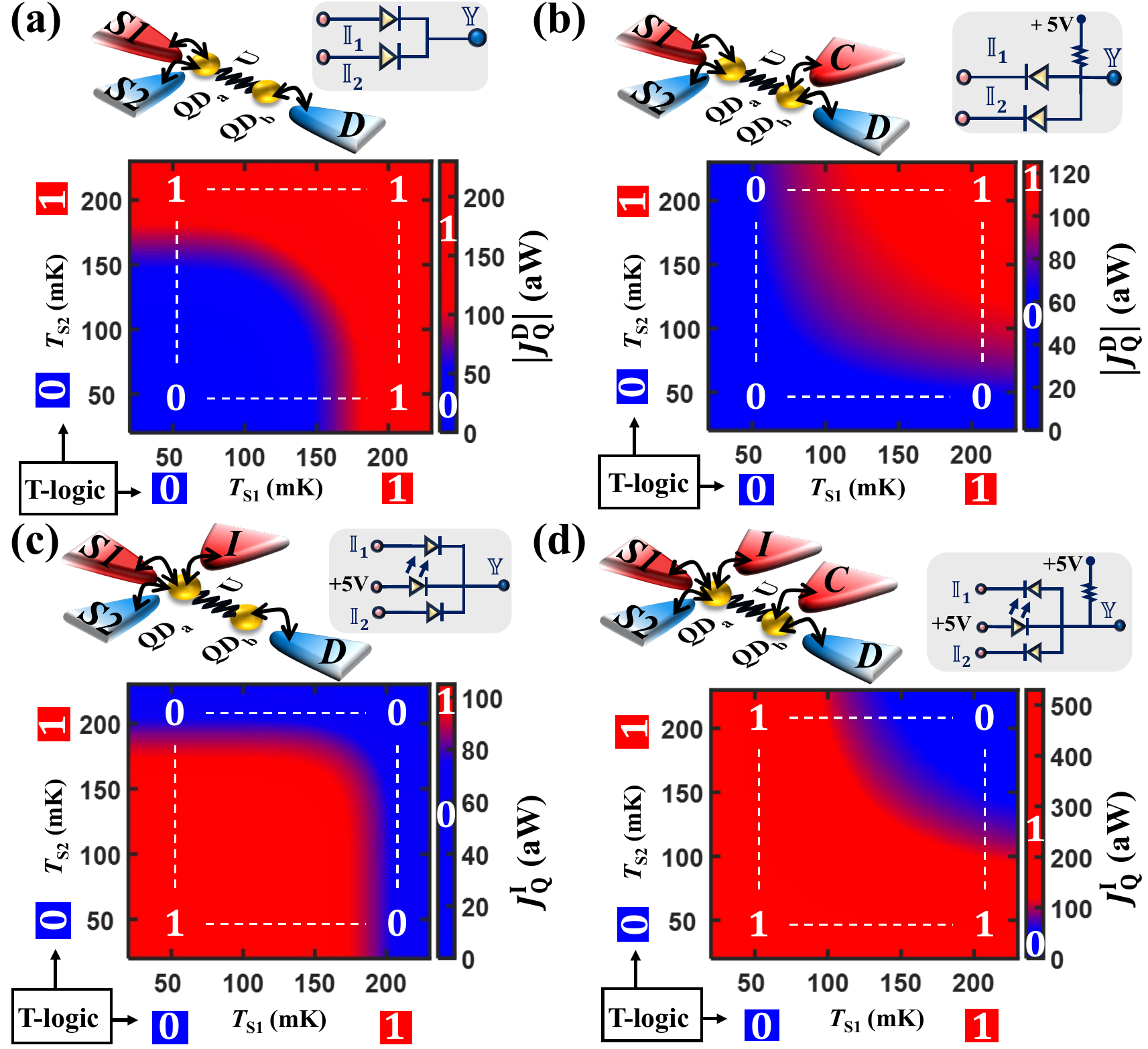}
    \caption{Double-input QTLG setup and the corresponding 2D thermal-current plot w.r.t. the input T-logic for all four gates. Common parameters: $\gamma=3\;\rm{GHz}$,
    $\varepsilon_{\rm{a}}=10\;\mu\rm{eV}$, $\varepsilon_{\rm{b}}=15\;\mu\rm{eV}$, $\rm{U}=12\;\mu\rm{eV}$,
    $T_{\rm{D}}=20~$mK. Additional parameters for (a) OR gate $\mu_{\rm{S1}}=\mu_{\rm{S2}}=-25\;\mu\rm{eV}$,
    $\mu_{\rm{D}}=21\;\mu\rm{eV}$; (b) AND gate: $\mu_{\rm{S1}}=\mu_{\rm{S2}}=16\;\mu\rm{eV}$, $\mu_{\rm{C}}=-72\;\mu\rm{eV}$,
    $\mu_{\rm{D}}=14\;\mu\rm{eV}$; (c) NOR gate: $\mu_{\rm{S1}}=\mu_{\rm{S2}}=-95\;\mu\rm{eV}$, $\mu_{\rm{I}}=-90\;\mu\rm{eV}$,
    $\mu_{\rm{D}}=21\;\mu\rm{eV}$; and (d) NAND gate:
$\mu_{\rm{S1}}=\mu_{\rm{S2}}=16\;\mu\rm{eV}$, $\mu_{\rm{C}}=-16\;\mu\rm{eV}$, $\mu_{\rm{I}}=16\;\mu\rm{eV}$,
    $\mu_{\rm{D}}=9\;\mu\rm{eV}$. The equivalent electrical circuit for each gate is shown in the top inset, and output logic states are indicated on the color plots. 
        }
    \label{All Gates}
\end{figure}

Now, we construct double-input thermal logic gates, beginning with the OR and AND gates and their corresponding NOT operations, NOR and NAND. The proposed thermal-OR gate is shown in Fig.~\ref{All Gates}(a) (top) [inset: electrical equivalent], with two inputs S1, S2, and one output D, forming two thermal-diodes operating in parallel, using each S and the common D [Appendix-\ref{Appendix-C4}]. If either input satisfies $T_{\rm S1(S2)} \gtrsim 200$ mK (logic-1), $|J_{\rm Q}^{\rm D}| \gg 100$ aW, giving output logic-1. Only when both inputs satisfy $T_{\rm S1(S2)} \lesssim 50$ mK (logic-0), $|J_{\rm Q}^{\rm D}| \ll 20$ aW, resulting output logic-0. The calculated $|J_{\rm Q}^{\rm D}|$ in the $T_{\rm S1}$–$T_{\rm S2}$ plane [Fig.~\ref{All Gates}(a), bottom], generates the thermal-OR gate truth-table.

A thermal-AND gate is obtained by adding a C-lead with the OR gate [Fig.~\ref{All Gates}(b), top], keeping its temperature fixed at $T_{\rm C} \gtrsim 230$ mK, in analogy to fixed +5V in an electronic AND gate (Inset). The chemical potentials of the leads are chosen in such a way that the two thermal-diodes are
connected in reverse-bias w.r.t the \textit{cold} D, while they are in forward-bias w.r.t the \textit{hot} C, like in an electronic AND gate~\cite{millman1972integrated}. Since a hot C-lead is coupled to $\rm{QD_b}$, the heat-flow cycle can proceed in both directions $\circlearrowright$ and $\circlearrowleft$ ($\circlearrowleft$:~$\mathbb{A}$$\rightarrow$$\mathbb{B}$$\rightarrow$$\mathbb{D}$$\rightarrow$$\mathbb{C}$$\rightarrow$$\mathbb{A}$). When both inputs are 0 ($T_{\rm S1(S2)} \lesssim 50$ mK), the cold sources cannot populate $\mathrm{QD_a}$, blocking the $\circlearrowright$ cycle, but $\circlearrowleft$ cycle proceeds with a negligible heat-current below the threshold $J_{\rm Q}^{0}$, giving output logic-0 [See ~Appendix~\ref{Appendix-C5}~for~details]. When one input is 0, and the other is 1, both cycles can proceed; however, due to the position of the $\mu_{\rm C}$ and $\mu_{\rm D}$, $|J_{\rm Q}^{\rm D}|$ still remains significantly below threshold, and the output logic stays 0 [Appendix~\ref{Appendix-C5}]. When both input thermal logics are 1, the two hot sources jointly drive the heat-flow, causing $|J_{\rm Q}^{\rm D}|$ to exceed $J_{\rm Q}^{1}$. Consequently, the output logic becomes 1. The color plot of $|J_{\rm Q}^{\rm D}|$ vs $T_{\rm S1}$, $T_{\rm S2}$ [Fig. \ref{All Gates}(b), bottom] directly reproduces the truth-table of a thermal-AND gate.

A thermal-NOR gate can be implemented by combining the thermal-OR gate with the thermal-NOT gate [Fig. \ref{All Gates}(c), top]. Unlike the OR gate, the hot I-lead actively drives the cycle for the $(0,0)$ case, producing a detectable output (logic-1) when measured at the I-lead. However, if any input is 1, the heat-flow from the I-lead is suppressed, causing the output current to fall well below $J_{\rm Q}^{0}$ and yielding output logic-0. All logic operations follow directly from the heat-flow cycle under different input-logic conditions [Appendix~\ref{Appendix-C6}]. As no hot-lead is attached to $\rm{QD_b}$, the thermal NOR gate also operates via a net $\circlearrowright$ cycle. The calculated $J_{\rm Q}^{\rm I}$  [Fig. \ref{All Gates}(c), bottom] reproduces the complete NOR gate truth-table.

Similar to the NOR gate, a thermal-NAND gate is realized by combining the thermal-AND and thermal-NOT gates [Fig. \ref{All Gates}(d), top]. Here, unlike the AND gate, if any input is 0, the I-lead actively drives the cycle, and owing to the position of $\mu_{\rm C}$ and $\mu_{\rm I}$, the output heat current $J_{\rm Q}^{\rm I}$ exceeds $J_{\rm Q}^{\rm 1}$, giving output logic-1. However, for both input-1, two S-leads and I-lead participate equally in the heat-cycle, and thus $J_{\rm Q}^{\rm I}$  is reduced below $J_{\rm Q}^{\rm 0}$, resulting in logic-0 [Appendix~\ref{Appendix-C7}]. The calculated heat-current is plotted in the $T_{\rm S1}$–$T_{\rm S2}$ plane [Fig. \ref{All Gates}(d), bottom], producing the complete NAND gate truth-table.

\begin{figure}[t]
\centering    
\includegraphics[width=\columnwidth,height=1.35in]{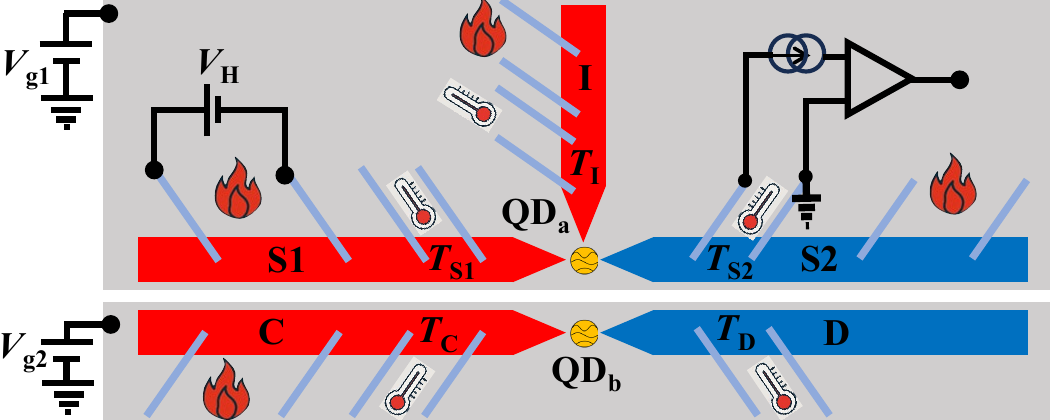}
    \caption{Schematic of the proposed device, with two QD-junctions, integrated with a local heater and thermometers (cyan). The QD levels can be tuned by the corresponding back-gates.}
    \label{Exp_device}
\end{figure}

An experimental proposal for the QTLG is presented in \mbox{Fig. \ref{Exp_device}}. The proposed device consists of two QD-junctions, with the nanoparticle QDs in the middle (yellow circles) tunnel-coupled to two or more metallic leads. Here, $\rm{QD_a}$ is coupled with three leads, serving as S1 (red), S2 (blue) and I (red), while $\rm{QD_b}$ is coupled to two leads, serving as the C (red) and D (blue) leads of the QTLG. Each metallic lead is connected to several superconducting contacts (cyan), which allow for heating (using Joule dissipation) and local thermometry with the use of superconducting junctions~\cite{Dubos2001Josephson}. The QD levels can be tuned by the gate voltages $V_{\rm{g1}}$ and $V_{\rm{g2}}$ to configure the forward/reverse bias condition of the thermal-diodes. The input-thermal-logic is determined by the heating voltage ($V_{\rm{H}}$) in S1 and S2: logic-1 for $V_{\rm{H}} > 0$, such that the local-temperature raises above $T_{\rm S}^{1}\sim200$ mK, and logic-0 for $V_{\rm{H}}=0$, with local-temperature lies below $T_{\rm S}^{0}\sim50$ mK. The I and C-leads are heated up to $\sim230$ mK using a constant $V_{\rm{H}}$, while the D-lead remains cold at the cryostat's base-temperature ($\sim10-20$ mK). With the help of local-thermometers based on superconducting-normal-metal junctions, one can measure the local-temperature of the leads and hence extract the heat-flow rate at the D (or I) and thereby determine the output thermal-logic. Thus, with the same device, one can build several QTLGs by properly selecting the input, output, and control leads. The fabrication of such QD-junctions integrated with local-heater and thermometer is already well-developed in experiments, thanks to the well-established nano-fabrication techniques \cite{dutta2020single,dutta2017thermal, dutta2019direct}. The parameters in our calculations are motivated by the currently available experimental data in such CQD-junction devices, e.g., $\gamma \sim$ 3 GHz, $U \sim$ 12 $\mu$eV \cite{thierschmann2015three}, and lowest measurable heat-current $J_{\rm{Q}}^{1} \sim$ 100 aW~\cite{dutta2020single}, etc. Thus, the experimental realization of the QTLGs with the proposed device is truly achievable.

In conclusion, the work describes a new concept for implementing thermal logic gates in quantum systems, involving a CQD system tunnel-coupled to Fermionic-leads. The formalism and analysis include realistic experimental parameters, along with an experimental device proposal based on well-established nano-fabrication techniques; thus, the proposed thermal logic device can be realized in future experiments. Moreover, the striking one-to-one correspondence between the QTLGs and conventional classical logic-circuits based on electronic-diodes, in turn, validates the proposal. The proposed device, owing to its components operating in the quantum regime, offers a wide range of opportunities for implementing thermal logic operations in futuristic quantum computation circuits.

We acknowledge Clemens B. Winkelmann for critical reading, comments, and stimulating discussions on the manuscript. SG and AG acknowledge the support from PMRF and IIT Kanpur, respectively. BD acknowledges the support funded by ANRF under the ANRF-PMECRG grant with the project number ANRF/ECRG/2024/003114/PMS, support from the IIT Indore YFRSG grant, and support from PAIR Network of SAKSHAM funded by ANRF, India, through the grant ANRF/PAIR/2025/000018/PAIR, administered and mentored by IIT Indore.

\onecolumngrid 
\appendix

\section{Lindblad Master Equation and its Steady-state solution}\label{Appendix-A}

To derive the quantum master equation for our quantum thermal logic gates (QTLG) model, we start with the von-Neumann equation in the interaction picture for the total density matrix $\rho_{\rm{tot}}$ of the system and bath combined
\begin{equation}\label{A2}
    \frac{d}{dt}\rho_{\rm{tot}}=-\frac{i}{\hbar}[H_{\rm{T}}(t),\rho_{\rm{tot}}(t)].
\end{equation}
Here, $H_{\rm{T}}$ is the tunneling Hamiltonian $H_{\rm{T}}=\sum_{{\lambda}} H_{\rm{T}}^{{\lambda}}$, where $\lambda = \{\rm{S1, S2, D, I, C} \}$. The $H_{\rm{T}}^{{\lambda}}$ represents the tunneling Hamiltonian between the $\alpha$-th ($\alpha=\rm{a,b}$) quantum dot (QD), and a $\lambda$-th bath and is given by $H_{{\rm{T}}}^{\lambda}\equiv H_{{\rm{T}}}^{\alpha\lambda}=\hbar\sum_k[t^{\alpha\lambda}_k c_{\lambda k}^\dagger d_{\alpha}+t^{\alpha\lambda*}_ kd_{\alpha}^{\dagger}c_{\lambda k}]$, where $t_{k}^{\alpha\lambda}$ represents the tunneling amplitude. Integrating the above equation, and	taking a trace over the bath degrees of freedom under the Born-Markov approximation~\cite{breuer2002book,gupt2022PRE,shuvadip2022univarsal}, one obtains
\begin{equation}\label{A4}
\begin{split}
\dot{\rho_{\rm{s}}}(t)=\frac{1}{(i\hbar)^2}\sum_{\lambda, \lambda'}
\int_0^\infty dt^\prime\ {\rm Tr}_{\lambda}[H_{\rm{T}}^{\lambda}(t),[H_{\rm{T}}^{\lambda'}(t-t^\prime),\rho_{\rm{s}}(t)\otimes \prod_{\lambda}\rho_{\lambda}]], \quad \text{where}, \quad \lambda, \lambda^{\prime}= \{\rm{S1, S2, D, I, C} \}.
\end{split}
\end{equation}
${\rm Tr}_{\lambda}$ refers to the trace over each bath degree of freedom and ${\rm Tr}_{\lambda}\{ \rho_{\rm{tot}}(t)\}=\rho_{\rm{s}}(t)$, being the reduced density opeartor of the system. Here, $\rho_{\lambda}$ is the initial density operator of the $\lambda$-th bath which is assumed to be always in equilibrium, and we use the fact that ${\rm Tr}_{\lambda}\{{c}_{\lambda}(t)\rho_{\lambda}\}=0={\rm Tr}_{\lambda}\{{c}^\dagger_{\lambda}(t)\rho_{\lambda}\}$
and ${\rm Tr}_{\lambda}\{ [H^{\lambda}_{\rm{T}}(t),[H^{\lambda^{\prime}}_{\rm{T}}(t-t'),{\rho_{\rm{s}}}(t)\otimes \prod_{\lambda}\rho_{\lambda}]]\}=0;{\lambda}\ne{\lambda^{\prime}}$. We further assume that ${\rm Tr}_{\lambda}[H_{\rm{T}}(t),\rho_{\rm{tot}}(0)]=0$. Finally, eliminating the high-frequency oscillating terms by the standard procedure of secular approximation, we arrive at the Lindblad master equation (LME) of the following form
\begin{equation}\label{A6}
\dot\rho_{\rm{s}}=\sum_{\lambda}\mathcal{L}_{\lambda}[\rho_{\rm{s}}],
\end{equation}
where, the Lindbladian $\mathcal{L}_{\lambda}[\rho_{\rm{s}}]$ is given by
\begin{eqnarray}\label{A7}
\mathcal{L}_{\lambda}[\rho_{\rm{s}}]=\sum_{\{\omega_{\alpha}\}>0}&\mathcal{G}_{{\lambda}}(\omega_{\alpha})&\left[d_{\alpha}^\dagger(\omega_{\alpha})\rho_{\rm{s}} d_{\alpha}(\omega_{\alpha})-\frac{1}{2}{\{\rho_{\rm{s}}}, d_{\alpha}(\omega_{\alpha})d_{\alpha}^\dagger(\omega_{\alpha})\}\right]
\nonumber\\
+&\mathcal{G}_{{\lambda}}(-\omega_{\alpha})&
\left[d_{\alpha}(\omega_{\alpha})\rho_{\rm{s}} d_{\alpha}^\dagger(\omega_{\alpha})-\frac{1}{2}{\{\rho_{\rm{s}}}, d_{\alpha}^\dagger(\omega_{\alpha})d_{\alpha}(\omega_{\alpha})\}\right].
\end{eqnarray}
Here $\{\omega_\alpha\}$ represent the energy associated with the transition between ${\rm QD_\alpha}$ with its coupled leads. In the present QTLG model, there are four allowed transitions and each lead guides two transitions each. The leads $\mathrm{S1,S2,I}$ drive the transitions between $\rvert\mathbb{A}\rangle\leftrightarrow\rvert\mathbb{B}\rangle$ and $\rvert \mathbb{D}\rangle\leftrightarrow\rvert \mathbb{C}\rangle$, while the $\mathrm{D,C}$ leads control the $\rvert \mathbb{A}\rangle\leftrightarrow\rvert \mathbb{C}\rangle$ and $\rvert \mathbb{D}\rangle\leftrightarrow\rvert\mathbb{B}\rangle$ transitions. To derive the forms of the Lindbladians, we express the creation and annihilation operators in terms of the system eigenstates in the following forms
\begin{equation}\label{A9}
\begin{split}
\sum_{\{\omega_{\rm{a}}\}}d^\dagger_{\rm{a}}(\omega_{\rm{a}})=d_{\rm{a}}^\dagger(\varepsilon_\mathbb{AB})+d^\dagger_{\rm{a}}(\varepsilon_\mathbb{CD})=|\mathbb{B}\rangle \langle\mathbb{A}|+|\mathbb{D}\rangle \langle\mathbb{C}|;\quad&\quad \sum_{\{\omega_{\rm{a}}\}}d_{\rm{a}}(\omega_{\rm{a}})=d_{\rm{a}}(\varepsilon_\mathbb{BA})+d_{\rm{a}}(\varepsilon_\mathbb{DC})=|\mathbb{A}\rangle \langle\mathbb{B}|+|\mathbb{C}\rangle \langle\mathbb{D}|,
\\\sum_{\{\omega_{\rm{b}}\}}d_{\rm{b}}^\dagger(\omega_{\rm{b}})=d_{\rm{b}}^\dagger(\varepsilon_\mathbb{AC})+d_{\rm{b}}^\dagger(\varepsilon_\mathbb{BD})=|\mathbb{C}\rangle \langle\mathbb{A}|+|\mathbb{D}\rangle \langle\mathbb{B}|;\quad&\quad
\sum_{\{\omega_{\rm{b}}\}}d_{\rm{b}}(\omega_{\rm{b}})=d_{\rm{b}}(\varepsilon_\mathbb{CA})+d_{\rm{b}}(\varepsilon_\mathbb{DB})=|\mathbb{A}\rangle \langle\mathbb{C}|+|\mathbb{B}\rangle \langle\mathbb{D}|.
\end{split}
\end{equation}
Furthermore, temperature-dependent bath spectral functions in Eq.~\eqref{A7} are defined as
\begin{equation}\label{A8}
\mathcal{G}_{{\lambda}}(\omega_{\alpha})=\gamma_{\lambda}(\omega_\alpha)f^+_{\lambda}(\omega_\alpha);\quad \text{and} \quad  \mathcal{G}_{{\lambda}}(-\omega_{\alpha})=\gamma_{\lambda}(\omega_\alpha) f^-_{\lambda}(\omega_\alpha),
\end{equation}
where $\gamma_\lambda(\omega_\alpha)$ is the bare tunneling rate between the reservoir $\lambda$ and the coupled ${\rm QD_\alpha}$, which can be calculated using Fermi's golden rule, as 
$\gamma_\lambda(\varepsilon_{\mathbb{ij}})=2\pi \sum_{k} |t_{k}^{\alpha\lambda}|^2 \delta\big(\varepsilon_{\mathbb{ij}}-\epsilon_{k}^{\lambda}\big)$ for $\omega_\alpha=\varepsilon_{\mathbb{ij}}$ and is abbreviated as $\gamma_\lambda$ for brevity. For simplicity of our analytical calculations, we later assume that all $\gamma_{{\lambda}}$ are equal. The $f^{\pm}_{\lambda}(\omega_\alpha)$ in Eq.~\eqref{A8} are the Fermi distribution functions  calculated via $f_{\lambda}^+\big(\omega_\alpha=\varepsilon_{\mathbb{ij}}\big)=\Tr_{\lambda}\big(c^\dagger_{\lambda} c_{\lambda} \rho_{\lambda}\big)$ and $f_{\lambda}^-\big(\omega_\alpha=\varepsilon_{\mathbb{ij}}\big)=\Tr_{\lambda}\big(c_{\lambda} c^\dagger_{\lambda} \rho_{\lambda}\big)$. The time evolution of the occupation probabilities of the reduced density matrix $\rho_\mathbb{i}=\langle\mathbb{i}|\rho_{\rm{s}}(t)|\mathbb{i}\rangle\; (\mathbb{i=A,B,C,D})$, is obtained using the LME [Eq.~\eqref{A7}], which leads to a simple loss-gain type rate equation for the steady-state occupation probabilities~\cite{breuer2002book,strasberg2022quantum}:
\begin{eqnarray}\label{rate-eqns}
\dot{\rho}_\mathbb{A}&=&\sum_{\lambda}\Gamma^{{\lambda}}_\mathbb{BA}-\sum_{\lambda}\Gamma^{{\lambda}}_\mathbb{AC}=0;\quad \nonumber
\dot{\rho}_\mathbb{B}=\sum_{\lambda}\Gamma^{{\lambda}}_\mathbb{AB}-\sum_{\lambda}\Gamma^{{\lambda}}_\mathbb{BD}=0; 
\\\dot{\rho}_\mathbb{C}&=&\sum_{\lambda}\Gamma^{{\lambda}}_\mathbb{DC}-\sum_{\lambda}\Gamma^{{\lambda}}_\mathbb{CA}=0;\quad 
\dot{\rho}_\mathbb{D}=\sum_{\lambda}\Gamma^{{\lambda}}_\mathbb{CD}-\sum_{\lambda}\Gamma^{{\lambda}}_\mathbb{DB}=0.
\end{eqnarray} 
From Eq.~\eqref{rate-eqns}, we find all $\Gamma^{\lambda}_\mathbb{ij}$ in the steady state are equal and are denoted by clockwise ($\Gamma_{\circlearrowright}$) or anti-clockwise ($\Gamma_{\circlearrowleft}$) transition rates:  
\begin{equation}\label{gamma-appendix}
\begin{split}
\sum_{\lambda}\Gamma^{{\lambda}}_\mathbb{AC}=\sum_{\lambda}\Gamma^{{\lambda}}_\mathbb{CD}=\sum_{\lambda}\Gamma^{{\lambda}}_\mathbb{DB}=\sum_{\lambda}\Gamma^{{\lambda}}_\mathbb{BA}\equiv\Gamma_{\circlearrowright};\qquad\qquad
\sum_{\lambda}\Gamma^{{\lambda}}_\mathbb{AB}=\sum_{\lambda}\Gamma^{{\lambda}}_\mathbb{BD}=\sum_{\lambda}\Gamma^{{\lambda}}_\mathbb{DC}=\sum_{\lambda}\Gamma^{{\lambda}}_\mathbb{CA}\equiv\Gamma_{\circlearrowleft},
\end{split}
\end{equation}
and satisfies the relation  $\Gamma_{\circlearrowright}=-\Gamma_{\circlearrowleft}$. In Eqs.~\eqref{rate-eqns}-\eqref{gamma-appendix}, the net transition rate $\Gamma^{\lambda}_\mathbb{ij}$ from state $|\mathbb{i}\rangle \rightarrow |\mathbb{j}\rangle$ induced by reservoir $\{\lambda\}$ is given by $\Gamma_\mathbb{ij}^{\lambda}=
{\rm{k}}_\mathbb{ij}^{\lambda}\rho_\mathbb{i}-{\rm{k}}_\mathbb{ji}^{\lambda}\rho_\mathbb{j}$; where, ${{\rm{k}}}_\mathbb{ij}^{{\lambda}}=\gamma_{{\lambda}}f^{\pm}_{{\lambda}}(\omega_\mathbb{ij})$. The Fermi distribution $f_{\lambda}(\varepsilon_\mathbb{ij})$ associated with the corresponding transition energy $\varepsilon_\mathbb{ij}=\varepsilon_\mathbb{j}-\varepsilon_\mathbb{i}$ satisfies the following relation $f_{\lambda}(\varepsilon_\mathbb{ij})=1-f_{\lambda}(\varepsilon_\mathbb{ji})=\big[{1+\exp {\big(\beta_{\lambda}(\varepsilon_\mathbb{ij}-\mu_{\lambda}}})\big)\big]^{-1}$~\cite{shuvadip2022univarsal,tesser2022heat}; $\beta_{\lambda}$ being the inverse temperature $\beta_{\lambda}=1/k_BT_{\lambda}$. Among the four allowed transitions, $\varepsilon_\mathbb{ij}$ are: $\varepsilon_{\mathbb{AB}}=\varepsilon_{\rm{a}}$, $\varepsilon_{\mathbb{AC}}=\varepsilon_{\rm{b}}$, $\varepsilon_{\mathbb{CD}}=\varepsilon_{\rm{a}}+U$, and $\varepsilon_{\mathbb{BD}}=\varepsilon_{\rm{b}}+U$. For brevity, we denote the Fermi distributions associated with the transition energy $\varepsilon_\alpha$ as $f_{\lambda}^1$, and $f_{\lambda}^2$ with transition energy $\varepsilon_\alpha+U$. Now, in order to determine the steady-state current, we need the expression of $\Gamma_{\circlearrowright}$ or $\Gamma_{\circlearrowleft}$. We rewrite Eq.~\eqref{rate-eqns} as
\begin{equation}\label{C1.9}
\mathcal{M}\begin{bmatrix}
\rho_{\mathbb{A}}
\\\rho_{\mathbb{B}}
\\\rho_{\mathbb{C}}
\\\rho_{\mathbb{D}}
\end{bmatrix}=\begin{bmatrix}
0
\\0
\\0
\\1
\end{bmatrix},
\end{equation}
where we use $\Tr{\rho}=1$, i.e., $\rho_{\mathbb{A}}+\rho_{\mathbb{B}}+\rho_{\mathbb{C}}+\rho_{\mathbb{D}}=1$. The matrix $\mathcal{M}$ is given by
\begin{equation}\label{matrix-M}
\begin{split}
\mathcal{M}=\begin{bmatrix}
-\sum_{\lambda}f^{1}_{\lambda}-\sum_{\lambda}f^{1}_{\lambda} & \sum_{\lambda}f^{1}_{\lambda} & \sum_{\lambda}f^{1}_{\lambda} & 0
\\\sum_{\lambda}f^{1}_{\lambda} & -\sum_{\lambda}f^{1}_{\lambda}-\sum_{\lambda}f^{2}_{\lambda} & 0 & \sum_{\lambda}f^{2}_{\lambda}
\\\sum_{\lambda}f^{1}_{\lambda} & 0 & -\sum_{\lambda}f^{2}_{\lambda}-\sum_{\lambda}f^{1}_{\lambda} & \sum_{\lambda}f^{2}_{\lambda}
\\ 1 & 1 & 1 & 1
\end{bmatrix}
\end{split}
\end{equation}
where, we substitute the forms of various $\Gamma_{\mathbb{ij}}^{\lambda}$ and assume all $\gamma_{\lambda}=\gamma$. Using Eqs.~\eqref{C1.9} and~\ref{matrix-M}, one can obtain the populations $\{\rho_{\mathbb{i}}\}$ in terms of which one can evaluate the final expression of $\Gamma_{\circlearrowright}$ or $\Gamma_{\circlearrowleft}$ for different gates.

\section{General expressions of steady-state currents}\label{Appendix-B}

As the composite system weakly interacts with multiple reservoirs and exchanges energy and particles, we derive the expressions for steady-state currents under the grand canonical formalism, following Refs.~\cite{esposito2010entropy,landi2021irreversible,strasberg2022quantum}. Since the system Hamiltonian $H_{\rm{s}}$ and the total number operator of the system $\mathcal{N}$ are time-independent, we immediately recognize the net energy flux ($J_{\rm{E}}$) and the particle flux ($J_{\rm{N}}$) as
\begin{equation}\label{B5}
\begin{split}
J_{\rm{E}}(t)=\Tr_{\rm{s}}[\Dot{\rho}_{\rm{s}}(t)H_{\rm{s}}]=\sum_{\lambda}\Tr_{\rm{s}}[\mathcal{L}_{\lambda}[\rho_{\rm{s}}(t)]H_{\rm{s}}];\quad\quad
J_{\rm{N}}(t)=\Tr_{\rm{s}}[\Dot{\rho}_{\rm{s}}(t)\mathcal{N}]=\sum_{\lambda}\Tr_{\rm{s}}[\mathcal{L}_{\lambda}[\rho_{\rm{s}}(t)]\mathcal{N}],
\end{split}
\end{equation}
where we use Eq.~\eqref{A6} for $\Dot{\rho}_{\rm{s}}(t)$. Identifying $J_{\rm{E}}(t)=\sum_{\lambda}J^{\lambda}_{\rm{E}}(t)$ and   
$J_{\rm{N}}(t)=\sum_{\lambda}J^{\lambda}_{\rm{N}}(t)$, we find 
\begin{equation}\label{B7}
J^{\lambda}_{\rm{E}}(t)=\Tr_{\rm{s}}[\mathcal{L}_{\lambda}[\rho(t)]H_{\rm{s}}];\quad\quad 
J^{\lambda}_{\rm{N}}(t)=\Tr_{\rm{s}}[\mathcal{L}_{\lambda}[\rho(t)]\mathcal{N}]; \quad \text{and} \quad J^{\lambda}_{\rm{Q}}(t)=J^{\lambda}_{\rm{E}}(t)-\mu_{\lambda}J^{\lambda}_{\rm{N}}(t),
\end{equation}
where $J^{\lambda}_{\rm{E(N)}}$ or $J^{\lambda}_{\rm{Q}}$ is positive if the energy (particle) or heat current flows from the reservoir $\lambda$ to the system. In the steady state, both the energy and the particle currents would be independent of time; thus, the steady-state heat current reduces to
\begin{equation}\label{B9}
J^{\lambda}_{\rm{Q}}=J^{\lambda}_{\rm{E}}-\mu_{\lambda}J^{\lambda}_{\rm{N}}=\Tr_{\rm{s}}[\mathcal{L}_{\lambda}[\rho_{ss}]H_{\rm{s}}] - \mu_{\lambda}\Tr_{\rm{s}}[\mathcal{L}_{\lambda}[\rho_{ss}]\mathcal{N}].
\end{equation}
Using Eq.~\eqref{A7} we obtain the formal expressions for the energy, particle, and heat currents as 
\begin{equation}\label{S15}
\begin{split}
J^{\lambda}_{\rm{E}}=\sum_{\{\varepsilon_{\mathbb{ij}}\}}{\varepsilon_{\mathbb{ij}}}\Gamma_{\mathbb{ij}}^{\lambda};\qquad
J^{\lambda}_{\rm{N}}=\sum_{\{\varepsilon_{\mathbb{ij}}\}}\Gamma_{\mathbb{ij}}^{\lambda};\qquad
J^{\lambda}_{\rm{Q}}=\sum_{\{\varepsilon_{\mathbb{ij}}\}}({\varepsilon_{\mathbb{ij}}}-\mu_{\lambda})\Gamma_{\mathbb{ij}}^{\lambda}.
\end{split}
\end{equation}
where, we note that $\varepsilon_{\mathbb{ij}}=-\varepsilon_{\mathbb{ji}}$ and $\Gamma_{\mathbb{ij}}^{\lambda}=-\Gamma_{\mathbb{ji}}^{\lambda}$.

\section{Steady-state transition rates and currents for all gates}\label{Appendix-C}

\subsection{Detailed Operation of thermal diode}\label{Appendix-C1}

\begin{figure}[h]
   \centering    
\includegraphics[width=\columnwidth,height=7.7cm]{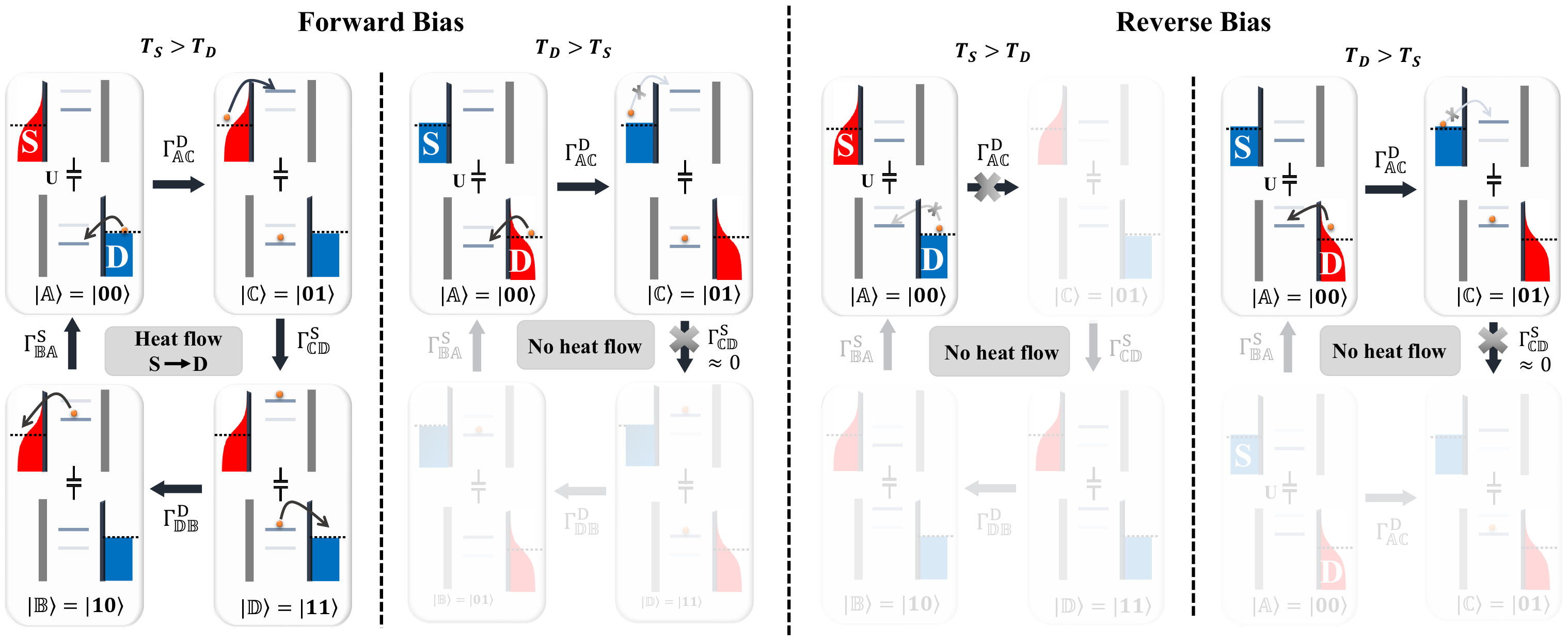}
  \caption{Operation of thermal diode: (Left) In the forward-bias configuration, heat flows from S to D when $T_{\rm S} > T_{\rm D}$, but no heat flows from D to S even when $T_{\rm D} > T_{\rm S}$, i.e., only a unidirectional heat-flow is allowed in a thermal diode. (Right) In the reverse-bias configuration, no heat flows regardless of whether $T_{\rm S}$ is greater than or less than $T_{\rm D}$. In both cases, the cycles are shown in the $\circlearrowright$ direction, corresponding to the thermodynamically preferred direction of heat flow.}
    \label{thermal-diode-SM}
\end{figure}

Let us now describe the detailed operation of the thermal diode shown in Fig.~1(b) of the main text. Here, two QDs, $\mathrm{QD_a}$ and $\mathrm{QD_b}$, are tunnel-coupled to the Source (S) and Drain (D) leads, respectively. Under forward-bias conditions, heat flows from the hot S lead to the cold D lead, whereas under reverse-bias conditions, no heat flow occurs [see Fig.~\ref{thermal-diode-SM}].

In the above figure, only the clockwise transition cycles are shown for both forward and reverse bias configurations, while the anticlockwise transition cycle doesn't proceed and is not shown. The reason for this is explained in the following section through the concept of a buffer gate, where we show that the thermal diode can function as a buffer gate.

\subsection{Thermal diode as Buffer Gate}\label{Appendix-C2}

For the above diode model, the steady-state transition rates for the clockwise and anti-clockwise transition cycles are illustrated via a diagrammatic picture in Fig.~\ref{fig:TC_Buffer}. According to Eq.~\eqref{gamma-appendix}, the transition rates satisfy the following relations

\noindent
\begin{minipage}{0.29\textwidth}
\centering 
\includegraphics[width=\linewidth]{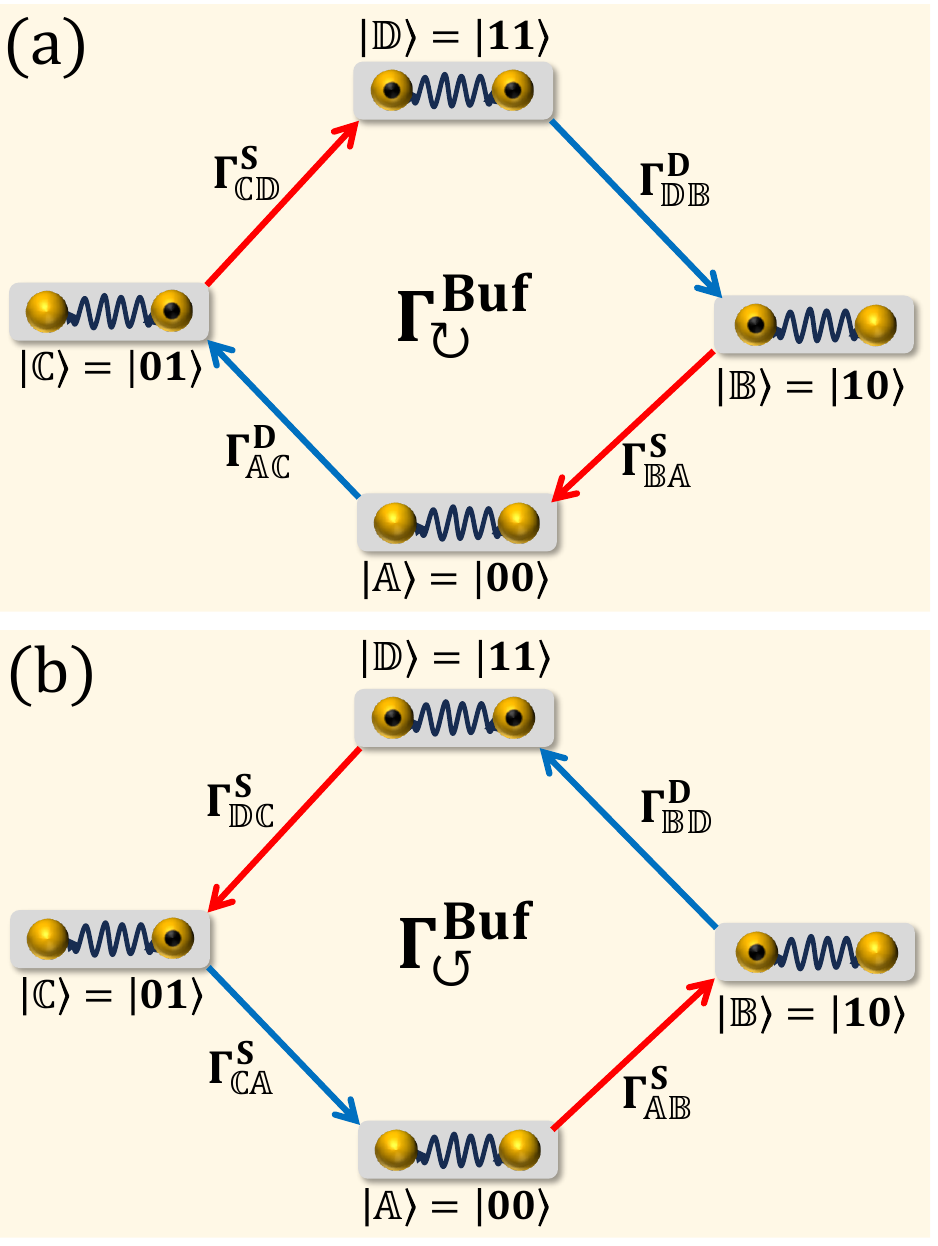}
\captionsetup{hypcap=false}
\captionof{figure}{Diagrammatic representation for clockwise and anti-clockwise transition cycles for buffer gate.}
\label{fig:TC_Buffer}
\end{minipage}
\hfill
\begin{minipage}{0.68\textwidth}
\begin{equation}\label{gamma1}
\begin{split}
\Gamma^{\rm{D}}_\mathbb{AC}=\Gamma^{\rm{S}}_\mathbb{CD}=\Gamma^{\rm{D}}_\mathbb{DB}=\Gamma^{\rm{S}}_\mathbb{BA}\equiv\Gamma^{\rm{Buf}}_{\circlearrowright}\quad;\quad
\Gamma^{\rm{S}}_\mathbb{AB}=\Gamma^{\rm{D}}_\mathbb{BD}=\Gamma^{\rm{S}}_\mathbb{DC}=\Gamma^{\rm{D}}_\mathbb{CA}\equiv\Gamma^{\rm{Buf}}_{\circlearrowleft};
\end{split}
\end{equation}
where
\begin{equation}\label{gamma_buf}
\Gamma^{\rm{Buf}}_{\circlearrowleft}=-\Gamma^{\rm{Buf}}_{\circlearrowright}=\frac{\gamma}{2}\Bigg[\frac{f_{\mathrm{D}}^{2}f_{\mathrm{D}}^{2}f_{\mathrm{S}}^{2}-f_{\mathrm{D}}^{2}f_{\mathrm{D}}^{2}f_{\mathrm{S}}^{1}+f_{\mathrm{S}}^{1}f_{\mathrm{S}}^{2}f_{\mathrm{D}}^{1}-f_{\mathrm{S}}^{1}f_{\mathrm{S}}^{2}f_{\mathrm{D}}^{2}+f_{\mathrm{S}}^{1}f_{\mathrm{D}}^{2}-f_{\mathrm{S}}^{2}f_{\mathrm{D}}^{1}}{f_{\mathrm{S}}^{1}f_{\mathrm{D}}^{1}+f_{\mathrm{S}}^{2}f_{\mathrm{D}}^{2}-f_{\mathrm{S}}^{1}f_{\mathrm{D}}^{2}-f_{\mathrm{S}}^{2}f_{\mathrm{D}}^{1}-1}\Bigg]. 
\end{equation}
Since the output current for the buffer gate is measured in the drain lead, substituting Eq.~\eqref{gamma1} into the general expressions for the steady-state currents given by Eq.~\eqref{S15}, we find the energy and particle currents at the drain lead are:
\begin{eqnarray}\label{J}
J_{\rm{E}}^{\rm{D}}&=&\varepsilon_{\rm{b}}\Gamma_\mathbb{AC}^{\rm{D}}+[-(\varepsilon_{\rm{b}}+U)]\Gamma_\mathbb{DB}^{\rm{D}}\nonumber\\
&=&\varepsilon_{\rm{b}}\Gamma^{\rm{Buf}}_{\circlearrowright}-(\varepsilon_{\rm{b}}+U)\Gamma^{\rm{Buf}}_{\circlearrowright}\nonumber\\
&=&-U\Gamma^{\rm{Buf}}_{\circlearrowright},
\end{eqnarray}
and
\begin{equation}
J_{\rm{N}}^{\rm{D}}=\Gamma_\mathbb{AC}^{\rm{D}}+\Gamma_\mathbb{BD}^{\rm{D}}=0,   
\end{equation}
respectively. Since inter-dot particle hopping is prohibited, $J_{\rm N}^{\rm D}=0$. Consequently, $J_{\rm Q}^{\rm D}$ becomes
\begin{equation}\label{JQ}
J_{\rm{Q}}^{\rm{D}}=J_{\rm{E}}^{\rm{D}}-\mu_{\rm D}J_{\rm{N}}^{\rm{D}}=-U\Gamma^{\rm{Buf}}_{\circlearrowright} < 0,
\end{equation}
\end{minipage}
\begin{figure}
   \centering    
\includegraphics[width=\columnwidth,height=6.5cm]{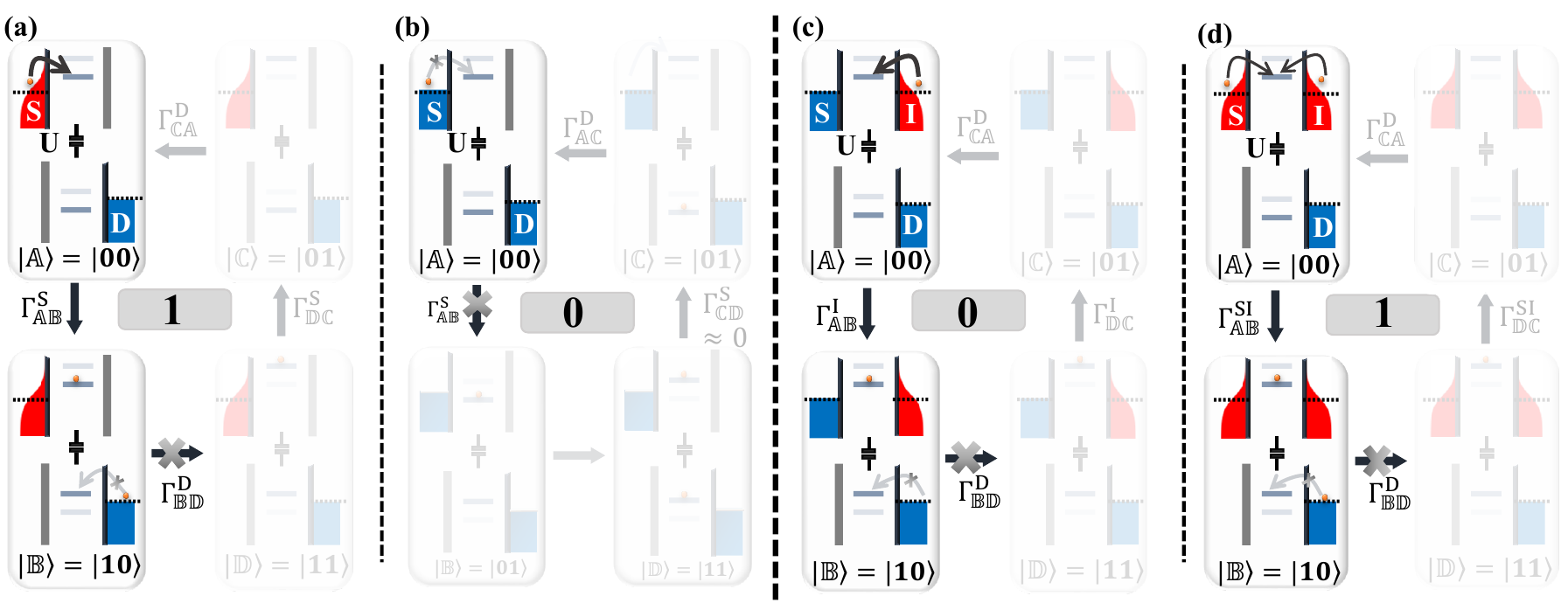}
  \caption{Thermal Buffer and NOT gate anticlockwise cycle: (a) and (b) correspond to the buffer gate with logic inputs 1 and 0, respectively, while (c) and (d) correspond to the NOT gate with logic inputs 0 and 1, respectively. In cases (a), (c), and (d), the drain lead does not provide sufficient thermal energy to excite an electron into the excited state of $\rm{QD_b}$ (i.e., $\Gamma_{\mathbb{BD}}^{\rm D} \approx 0$); hence, the cycle cannot be completed. In case (b), since the source (S) lead is cold, it does not have sufficient thermal energy to initiate the anti-clockwise cycle (i.e., $\Gamma_{\mathbb{AB}}^{\rm S} \approx 0$); therefore, none of the anti-clockwise cycles operate.}
    \label{Buffer_NOT_Anti}
\end{figure}
where $\Gamma^{\rm{Buf}}_{\circlearrowright}$ is given by Eq.~\eqref{gamma_buf}. Here, due to the temperature gradient between S and D with $T_{\rm{S}}>T_{\rm{D}}$, the heat-current $J_{\rm Q}^{\rm D}$ enters into the D-lead, thus the $J_{\rm Q}^{\rm D}$ is negative. Therefore, the clockwise cycle rate $\Gamma^{\rm Buf}_{\circlearrowright}$ must be positive for the heat to flow from the hot source to cold D, favorable by laws of thermodynamics. Since $\Gamma^{\rm Buf}_{\circlearrowleft} = -\Gamma^{\rm Buf}_{\circlearrowright}$, this further implies that $\Gamma^{\rm Buf}_{\circlearrowleft} < 0$ and hence the heat-current at D in the anticlockwise cycle is positive, which indicates the heat-flow direction from cold D to hot S. However, this violates the laws of thermodynamics. Therfore, the anti-clockwise cycle (with $\Gamma^{\rm Buf}_{\circlearrowleft} < 0$) is highly improbable on average. As a result, for the buffer gate, only the clockwise cycle operates in accordance with the laws of thermodynamics, with a \textit{net clockwise cycle rate} $\Gamma^{\rm Buf}_{\circlearrowright}>0$. As a result, it is equivalent to a thermal diode which operates only under forward bias conditions, where heat flows from hot S to cold D lead, and remains inactive under reverse bias~[Fig.~\ref{thermal-diode-SM}].

As discussed above, the buffer gate operates only through the clockwise cycle under forward bias. This behavior is not restricted to the buffer gate only but also applies to the NOT, OR, and NOR gates. In all these cases, the $\rm QD_b$ is coupled only to a cold drain lead, maintained at the lowest temperature. In the anti-clockwise cycle, the cycle initiates with the tunneling from S to $\rm{QD_{\rm{a}}}$, which raises the $\rm{QD_{\rm{b}}}$ level to its high energy state ($\varepsilon_{\rm b}+U$). However, the cold drain can not provide sufficient energy to overcome this barrier and tunnel electrons into $\rm{QD_{\rm{b}}}$; as a result, the cycle can not proceed further to produce any heat current~[Fig.~\ref{Buffer_NOT_Anti}]. This remains valid as long as $\rm QD_b$ is coupled only to a single cold D lead. Accordingly, in the discussion of NOT, OR, and NOR gates, we consider only the clockwise transition cycle in both the diagrammatic representation in the Appendix and the main text. However, for the sake of completeness, we also provide the anticlockwise cycles for all four above gates [see Figs.~\ref{Buffer_NOT_Anti}, ~\ref{OR-Gate-Anti}, and ~\ref{NOR-Gate-Anti}]. This situation, however, changes for the AND and NAND gates, where $\rm QD_b$ is additionally coupled to a hot control (C) lead maintained at a fixed high temperature. In this case, both cycles can, in principle, occur, and we discuss them in detail in the analysis of the AND and NAND gate operations.

\subsection{Thermal-NOT Gate}\label{Appendix-C3}
By coupling an additional Invert lead (I) to ${\rm QD_a}$, characterized by temperature $T_{\mathrm{I}}$ and chemical potential $\mu_{\rm{I}}$, we control the output thermal current $J_{\rm{Q}}^{\rm{I}}$ at the I lead. This upgrades the thermal-diode model to a thermal-NOT gate: ${\rm QD_a}$ is coupled to the source (S) and invert (I) leads, while ${\rm QD_b}$ is coupled to the drain lead D; therefore, applying the same analysis as before, we obtain [Fig.~\ref{TC_NOT}]: 
\begin{equation}\label{gamma2}
\begin{split}
\Gamma^{\rm{D}}_\mathbb{AC}=\Gamma^{\rm{SI}}_\mathbb{CD}=\Gamma^{\rm{D}}_\mathbb{DB}=\Gamma^{\rm{SI}}_\mathbb{BA}\equiv\Gamma^{\rm{NOT}}_{\circlearrowright}
\end{split}
\end{equation}
where $\Gamma^{\rm{SI}}_\mathbb{ij}\equiv\Gamma^{\rm{S}}_\mathbb{ij}+\Gamma^{\rm{I}}_\mathbb{ij}$.

\noindent
\begin{minipage}{0.68\textwidth}
Note that Fig.~\ref{TC_NOT} is very similar to Fig.~\ref{fig:TC_Buffer} and shows the clockwise cycle operating under forward bias. However, the only difference is that the $|\mathbb{A}\rangle \leftrightarrow |\mathbb{B}\rangle$ and $|\mathbb{C}\rangle \leftrightarrow |\mathbb{D}\rangle$ transitions are now controlled by both the S and I leads. As a result, when we use the above relations in Eq.~\eqref{S15}, the steady-state energy and particle currents associated with the $\rm{I}$ lead take the following form
\begin{equation}\label{JNOT}
\begin{split}
J_{\rm{E}}^{\rm{I}}=\varepsilon_{\rm{a}}\Gamma_\mathbb{AB}^{\rm{I}}+(\varepsilon_{\rm{a}}+U)\Gamma_\mathbb{CD}^{\rm{I}};\qquad J_{\rm{N}}^{\rm{I}}=\Gamma_\mathbb{AB}^{\rm{I}}+\Gamma_\mathbb{CD}^{\rm{I}}.
\end{split}
\end{equation}
Therefore, the heat current associated with the I lead reads,
\begin{equation}\label{JQNOT}
J_{\rm{Q}}^{\rm{I}}=J_{\rm{E}}^{\rm{I}}-\mu_{\rm{I}}J_{\rm{N}}^{\rm{I}}=(\varepsilon_{\rm{a}}-\mu_{\rm{I}})\Gamma_\mathbb{AB}^{\rm{I}}+(\varepsilon_{\rm{a}}+U-\mu_{\rm{I}})\Gamma_\mathbb{CD}^{\rm{I}}=(\varepsilon_{\rm{a}}-\mu_{\rm{I}})(\Gamma_\mathbb{CD}^{\rm{I}}-\Gamma_\mathbb{BA}^{\rm{I}})+U\Gamma_\mathbb{CD}^{\rm{I}}.
\end{equation}
The output logic value of the thermal NOT gate is determined through the magnitude of the heat current $J_{\rm{Q}}^{\rm{I}}$, obtained from the above analytical expression, where $\Gamma_\mathbb{ij}^{\lambda}=
{\rm{k}}_\mathbb{ij}^{\lambda}\rho_\mathbb{i}-{\rm{k}}_\mathbb{ji}^{\lambda}\rho_\mathbb{j}$ and ${{\rm{k}}}_\mathbb{ij}^{{\lambda}}=\gamma f^{\pm}_{{\lambda}}(\omega_\mathbb{ij})$.
\end{minipage}
\hfill
\begin{minipage}{0.29\textwidth}
\centering
\includegraphics[width=\linewidth]{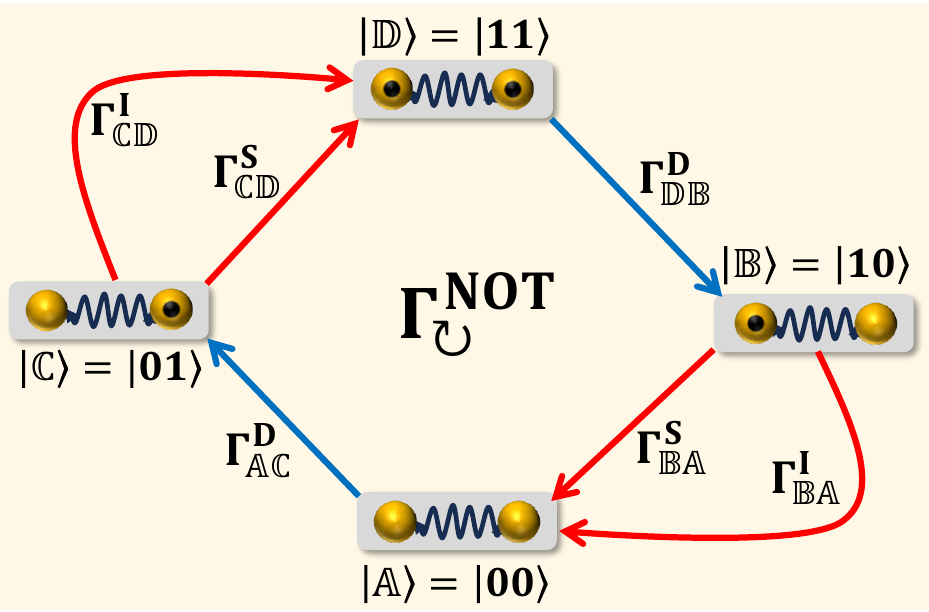}
\captionof{figure}{Diagrammatic representation for the clockwise transition cycle for the NOT gate.}
\label{TC_NOT}
\end{minipage}

\vspace{1em}
\noindent
\\
\\
$\bullet$ \textbf{For input logic 0}, it is evident from the transition cycle of the NOT gate [Figure 3(c) of the main text] that $\Gamma_\mathbb{CD}^{\rm{I}}>\Gamma_\mathbb{BA}^{\rm{I}}$, since the transition from $|\mathbb{C}\rangle$ to $|\mathbb{D}\rangle$ is predominantly driven by the I lead, whereas the de-excitation process from $|\mathbb{B}\rangle$ to $|\mathbb{A}\rangle$ is shared between the I and S leads, where cold S lead dominates the de-excitation process. Furthermore, the chosen parameters satisfy $\mu_{\mathrm{I}}<0$ and $\varepsilon_{\mathbb{a}},U>0$. Substituting these parameters into Eq.~\eqref{JQNOT}, it follows that the output current not only satisfies $J_{\rm{Q}}^{\rm{I}}>0$, but also it exceeds the threshold value $J_{\rm{Q}}^{1}$ and therefore corresponds to the output logic-1.
\\
\\
$\bullet$ \textbf{For input logic 1}, the transition cycle of the NOT gate [Figure 3(d) of the main text] indicates that $\Gamma_\mathbb{CD}^{\rm{I}}$ decreases (compared to the logic-0 case), since under this condition the excitation process from $|\mathbb{C}\rangle$ to $|\mathbb{D}\rangle$ also receives contribution from the hot S lead. In contrast, $\Gamma_\mathbb{BA}^{\rm{I}}$ increases (compared to the logic-0 case) due to the absence of cold S lead. According to Eq.~\eqref{JQNOT}, these conditions drive the output current $J_{\rm{Q}}^{\rm{I}}$ below the threshold value $J_{\rm{Q}}^{0}$, corresponding to the output logic-0.

\subsection{Thermal-OR Gate}\label{Appendix-C4}

\noindent
\begin{minipage}{0.29\textwidth}
\centering 

\includegraphics[width=\linewidth]{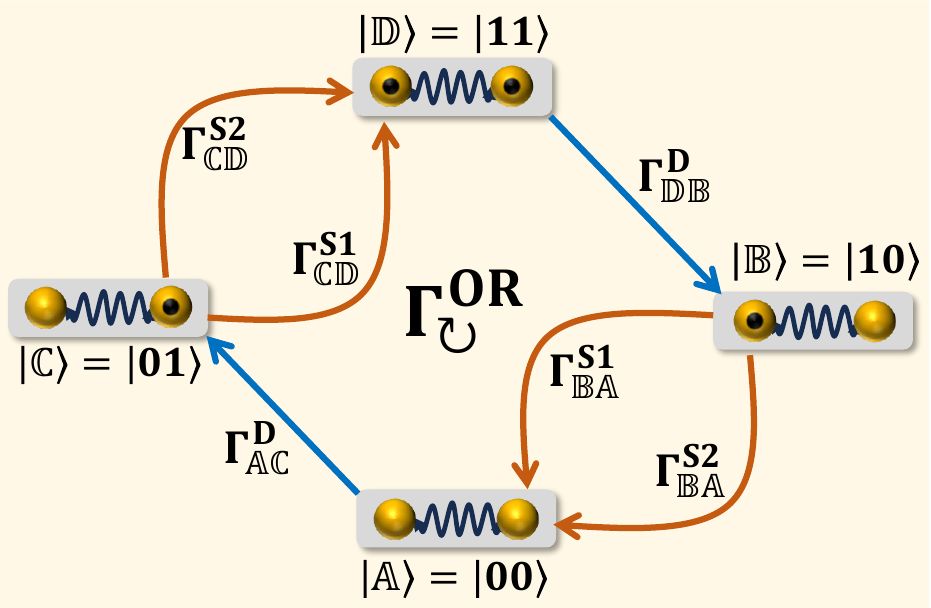}
\captionof{figure}{Diagrammatic representation for the clockwise transition cycle for the OR gate.}
\label{TC_OR}
\end{minipage}
\hfill
\begin{minipage}{0.68\textwidth}
The thermal-OR gate is a two-input gate, so we couple two source leads S1 and S2 to ${\rm QD_a}$. ${\rm QD_b}$ is tunnel-coupled to a single lead D. The input logic is defined by the temperature of the S1 and S2 leads, whereas the magnitude of the heat current $|J_{\rm{Q}}^{\rm{D}}|$ measured at the drain lead determines the output logic. At the steady-state, all transition rates $\sum_{\lambda}\Gamma_{\mathbb{ij}}^{\lambda}$ [Fig.~\ref{TC_OR}] associated with the transition $|\mathbb{i}\rangle \leftrightarrow |\mathbb{j}\rangle$ are equal [cf.~Eq.~\eqref{gamma-appendix}]. For the OR gate, this results in the following expressions of the transition rates under a clockwise cycle:
\begin{equation}\label{gamma3}
\begin{split}
\Gamma^{\rm{D}}_\mathbb{AC}=\Gamma^{\rm{S1S2}}_\mathbb{CD}=\Gamma^{\rm{D}}_\mathbb{DB}=\Gamma^{\rm{S1S2}}_\mathbb{BA}\equiv\Gamma^{{\rm{OR}}}_{\circlearrowright}
\end{split}
\end{equation}

where, $\Gamma^{\rm{S1S2}}_\mathbb{ij}\equiv\Gamma^{\rm{S1}}_\mathbb{ij}+\Gamma^{\rm{S2}}_\mathbb{ij}$.
\end{minipage}

\vspace{1em}
\noindent

\begin{figure}[h]
   \centering    
\includegraphics[width=0.96\columnwidth,height=8cm]{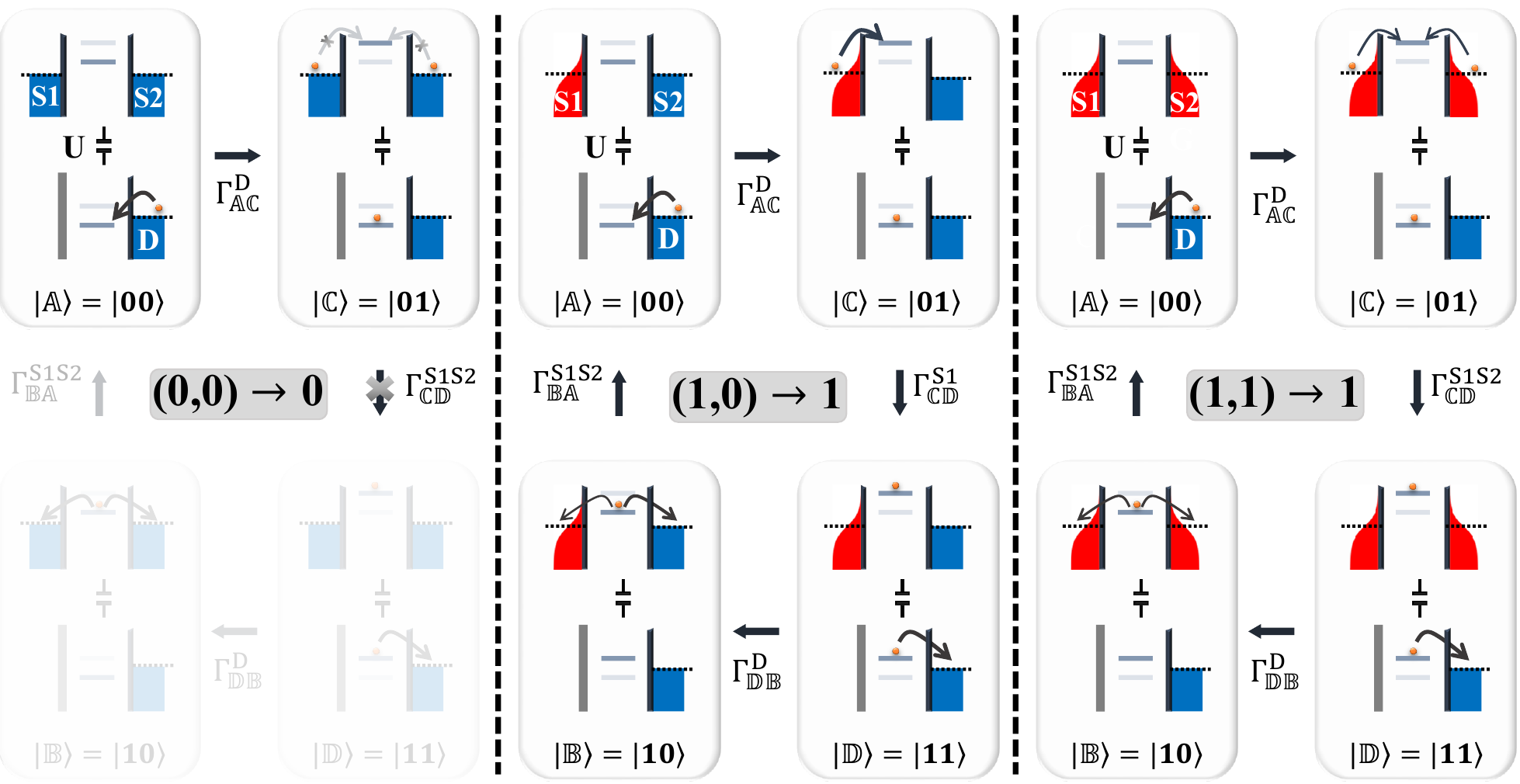}
  \caption{Thermal-OR gate heat-flow cycle is equivalent to two thermal-diodes operating in parallel, formed out of each S and the common D. For input logic $(0,0)$, both S leads are cold and can not provide sufficient energy for the electrons to tunnel into $\rm{QD_a}$, and thus the heat-flow cycle can not proceed, like in QTBGs, yielding thermal output logic-$0$. While for $(1,0)$, $(0,1)$, and $(1,1)$, at least one of the S leads is hot enough to tunnel electrons and drive the $\circlearrowright$ cycle under forward-bias, thereby yielding output logic-$1$. Parameters are: $\mu_{\rm{S1}}=\mu_{\rm{S2}}=-25\;\mu\rm{eV}$, and $\mu_{\rm{D}}=21\;\mu\rm{eV}$.}
    \label{OR Gate}
\end{figure}

Similar to the buffer gate, the steady-state energy and particle currents at the drain lead are obtained upon inserting Eq.~\eqref{gamma3} into Eq.~\eqref{S15}, and given by
\begin{eqnarray}\label{J}
J_{\rm{E}}^{\rm{D}}&=&\varepsilon_{\rm{b}}\Gamma_\mathbb{AC}^{\rm{D}}+(\varepsilon_{\rm{b}}+U)\Gamma_\mathbb{BD}^{\rm{D}}=\varepsilon_{\rm{b}}\Gamma^{\rm{OR}}_{\circlearrowright}-(\varepsilon_{\rm{b}}+U)\Gamma^{\rm{OR}}_{\circlearrowright}=-U\Gamma^{\rm{OR}}_{\circlearrowright},
\end{eqnarray}
and
\begin{equation}
J_{\rm{N}}^{\rm{D}}=\Gamma_\mathbb{AC}^{\rm{D}}+\Gamma_\mathbb{BD}^{\rm{D}}=0.   
\end{equation}
Hence, the heat current associated with the drain lead follows
\begin{equation}\label{JQOR}
J_{\rm{Q}}^{\rm{D}}=J_{\rm{E}}^{\rm{D}}-\mu_{\rm D}J_{\rm{N}}^{\rm{D}}=-U\Gamma^{\rm{OR}}_{\circlearrowright}.
\end{equation}
where
\begin{equation}
\Gamma^{\rm{OR}}_{\circlearrowleft}=-\Gamma^{\rm{OR}}_{\circlearrowright}=\gamma\Bigg[\frac{f_{\mathrm{S1S2}}^{1}[f_{\mathrm{S1S2}}^{2}(f_{\mathrm{D}}^{2}-f_\mathrm{D}^{1})+2f_\mathrm{D}^{2}(f_\mathrm{D}^{1}-1)]-2f_\mathrm{D}^{1}f_{\mathrm{S1S2}}^{2}(f_\mathrm{D}^{2}-1)}{3f_{\mathrm{S1S2}}^{1}(f_\mathrm{D}^{1}-f_\mathrm{D}^{2})-6+3f_{\mathrm{S1S2}}^{2}(f_\mathrm{D}^{2}-f_\mathrm{D}^{1})}\Bigg].   \end{equation}
where, $f_{\mathrm{S1S2}}^{1(2)}=f_{\mathrm{S1}}^{1(2)}+f_{\mathrm{S12}}^{1(2)}$. The magnitude of $J_{\rm{Q}}^{\rm{D}}$ determines the output logic value: if it is above the $J_{\rm{Q}}^{1}$$\sim 100$ aW, logic is $1$, and if it is less than $J_{\rm{Q}}^{0}$ $\sim 65$ aW, logic is $0$.
\\
\\
$\bullet$ \textbf{For input logic (0,0)}, it is evident that the transition cycle becomes frozen, since the transition from $|\mathbb{C}\rangle$ to $|\mathbb{D}\rangle$ is not driven by either of the source leads due to the absence of hot electrons [Fig.~\ref{OR Gate}]. Consequently, the steady-state transition rate $\Gamma^{\rm{OR}}_{\circlearrowright}$ approaches zero. Therefore, from Eq.~\eqref{JQOR}, the output current $|J_{\rm{Q}}^{\rm{D}}|$ becomes negligible, corresponding to the output logic 0.
\\
\\
$\bullet$ \textbf{For input logic (1,0) or (0,1)}, the transition cycle proceeds in the clockwise direction, since the transition from $|\mathbb{C}\rangle$ to $|\mathbb{D}\rangle$ is facilitated by the hot electron originating from one of the hot S leads [Fig.~\ref{OR Gate}]. As a result, a significant non-zero value of $\Gamma^{\rm{OR}}_{\circlearrowright}$ is generated. Substituting this into Eq.~\eqref{JQOR}, we obtain a considerable output heat current $|J_{\rm{Q}}^{\rm{D}}|$ that exceeds the threshold value $J_{\rm{Q}}^{1}$, thereby corresponding to the output logic 1.
\\
\\
$\bullet$ \textbf{For input logic (1,1)}, it indicates a significant enhancement of the steady-state transition rate $\Gamma^{\rm{OR}}_{\circlearrowright}$. This increase arises from the participation of both hot S leads in facilitating the transition from $|\mathbb{C}\rangle$ to $|\mathbb{D}\rangle$ [Fig.~\ref{OR Gate}]. Consequently, as follows from Eq.~\eqref{JQOR}, the output heat current $|J_{\rm{Q}}^{\rm{D}}|$ also increases substantially and remains well above the threshold value $J_{\rm{Q}}^{1}$, thereby corresponding to the output logic 1.

\begin{figure}[t]
   \centering    
\includegraphics[width=\columnwidth,height=8.6cm]{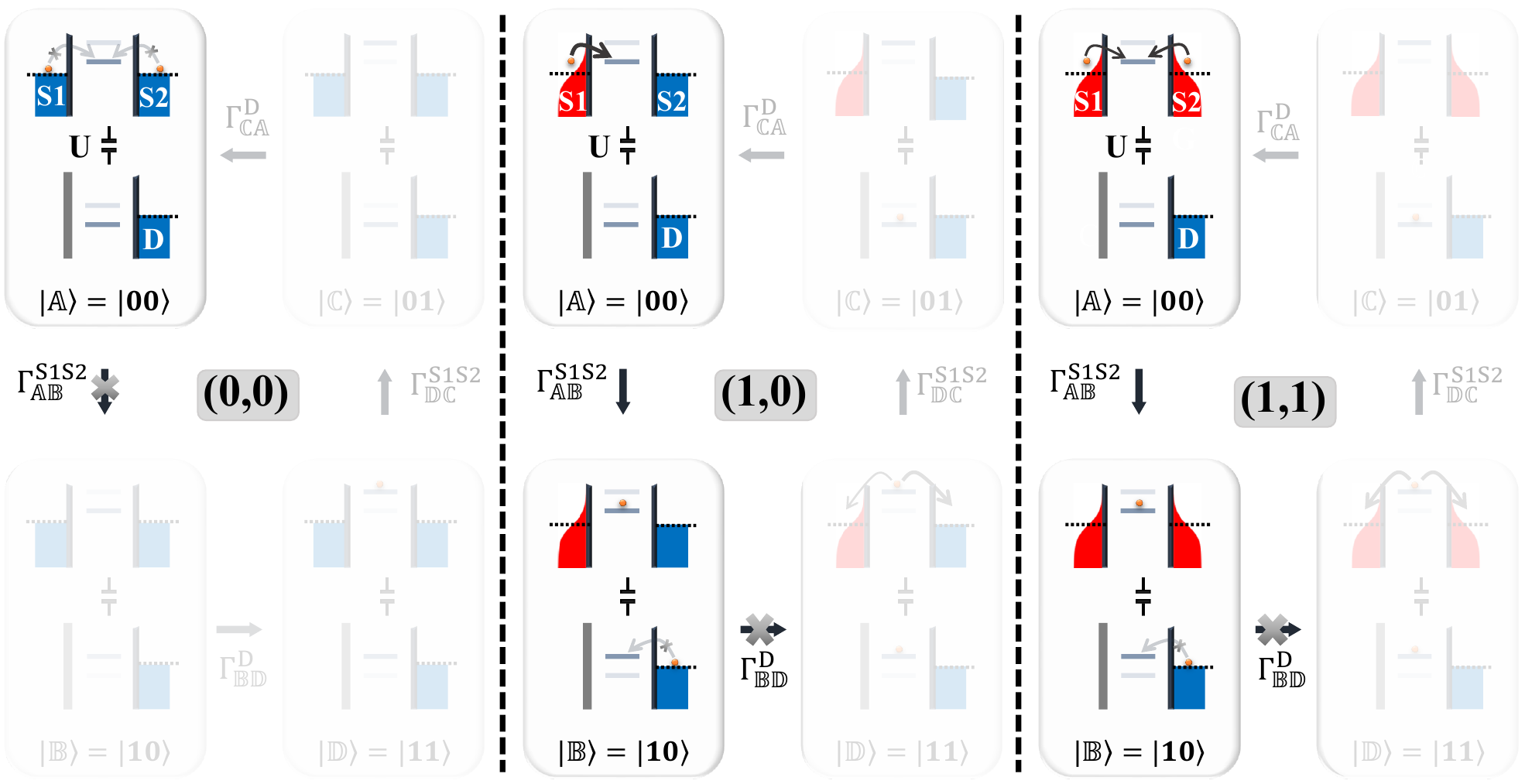}
  \caption{Thermal-OR gate anticlockwise cycle: Since no hot lead is attached to $\rm{QD}_{\rm b}$, the cold D lead cannot supply sufficient thermal energy to pump an electron into $\rm{QD}_{\rm b}$; thus, $\Gamma_{\mathbb{BD}}^{\rm D} \approx 0$, and the anti-clockwise cycle cannot be completed. For the input state (0,0), the anti-clockwise cycle cannot even be initiated because both S leads are cold (i.e., $\Gamma_{\mathbb{AB}}^{\rm S1S2} \approx 0$). Parameters for OR gate are: $\mu_{\rm{S1}}=\mu_{\rm{S2}}=-25\;\mu\rm{eV}$, and $\mu_{\rm{D}}=21\;\mu\rm{eV}$.}
    \label{OR-Gate-Anti}
\end{figure}

\subsection{Thermal-AND Gate}\label{Appendix-C5}

\noindent

The thermal AND gate can be constructed from the OR-gate by introducing an additional control (C) lead attached to ${\rm QD_b}$. In this configuration, ${\rm QD_a}$ is tunnel-coupled to the source leads S1 and S2. In contrast, ${\rm QD_b}$ is tunnel-coupled to one drain lead D and a control lead C. Under these conditions, the steady-state transition rates [Fig.~\ref{TC_AND}] obey the following relation
\begin{equation}\label{gamma4}
\begin{split}
\Gamma^{\rm{DC}}_\mathbb{AC}=\Gamma^{\rm{S1S2}}_\mathbb{CD}=\Gamma^{\rm{DC}}_\mathbb{DB}=\Gamma^{\rm{S1S2}}_\mathbb{BA}\equiv\Gamma^{{\rm{AND}}}_{\circlearrowright}\quad;\quad\Gamma^{\rm{S1S2}}_\mathbb{AB}=\Gamma^{\rm{DC}}_\mathbb{BD}=\Gamma^{\rm{S1S2}}_\mathbb{DC}=\Gamma^{\rm{DC}}_\mathbb{CA}\equiv\Gamma^{{\rm{AND}}}_{\circlearrowleft}
\end{split}
\end{equation}
where $\Gamma^{\rm{S1S2}}_\mathbb{ij}\equiv\Gamma^{\rm{S1}}_\mathbb{ij}+\Gamma^{\rm{S2}}_\mathbb{ij}$ and $\Gamma^{\rm{DC}}_\mathbb{ij}\equiv\Gamma^{\rm{D}}_\mathbb{ij}+\Gamma^{\rm{C}}_\mathbb{ij}$, with $\Gamma^{\rm{AND}}_{\circlearrowright}=-\Gamma^{\rm{AND}}_{\circlearrowleft}$. Since the $\rm{QD_b}$ is coupled to both cold (D) and hot (C) leads, the transition cycle can run in both clockwise and anti-clockwise directions [Fig.~\ref{TC_AND}]. Like the buffer and OR gate, the AND gate output is measured at the D-lead. So, substituting the above relations into Eq.~\eqref{S15}, the steady-state energy and particle currents associated with the drain lead are obtained as
\begin{equation}\label{JAND}
\begin{split}
J_{\rm{E}}^{\rm{D}}=\varepsilon_{\rm{b}}\Gamma_\mathbb{AC}^{\rm{D}}+(\varepsilon_{\rm{b}}+U)\Gamma_\mathbb{BD}^{\rm{D}};\qquad
J_{\rm{N}}^{\rm{D}}=\Gamma_\mathbb{AC}^{\rm{D}}+\Gamma_\mathbb{BD}^{\rm{D}}.
\end{split}
\end{equation}
Consequently, the heat current measured at the drain lead can be calculated as
\begin{equation}\label{JQAND}
\begin{split}
J_{\rm{Q}}^{\rm{D}}&=(\varepsilon_{\rm{b}}-\mu_{\rm{D}})\Gamma_\mathbb{AC}^{{\rm{D}}}+(\varepsilon_{\rm{b}}+U-\mu_{\rm{D}})\Gamma_\mathbb{BD}^{{\rm{D}}}\\
&=(\varepsilon_{\rm{b}}-\mu_{\rm{D}})(\Gamma_\mathbb{AC}^{{\rm{D}}}-\Gamma_\mathbb{DB}^{{\rm{D}}})-U\Gamma_\mathbb{DB}^{{\rm{D}}}\\
&=(\varepsilon_{\rm{b}}-\mu_{\rm{D}})(\Gamma_\mathbb{BD}^{{\rm{D}}}-\Gamma_\mathbb{CA}^{{\rm{D}}})+U\Gamma_\mathbb{BD}^{{\rm{D}}}.
\end{split}
\end{equation}

\begin{figure}[b]
   \centering    
\includegraphics[width=0.96\columnwidth,height=8cm]{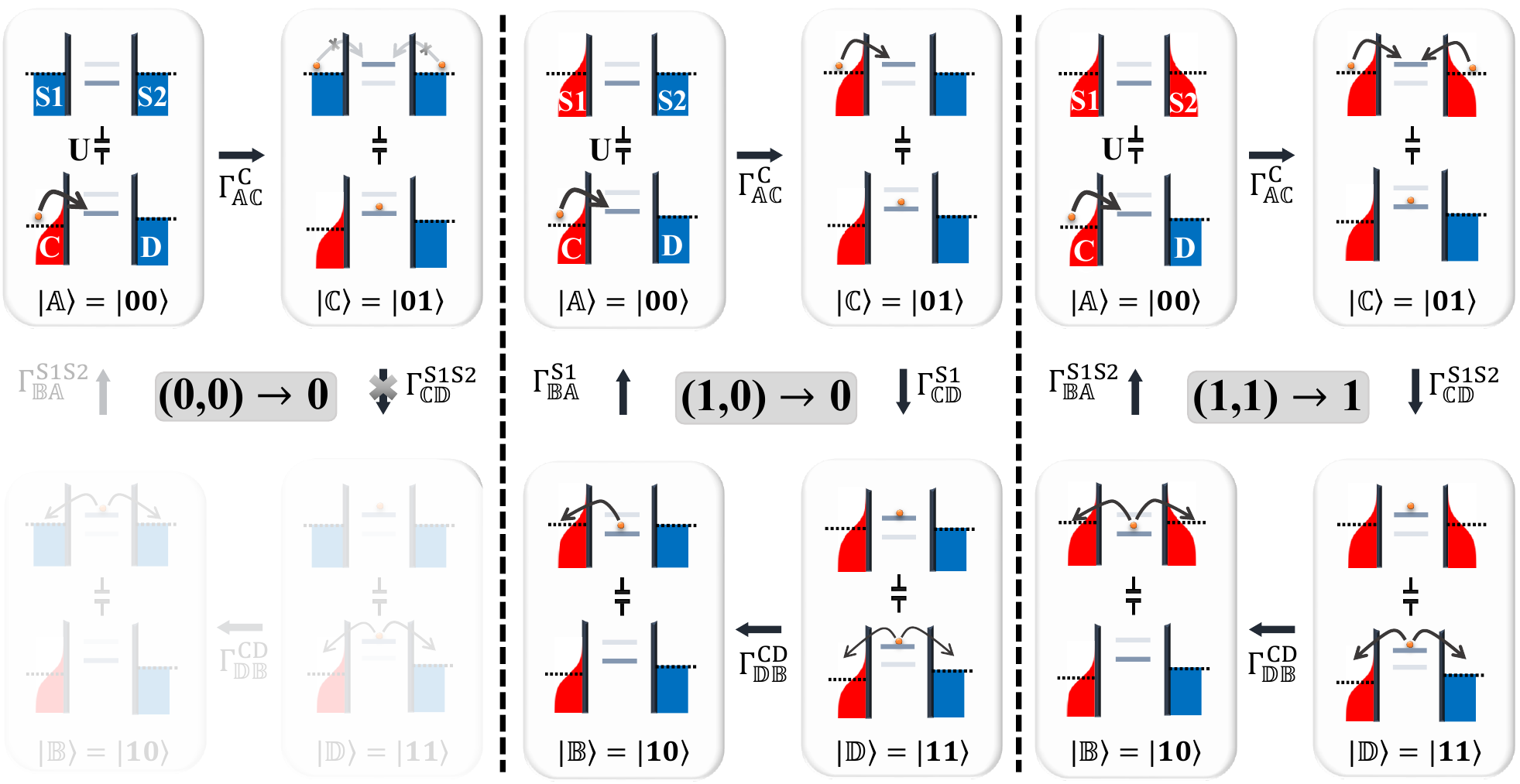}
  \caption{Thermal-AND gate heat-flow cycle can proceed in both directions; $\circlearrowright$ cycle is shown here (cf.~Fig.~\ref{AND-Gate-Anti} for $\circlearrowleft$ cycle). The chemical potentials are tuned to keep both thermal-diodes in reverse-bias w.r.t the cold D and forward-bias w.r.t the hot C. For $(0,0)$, $\circlearrowright$ cycle is blocked, although it yields negligible heat-current due to $\circlearrowleft$ with net output logic-0. For $(1,0)$, both cycles proceed, but position of $\mu_{\rm{C}}$ and $\mu_{\rm{D}}$, suppress the tunnel rates and the heat-current (thin arrow), yielding logic-0. For both inputs 1, both sources drive the cycle, and the net output exceeds the threshold (thick arrow), yielding logic-1. Parameters are: $\mu_{\rm{S1}}=\mu_{\rm{S2}}=16\;\mu\rm{eV}$, $\mu_{\rm{C}}=-72\;\mu\rm{eV}$,and $\mu_{\rm{D}}=14\;\mu\mathrm{eV}$.}
    \label{AND Gate}
\end{figure}

\begin{figure}[b]
   \centering    
\includegraphics[width=\columnwidth,height=8.6cm]{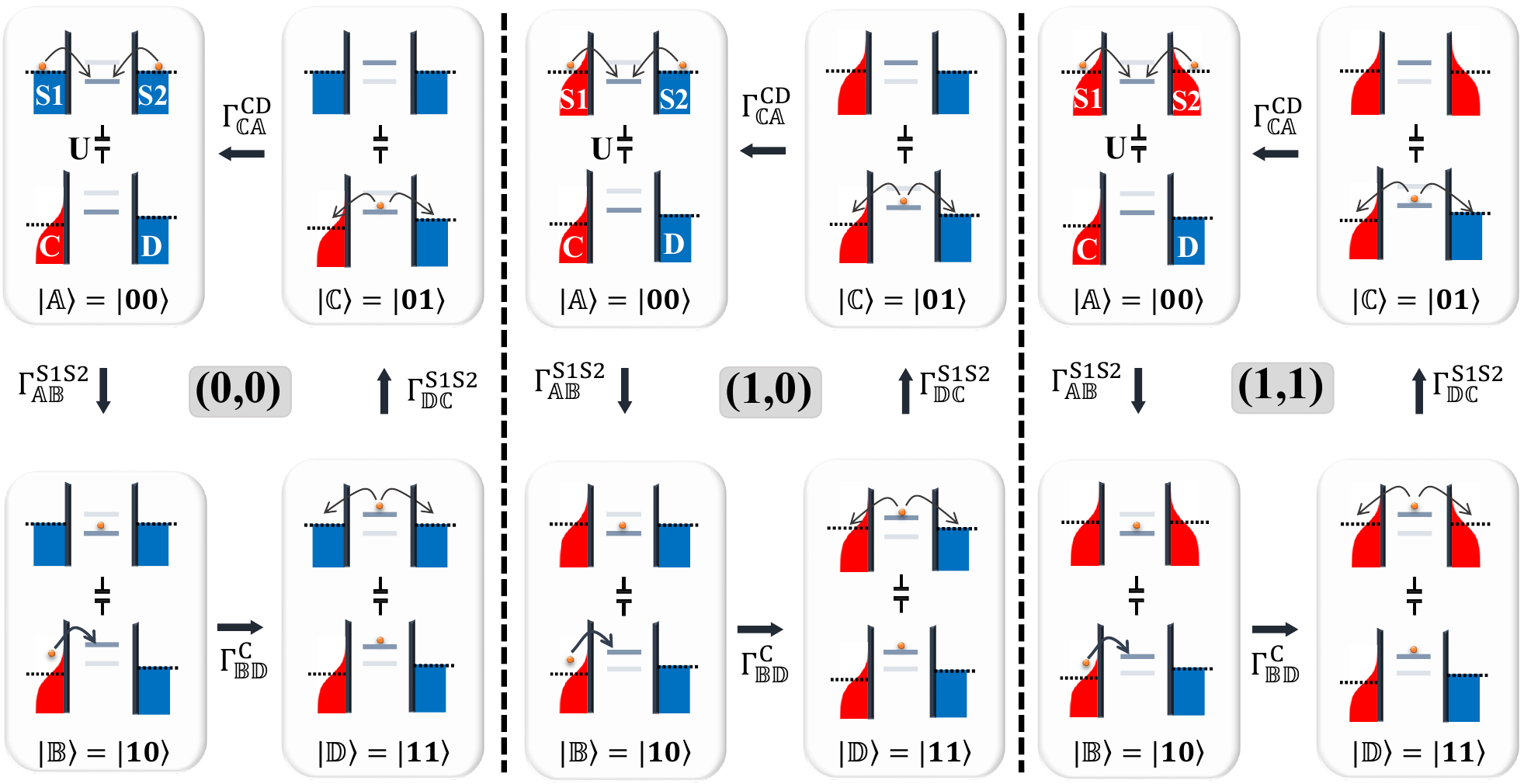}
  \caption{Anti-clockwise cycle of thermal-AND gate: The anti-clockwise cycle operates for all input logic states. However, the output current arising solely from the anti-clockwise cycle is small in magnitude for all three input configurations. This is because the Fermi-level of the C lead is kept very low; as a result, it pumps electrons into the excited state of $\rm{QD_b}$ at a low rate, $\Gamma_{\mathbb{BD}}^{\rm C}$ (shown by the thin arrow). However, when this contribution is added to the clockwise cycle currents, the resulting total current produces the AND gate output logic values. The parameters for the AND gate are: $\mu_{\rm{S1}}=\mu_{\rm{S2}}=16\;\mu\rm{eV}$, $\mu_{\rm{C}}=-72\;\mu\rm{eV}$, and $\mu_{\rm{D}}=14\;\mu\rm{eV}$.}
    \label{AND-Gate-Anti}
\end{figure}

\begin{minipage}{0.68\textwidth}
The output logic of the thermal AND gate is determined by the magnitude of $|J_{\rm{Q}}^{\rm{D}}|$, obtained from the above analytical expression, while the temperatures of the S1 and S2 leads define the thermal input logic. Unlike the previously discussed gates, both clockwise and anti-clockwise transition cycles can occur in the AND gate configuration~[Fig.~\ref{AND-Gate-Anti}], since $\mathrm{QD_b}$ is coupled to the hot C lead. Consequently, the logical operation of this gate is comparatively more intricate, as the contributions from both transition cycles must be taken into account.

However, the crucial contribution arises from the participation of the drain lead in the transitions between $|\mathbb{B}\rangle$ and $|\mathbb{D}\rangle$, and between $|\mathbb{C}\rangle$ and $|\mathbb{A}\rangle$, as indicated by Eq.~\eqref{JQAND}. Since the configurations of the C and D leads remain unchanged with varying input logic, a finite heat current is always driven from the hot C lead to the cold D lead. For the clockwise transition cycle [Fig.~\ref{AND Gate}], the transition from $|\mathbb{A}\rangle$ to $|\mathbb{C}\rangle$ is predominantly driven by the C lead, whereas both leads participate in the transition from $|\mathbb{D}\rangle$ to $|\mathbb{B}\rangle$. Consequently, a significant amount of heat current flows into the drain lead during the clockwise cycle. In contrast, for the anti-clockwise transition cycle~[Fig.~\ref{AND-Gate-Anti}], the C lead must facilitate the transition between $|\mathbb{B}\rangle$ and $|\mathbb{D}\rangle$, which requires more energy (as $\mu_{\rm{C}}$ is placed much below $\mathrm{QD_b}$ level) and is therefore less favorable than the excitation between $|\mathbb{A}\rangle$ and $|\mathbb{C}\rangle$ (in the clockwise cycle), which has a lower energy cost. Moreover, both leads contribute to the de-excitation transition from $|\mathbb{C}\rangle$ to $|\mathbb{A}\rangle$. Therefore, the anti-clockwise cycle~[Fig.~\ref{AND-Gate-Anti}] generates a comparatively smaller heat current in the drain lead than the clockwise cycle.
\end{minipage}
\hfill
\begin{minipage}{0.28\textwidth}
\centering
\includegraphics[width=\linewidth]{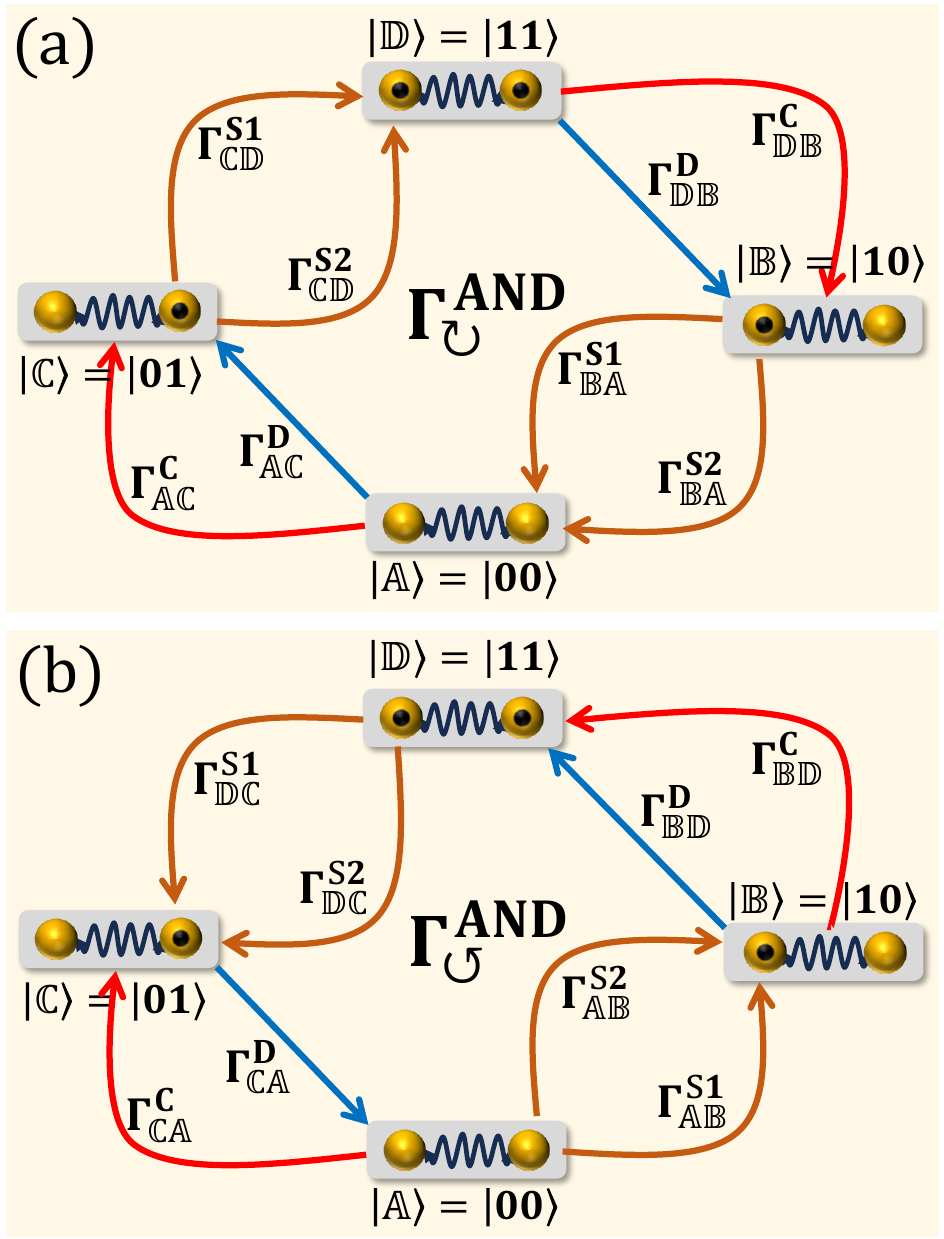}
\captionof{figure}{Diagrammatic representation for clockwise and anti-clockwise transition cycles for the AND gate.}
\label{TC_AND}
\end{minipage}

\vspace{1em}
\noindent

Thus, it is evident that for both transition cycles, the heat current flows into the drain lead. Hence, the simultaneous occurrence of both cycles does not cancel the drain heat current; rather, it enhances the overall heat flow. However, the magnitude of the output heat current contribution from the clockwise transition cycle becomes comparatively larger, whereas the anti-clockwise cycle contributes only a small amount (which remains below the threshold $J_{\rm{Q}}^{0}$) to the overall current.
\\
\\
$\bullet$ \textbf{For input logic (0,0)}, the clockwise transition cycle [Fig.~\ref{AND Gate}] becomes frozen, since the transition from $|\mathbb{C}\rangle$ to $|\mathbb{D}\rangle$ is not facilitated by either of the source leads due to the absence of hot electrons. Nevertheless, the anti-clockwise cycle continues to operate in the opposite direction, resulting in a small amount of heat current flowing into the drain lead. However, the output current $|J_{\rm{Q}}^{\rm{D}}|$ remains below the threshold value $J_{\rm{Q}}^{0}$, thus corresponds to the output logic-0.
\\
\\
$\bullet$ \textbf{For input logic (1,0) or (0,1)}, the clockwise transition cycle operates simultaneously with the anti-clockwise cycle~[Fig.~\ref{AND-Gate-Anti}] due to the presence of one hot source lead, which facilitates the transition from $|\mathbb{A}\rangle$ to $|\mathbb{C}\rangle$ (in clockwise cycle). Despite the activation of the clockwise cycle [Fig.~7], due to the chosen position of the chemical potentials of the drain, control, and source leads, in order to keep the two thermal-diodes in reverse bias condition w.r.t the D lead and in forward bias condition w.r.t the C lead, the output heat current $|J_{\rm{Q}}^{\rm{D}}|$ still remains below threshold $J_{\rm{Q}}^{0}$, corresponding to the output logic-0.
\\
\\
$\bullet$ \textbf{For input logic (1,1)}, the contribution of the clockwise transition cycle increases significantly due to the presence of both hot S leads [Fig.~\ref{AND Gate}], as the transition from $|\mathbb{A}\rangle$ to $|\mathbb{C}\rangle$ (in the clockwise cycle) is facilitated by both leads. Consequently, the overall output current $|J_{\rm{Q}}^{\rm{D}}|$ increases and exceeds the threshold value $J_{\rm{Q}}^{1}$, thereby corresponding to the output logic-1.

\subsection{Thermal-NOR Gate}\label{Appendix-C6}

\noindent
\begin{minipage}{0.29\textwidth}
\centering 
\includegraphics[width=\linewidth]{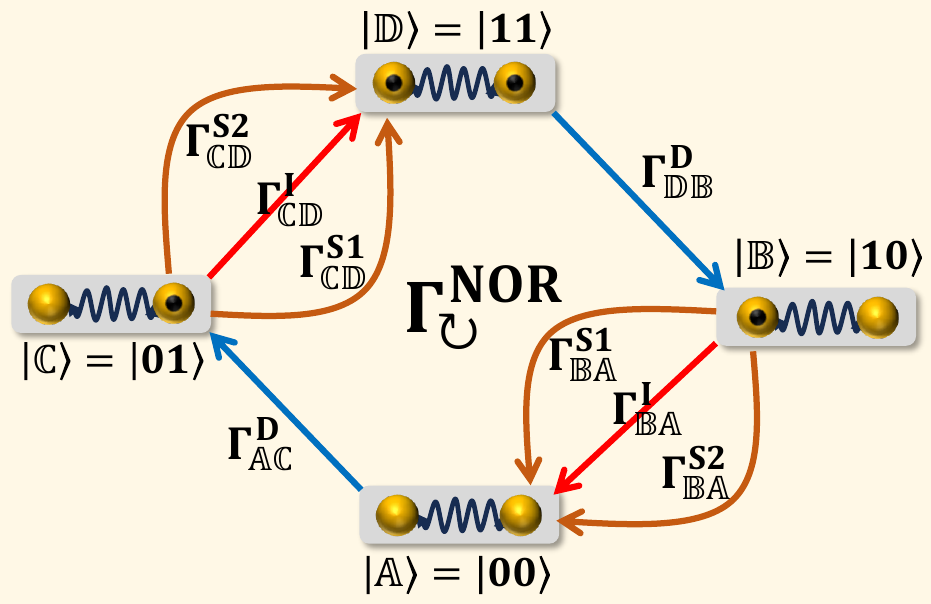}
\captionof{figure}{Diagrammatic representation of the clockwise transition cycle for the NOR gate.}
\label{TC_NOR}
\end{minipage}
\hfill
\begin{minipage}{0.68\textwidth}
To construct the thermal-NOR gate, the thermal-NOT gate setup is combined with the thermal-OR gate circuit. Hence, ${\rm QD_a}$ is coupled to an additional inverting lead I, along with S1 and S2. This configuration functions as a quantum thermal NOR gate where the input logics are governed by the temperatures of the S1 and S2-leads, whereas, like the thermal-NOT gate, the heat-current at the I-lead $J_{\rm{Q}}^{\rm{I}}$ determines the output logic. In this setup, the steady-state rates [Fig.~\ref{TC_NOR}] follow [cf.~Eq.~\eqref{gamma-appendix}]
\begin{equation}\label{gamma4}
\begin{split}
\Gamma^{\rm{D}}_\mathbb{AC}=\Gamma^{\rm{IS1S2}}_\mathbb{CD}=\Gamma^{\rm{D}}_\mathbb{DB}=\Gamma^{\rm{IS12}}_\mathbb{BA}\equiv\Gamma^{{\rm{NOR}}}_{\circlearrowright}
\end{split}
\end{equation}

where $\Gamma^{\rm{IS1S2}}_\mathbb{ij}\equiv\Gamma^{\rm{I}}_\mathbb{ij}+\Gamma^{\rm{S1}}_\mathbb{ij}+\Gamma^{\rm{S2}}_\mathbb{ij}$.
\end{minipage}
\vspace{1em}
\noindent

\begin{figure}[b]
   \centering    
\includegraphics[width=0.96\columnwidth,height=8.4cm]{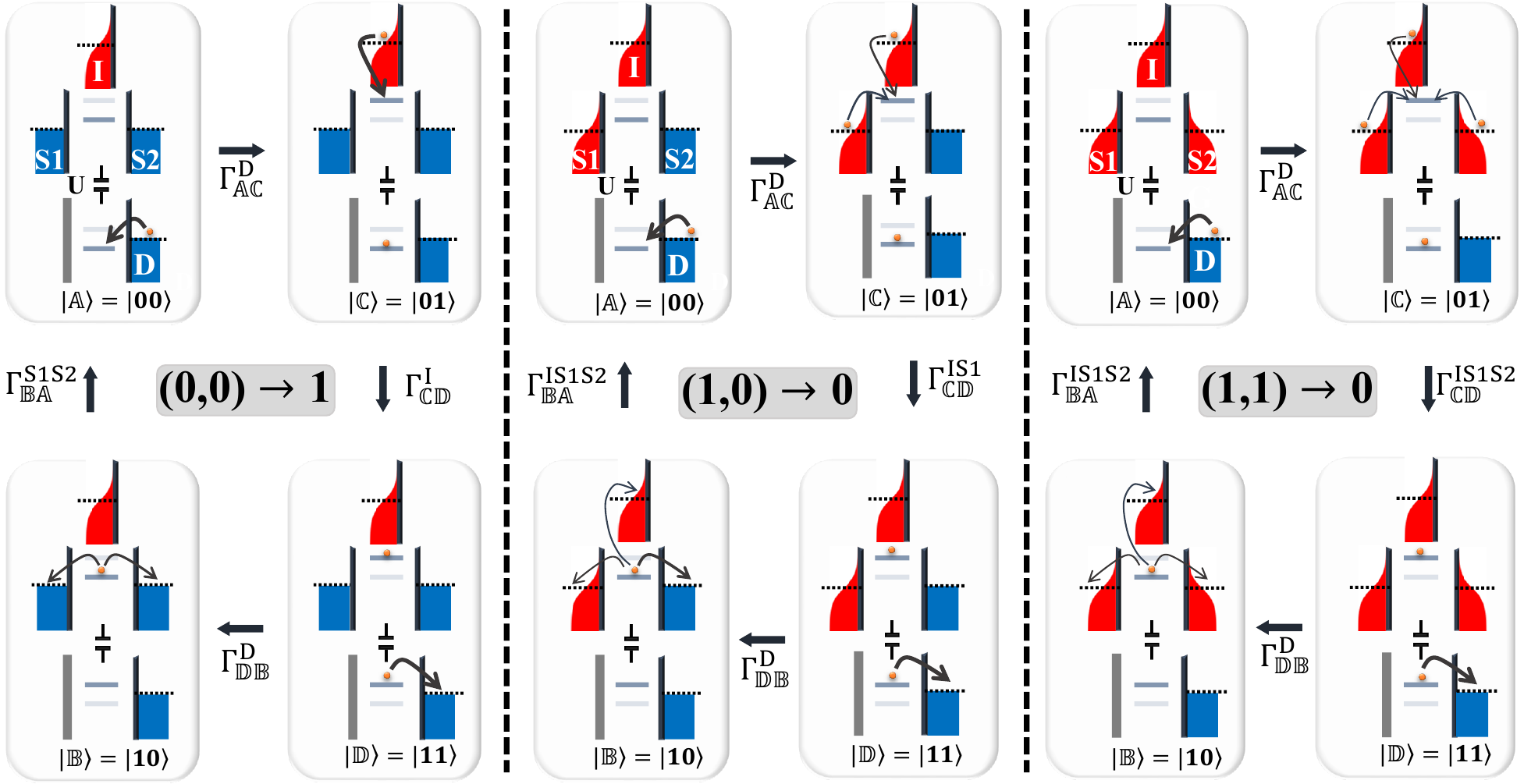}
  \caption{Thermal-NOR gate is implemented by applying a NOT operation to an OR gate, achieved by introducing an additional I lead coupled to $\rm{QD_a}$. In contrast to the $(0,0)$ case of the OR gate, the presence of the hot I-lead drives the cycle. Since the output current is measured at the I-lead, this configuration yields output logic-1. When one of the inputs is 1, the heat-flow rate from the I-lead is suppressed, and the measured current drops below threshold, resulting in output logic-0. The parameters for the NOR gate are: $\mu_{\rm{S1}}=\mu_{\rm{S2}}=-95\;\mu\rm{eV}$, $\mu_{\rm{I}}=-90\;\mu\rm{eV}$, and $\mu_{\rm{D}}=21\;\mu\rm{eV}$.}
    \label{NOR Gate}
\end{figure}

Now, just like the NOT gate, using the above relations in Eq.~\eqref{S15}, the steady-state energy and particle currents associated with the $\rm{I}$ lead can be obtained as
\begin{equation}\label{JNOR}
\begin{split}
J_{\rm{E}}^{\rm{I}}=\varepsilon_{\rm{a}}\Gamma_\mathbb{AB}^{\rm{I}}+(\varepsilon_{\rm{a}}+U)\Gamma_\mathbb{CD}^{\rm{I}};\qquad J_{\rm{N}}^{\rm{I}}=\Gamma_\mathbb{AB}^{\rm{I}}+\Gamma_\mathbb{CD}^{\rm{I}}.
\end{split}
\end{equation}
Therefore, the heat current $J_{\rm{Q}}^{\rm{I}}$ with the invert lead reads,
\begin{equation}\label{JQNOR}
J_{\rm{Q}}^{\rm{I}}=J_{\rm{E}}^{\rm{I}}-\mu_{\rm{I}}J_{\rm{N}}^{\rm{I}}=(\varepsilon_{\rm{a}}-\mu_{\rm{I}})\Gamma_\mathbb{AB}^{\rm{I}}+(\varepsilon_{\rm{a}}+U-\mu_{\rm{I}})\Gamma_\mathbb{CD}^{\rm{I}}=(\varepsilon_{\rm{a}}-\mu_{\rm{I}})(\Gamma_\mathbb{CD}^{\rm{I}}-\Gamma_\mathbb{BA}^{\rm{I}})+U\Gamma_\mathbb{CD}^{\rm{I}}.
\end{equation}
The above analytical expression gives the output heat-current $J_{\rm{Q}}^{\rm{I}}$ and hence the output logic values. Now, the logical operations of the thermal-NOR gate can be explained by combining the thermal-NOT gate operations with the thermal-OR gate. 
\\
\\
$\bullet$ \textbf{For input logic (0,0)}, the logical operation follows a pathway similar to that of the input logic-0 of the NOT gate. As illustrated in Fig.~\ref{NOR Gate}, the transition from $|\mathbb{C}\rangle$ to $|\mathbb{D}\rangle$ is predominantly driven by the I-lead due to the absence of hot electrons from both cold S leads, whereas the de-excitation process from $|\mathbb{B}\rangle$ to $|\mathbb{A}\rangle$ is shared between the I and both S leads, where cold source leads take the advantage. Furthermore, the  parameters of thermal-NOR gate satisfy $\mu_{\mathrm{I}}<0$ and $\varepsilon_{\mathbb{a}},U>0$. Substituting these conditions into Eq.~\eqref{JQNOR}, it follows that the output current not only satisfies $J_{\rm{Q}}^{\rm{I}}>0$, but also exceeds the threshold value $J_{\rm{Q}}^{1}$, thereby corresponding to the output logic-1.
\\
\\
$\bullet$ \textbf{For input logic (1,0) or (0,1)}, one of the source leads provides hot electrons to facilitate the transition from $|\mathbb{C}\rangle$ to $|\mathbb{D}\rangle$ along with the I-lead, thereby effectively reducing the dominant contribution of the I-lead to this transition. On the other hand, the de-excitation rate corresponding to the I-lead $\Gamma_\mathbb{BA}^{\rm{I}}$ increases, since the contribution from the hot S-lead to the de-excitation process decreases [Fig.~\ref{NOR Gate}]. Consequently, following Eq.~\eqref{JQNOR}, above conditions drive the output current $J_{\rm{Q}}^{\rm{I}}$ below the threshold value $J_{\rm{Q}}^{0}$, corresponding to output logic-0.
\\
\\
$\bullet$ \textbf{For input logic (1,1)}, the participation of the I-lead in the transition from $|\mathbb{C}\rangle$ to $|\mathbb{D}\rangle$ is significantly reduced because both hot source leads actively facilitate this transition, whereas $\Gamma_\mathbb{BA}^{\rm{I}}$ becomes maximal due to the absence of any cold source lead [Fig.~\ref{NOR Gate}]. Consequently, as dictated by Eq.~\eqref{JQNOR}, the output current $J_{\rm{Q}}^{\rm{I}}$ decreases substantially and remains well below the threshold value $J_{\rm{Q}}^{0}$, thereby corresponding to the output logic-0.

\begin{figure}[b]
   \centering    
\includegraphics[width=\columnwidth,height=8.6cm]{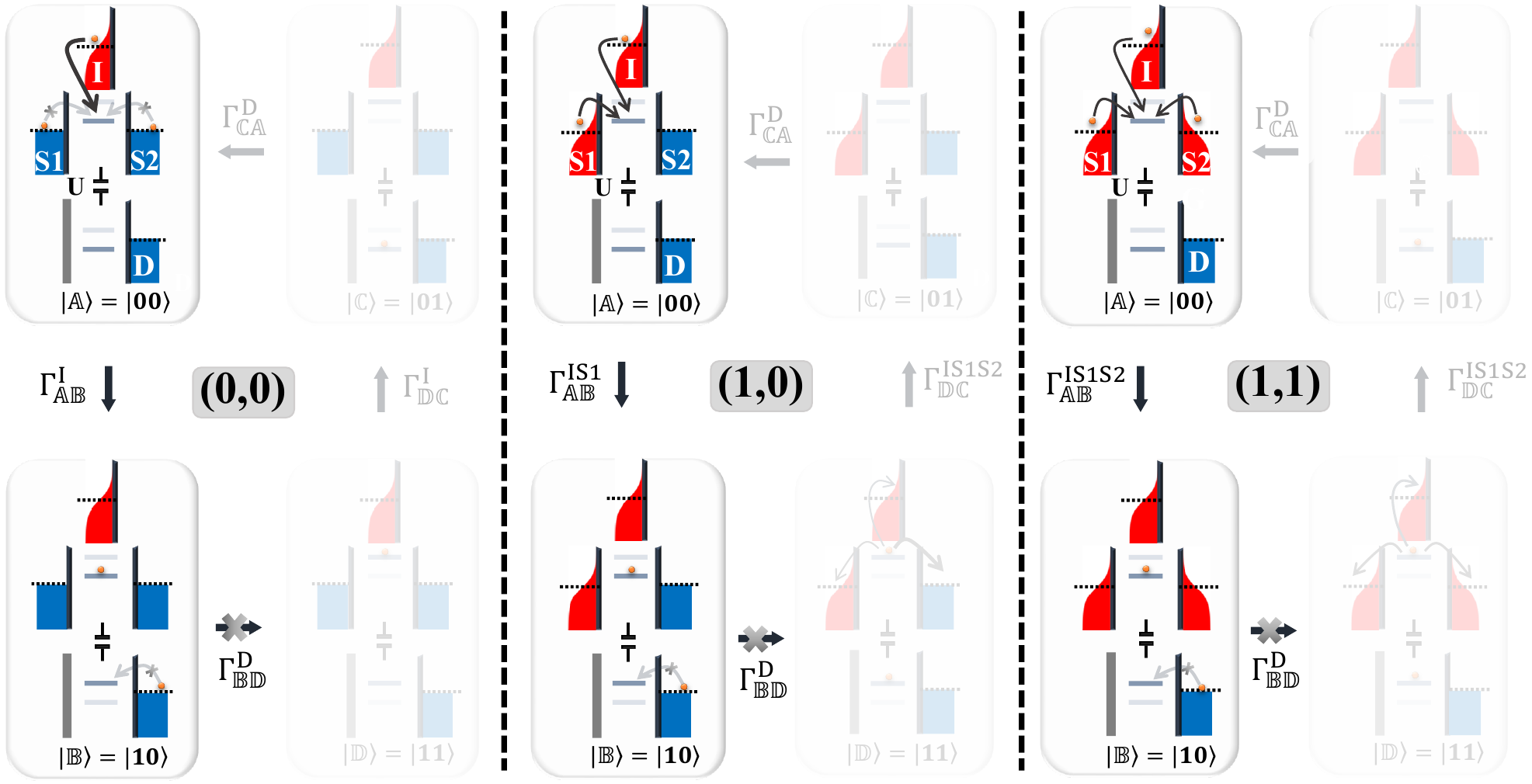}
  \caption{Thermal-NOR gate anticlockwise cycle: Since no hot lead is attached to $\rm{QD_b}$, during the second step of each input logic operation, the cold D lead cannot inject an electron into the excited state of $\rm{QD_b}$. Thus, none of the anti-clockwise cycles can be completed, and similar to the buffer, NOT, and OR gates, the output current arising from the anti-clockwise cycles in the NOR gate is ultimately negligible. The parameters for the NOR gate are: $\mu_{\rm{S1}}=\mu_{\rm{S2}}=-95;\mu\rm{eV}$, $\mu_{\rm{I}}=-90;\mu\rm{eV}$, and $\mu_{\rm{D}}=21;\mu\rm{eV}$.}
    \label{NOR-Gate-Anti}
\end{figure}

\subsection{Thermal-NAND Gate}\label{Appendix-C7}

Following a similar idea, a thermal-NAND gate can be constructed by combining a thermal-NOT gate with the thermal-AND gate circuit. In this configuration, ${\rm QD_a}$ is coupled to an additional inverting lead I, along with the two input reservoirs S1 and S2, whereas ${\rm QD_b}$ is coupled to D and C-lead, as in the thermal-AND gate.

\noindent
\begin{minipage}{0.68\textwidth}
As before, the input logic states are controlled by the temperatures of the leads S1 and S2. Meanwhile, similar to the NOT gate, the output logic is determined by the magnitude of the heat current associated with the inverting lead, i.e., $J_{\rm{Q}}^{\rm{I}}$. Under these conditions, the steady-state transition rates [Fig.~\ref{TC_NAND}] satisfy relations analogous to those given in Eq.~\eqref{gamma-appendix}:
\begin{equation}\label{gamma5}
\begin{split}
\Gamma^{\rm{DC}}_\mathbb{AC}=\Gamma^{\rm{IS1S2}}_\mathbb{CD}=\Gamma^{\rm{DC}}_\mathbb{DB}=\Gamma^{\rm{IS1S2}}_\mathbb{BA}\equiv\Gamma^{{\rm{NAND}}}_{\circlearrowright};\;\Gamma^{\rm{IS1S2}}_\mathbb{AB}=\Gamma^{\rm{DC}}_\mathbb{BD}=\Gamma^{\rm{IS1S2}}_\mathbb{DC}=\Gamma^{\rm{DC}}_\mathbb{CA}\equiv\Gamma^{{\rm{NAND}}}_{\circlearrowleft}
\end{split}
\end{equation}
where $\Gamma^{\rm{IS1S2}}_\mathbb{ij}\equiv\Gamma^{\rm{I}}_\mathbb{ij}+\Gamma^{\rm{S1}}_\mathbb{ij}+\Gamma^{\rm{S2}}_\mathbb{ij}$ and $\Gamma^{\rm{DC}}_\mathbb{ij}\equiv\Gamma^{\rm{D}}_\mathbb{ij}+\Gamma^{\rm{C}}_\mathbb{ij}$, with $\Gamma^{\rm{NAND}}_{\circlearrowright}=-\Gamma^{\rm{NAND}}_{\circlearrowleft}$.
Substituting these relations into Eq.~\eqref{S15}, the steady-state energy and particle currents associated with the inverting lead are obtained as
\begin{equation}\label{JNAND}
\begin{split}
J_{\rm{E}}^{\rm{I}}=\varepsilon_{\rm{a}}\Gamma_\mathbb{AB}^{\rm{I}}+(\varepsilon_{\rm{a}}+U)\Gamma_\mathbb{CD}^{\rm{I}};\qquad
J_{\rm{N}}^{\rm{I}}=\Gamma_\mathbb{AB}^{\rm{I}}+\Gamma_\mathbb{CD}^{\rm{I}}.
\end{split}
\end{equation}
Accordingly, the heat current associated with the inverting lead can be written as
\begin{equation}\label{JQNAND}
\begin{split}
J_{\rm{Q}}^{\rm{I}}=J_{\rm{E}}^{\rm{I}}-\mu_{\rm{I}}J_{\rm{N}}^{\rm{I}}=&(\varepsilon_{\rm{a}}-\mu_{\rm{I}})\Gamma_\mathbb{AB}^{\rm{I}}+(\varepsilon_{\rm{a}}+U-\mu_{\rm{I}})\Gamma_\mathbb{CD}^{\rm{I}}\\
=&(\varepsilon_{\rm{a}}-\mu_{\rm{I}})(\Gamma_\mathbb{AB}^{\rm{I}}-\Gamma_\mathbb{DC}^{\rm{I}})-U\Gamma_\mathbb{DC}^{\rm{I}}\\
=&(\varepsilon_{\rm{a}}-\mu_{\rm{I}})(\Gamma_\mathbb{CD}^{\rm{I}}-\Gamma_\mathbb{BA}^{\rm{I}})+U\Gamma_\mathbb{CD}^{\rm{I}}.
\end{split}
\end{equation}
\end{minipage}
\hfill
\begin{minipage}{0.30\textwidth}
\centering
\includegraphics[width=\linewidth]{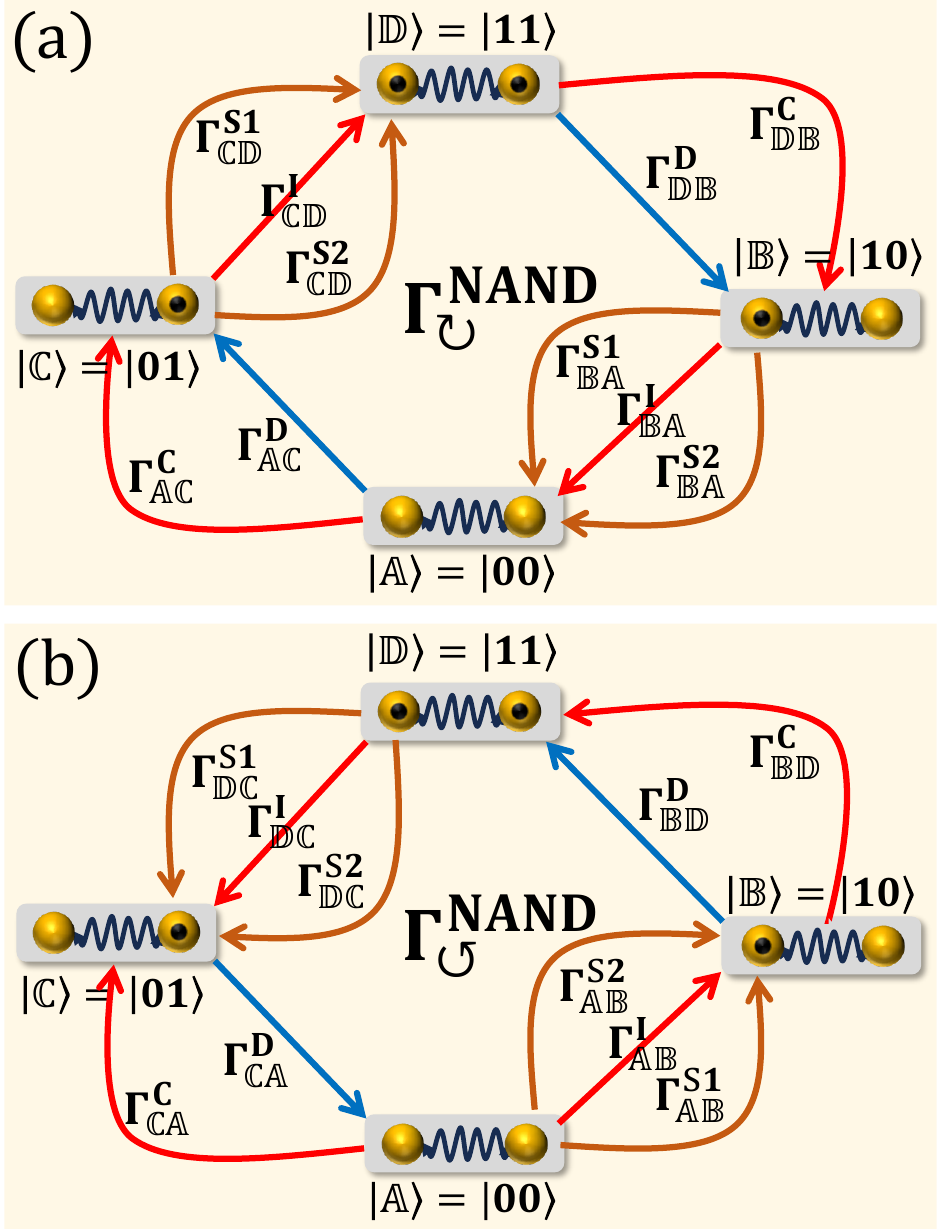}
\captionof{figure}{Diagrammatic representation for clockwise and anti-clockwise transition cycles for NAND gate.}
\label{TC_NAND}
\end{minipage}
\vspace{1em}
\noindent

The above analytical expression of $J_{\rm{Q}}^{\rm{I}}$ determines the thermal logic values. Similar to the NOR gate, the logical operation of the NAND gate can be understood as a combination of the NOT and AND gate operations. Like in the AND gate configuration, both clockwise and anti-clockwise transition cycles may occur due to the coupling of $\mathrm{QD_b}$ with the hot C lead. Therefore, as indicated by Eq.~\eqref{JQNAND}, the relevant transitions are those between $|\mathbb{A}\rangle$ and $|\mathbb{B}\rangle$, and between $|\mathbb{C}\rangle$ and $|\mathbb{D}\rangle$.
\begin{figure}[h]
   \centering    
\includegraphics[width=0.96\columnwidth,height=8.4cm]{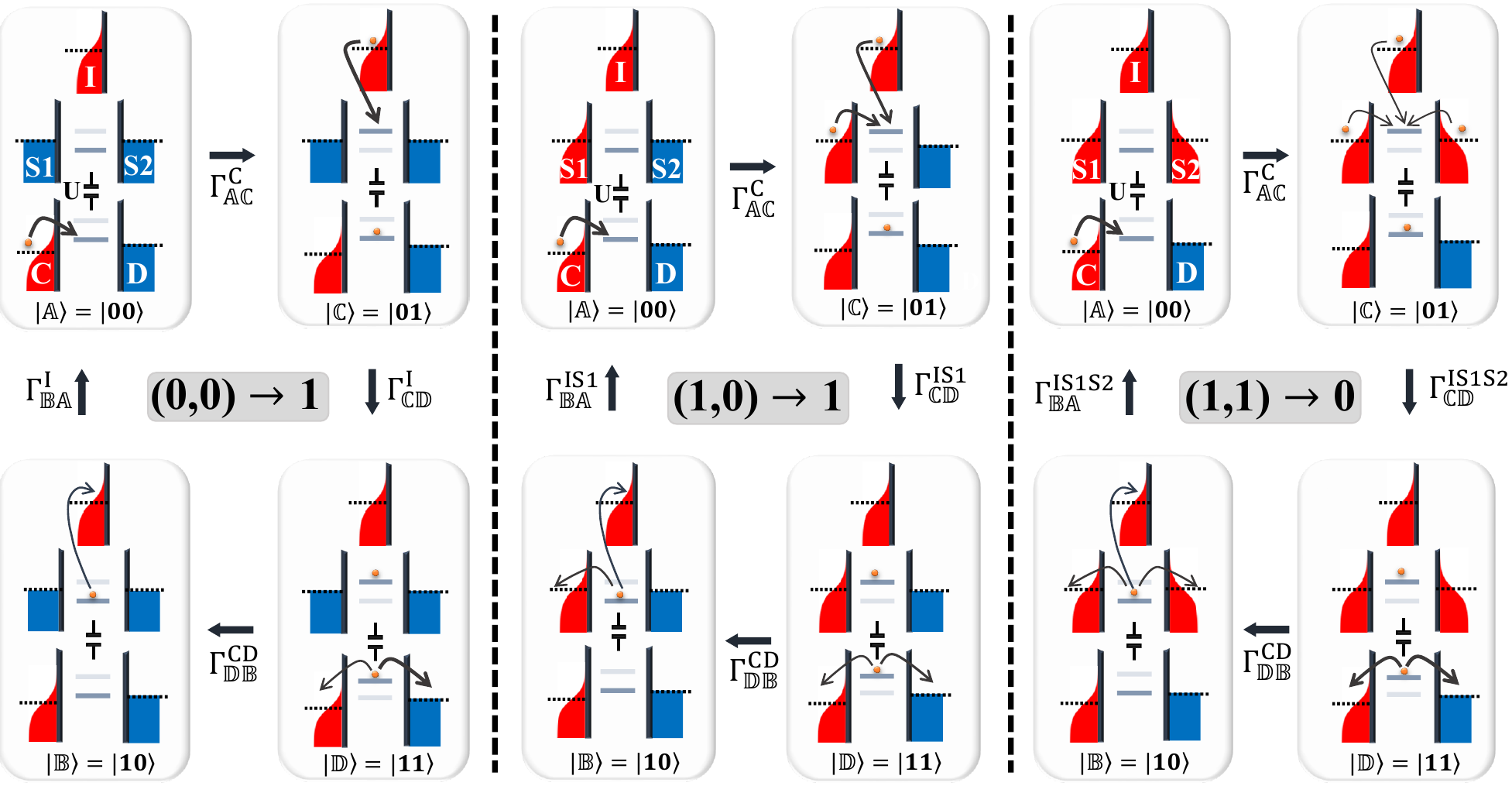}
  \caption{A thermal-NAND gate is realized by applying a NOT operation to an AND gate. Like the AND gate, the NAND gate cycle can also run in both directions. Only the $\circlearrowright$ cycle is shown here; cf.~Fig.~\ref{NAND-Gate-Anti} for the $\circlearrowleft$ cycle. Since the output is measured at the I-lead, the presence of any cold input (logic-0) ensures sufficient heat current at I (due to both $\circlearrowright$  and $\circlearrowleft$ cycles), resulting in output logic-1. Only when both inputs are 1, S and I equally contribute in heat-flow, thus heat-current at I drops significantly, resulting in logic-0. Here, arrow widths indicate relative heat-flow strengths.  The parameters for the NAND gate are: $\mu_{\rm{S1}}=\mu_{\rm{S2}}=16\;\mu\rm{eV}$, $\mu_{\rm{C}}=-16\;\mu\rm{eV}$, $\mu_{\rm{I}}=16\;\mu\rm{eV}$, and $\mu_{\rm{D}}=9\;\mu\rm{eV}$.}
    \label{NAND Gate}
\end{figure}
In the clockwise transition cycle [See Fig.~\ref{NAND Gate}], the I-lead participates in both the excitation process from $|\mathbb{C}\rangle$ to $|\mathbb{D}\rangle$ and the de-excitation process from $|\mathbb{B}\rangle$ to $|\mathbb{A}\rangle$, resulting in a finite heat current flowing from the I-lead. The magnitude of this current depends on the input logic states; it gradually decreases as the number of hot input-sources increases.

\begin{figure}[h!]
   \centering    
\includegraphics[width=\columnwidth,height=8.6cm]{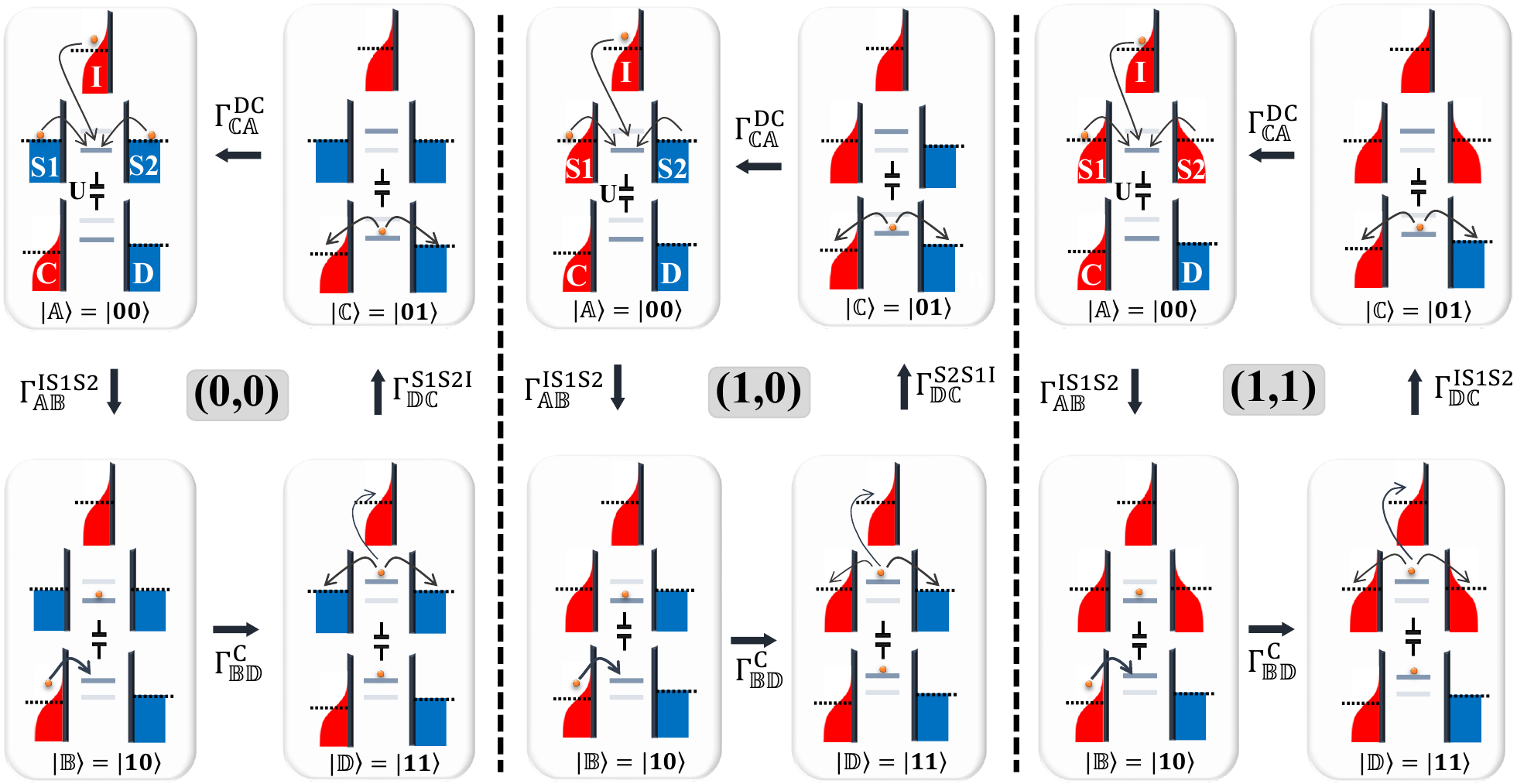}
  \caption{Thermal-NAND gate anticlockwise cycle: Unlike the AND gate, in this case, the Fermi level of the C lead is kept significantly higher than that in the AND gate configuration. As a result, the C lead can supply electrons into $\rm{QD_b}$ with a nonzero transition rate, $\Gamma_{\mathbb{BD}}^{\rm C}$, resulting in a finite output current measured at the I lead. When this current is combined with the clockwise cycle current, it produces a large output current at the I lead. The parameters for the NAND gate are: $\mu_{\rm{S1}}=\mu_{\rm{S2}}=16;\mu\rm{eV}$, $\mu_{\rm{C}}=-16\;\mu\rm{eV}$, $\mu_{\rm{I}}=16\;\mu\rm{eV}$, and $\mu_{\rm{D}}=9\;\mu\rm{eV}$.}
    \label{NAND-Gate-Anti}
\end{figure}

On the other hand, in the anti-clockwise transition cycle~[Fig.~\ref{NAND-Gate-Anti}], a finite heat current is also associated with the I-lead, as it contributes to both the excitation from $|\mathbb{A}\rangle$ to $|\mathbb{B}\rangle$ and the de-excitation from $|\mathbb{D}\rangle$ to $|\mathbb{C}\rangle$. However, in this case, the heat current is comparatively small and insensitive to changes in the input logic states.
\\
\\
$\bullet$ \textbf{For input logic (0,0)}, both the excitation process from $|\mathbb{C}\rangle$ to $|\mathbb{D}\rangle$ and the de-excitation process from $|\mathbb{B}\rangle$ to $|\mathbb{A}\rangle$ are predominantly governed by the I-lead due to the presence of both cold source leads [See Fig.~\ref{NAND Gate}]. Consequently, the clockwise transition cycle generates a substantial output current. In addition to the clockwise cycle [Fig.~\ref{NAND Gate}], the anti-clockwise cycle [Fig.~\ref{NAND-Gate-Anti}] also operates simultaneously and contributes to the overall output current. Owing to the combined contribution of both cycles, particularly the dominant clockwise cycle, a significantly large output heat current $J_{\rm{Q}}^{\rm{I}}$ flows from the I-lead. This current remains well above the threshold value $J_{\rm{Q}}^{1}$, thereby corresponding to the output logic-1.
\\
\\
$\bullet$ \textbf{For input logic (1,0) or (0,1)}, the presence of one hot source lead reduces the contribution of the I-lead in the transitions from $|\mathbb{C}\rangle$ to $|\mathbb{D}\rangle$ and from $|\mathbb{B}\rangle$ to $|\mathbb{A}\rangle$. Consequently, the heat current associated with the I-lead in the clockwise transition cycle [Fig.~\ref{NAND Gate}] decreases, although it remains appreciable since one of the source leads still acts as a cold bath. Alongside the clockwise cycle, the anti-clockwise cycle~[Fig.~\ref{NAND-Gate-Anti}] also persists; however, as discussed earlier, its contribution is not significantly affected by changes in the input logic states. Therefore, due to the combined contributions from both cycles, the output current $J_{\rm{Q}}^{\rm{I}}$ still remains above the threshold value $J_{\rm{Q}}^{1}$, thereby corresponding to the output logic-1.
\\
\\
$\bullet$ \textbf{For input logic (1,1)}, due to the absence of any cold source lead, the contribution of the I-lead to the transitions from $|\mathbb{C}\rangle$ to $|\mathbb{D}\rangle$ and from $|\mathbb{B}\rangle$ to $|\mathbb{A}\rangle$ decreases significantly [Fig.~\ref{NAND Gate}]. Consequently, the output current associated with the clockwise transition cycle is substantially reduced. Although the anti-clockwise cycle~[Fig.~\ref{NAND-Gate-Anti}] still contributes to the output current to some extent, the dominant contribution originates from the clockwise cycle. Therefore, as this contribution decreases markedly, the overall output current falls below the lower threshold value $J_{\rm{Q}}^{0}$, thereby corresponding to the output logic-0.


\begin{thebibliography}{70}%
	\makeatletter
	\providecommand \@ifxundefined [1]{%
		\@ifx{#1\undefined}
	}%
	\providecommand \@ifnum [1]{%
		\ifnum #1\expandafter \@firstoftwo
		\else \expandafter \@secondoftwo
		\fi
	}%
	\providecommand \@ifx [1]{%
		\ifx #1\expandafter \@firstoftwo
		\else \expandafter \@secondoftwo
		\fi
	}%
	\providecommand \natexlab [1]{#1}%
	\providecommand \enquote  [1]{``#1''}%
	\providecommand \bibnamefont  [1]{#1}%
	\providecommand \bibfnamefont [1]{#1}%
	\providecommand \citenamefont [1]{#1}%
	\providecommand \href@noop [0]{\@secondoftwo}%
	\providecommand \href [0]{\begingroup \@sanitize@url \@href}%
	\providecommand \@href[1]{\@@startlink{#1}\@@href}%
	\providecommand \@@href[1]{\endgroup#1\@@endlink}%
	\providecommand \@sanitize@url [0]{\catcode `\\12\catcode `\$12\catcode
		`\&12\catcode `\#12\catcode `\^12\catcode `\_12\catcode `\%12\relax}%
	\providecommand \@@startlink[1]{}%
	\providecommand \@@endlink[0]{}%
	\providecommand \url  [0]{\begingroup\@sanitize@url \@url }%
	\providecommand \@url [1]{\endgroup\@href {#1}{\urlprefix }}%
	\providecommand \urlprefix  [0]{URL }%
	\providecommand \Eprint [0]{\href }%
	\providecommand \doibase [0]{https://doi.org/}%
	\providecommand \selectlanguage [0]{\@gobble}%
	\providecommand \bibinfo  [0]{\@secondoftwo}%
	\providecommand \bibfield  [0]{\@secondoftwo}%
	\providecommand \translation [1]{[#1]}%
	\providecommand \BibitemOpen [0]{}%
	\providecommand \bibitemStop [0]{}%
	\providecommand \bibitemNoStop [0]{.\EOS\space}%
	\providecommand \EOS [0]{\spacefactor3000\relax}%
	\providecommand \BibitemShut  [1]{\csname bibitem#1\endcsname}%
	\let\auto@bib@innerbib\@empty
	\bibitem [{\citenamefont {Maurand}\ and\ \citenamefont
		{Jehl}(2022)}]{maurand2022transistor}%
	\BibitemOpen
	\bibfield  {author} {\bibinfo {author} {\bibfnamefont {R.}~\bibnamefont
			{Maurand}}\ and\ \bibinfo {author} {\bibfnamefont {X.}~\bibnamefont {Jehl}},\
	}\bibfield  {title} {\bibinfo {title} {Transistor qubits heat up},\ }\href
	{https://doi.org/10.1038/s41928-022-00736-8} {\bibfield  {journal} {\bibinfo
			{journal} {Nature Electronics}\ }\textbf {\bibinfo {volume} {5}},\ \bibinfo
		{pages} {131} (\bibinfo {year} {2022})}\BibitemShut {NoStop}%
	\bibitem [{\citenamefont {Aamir}\ \emph {et~al.}(2025)\citenamefont {Aamir},
		\citenamefont {Jamet~Suria}, \citenamefont {Mar{\'i}n~Guzm{\'a}n},
		\citenamefont {Castillo-Moreno}, \citenamefont {Epstein}, \citenamefont
		{Yunger~Halpern},\ and\ \citenamefont {Gasparinetti}}]{aamir2025thermally}%
	\BibitemOpen
	\bibfield  {author} {\bibinfo {author} {\bibfnamefont {M.~A.}\ \bibnamefont
			{Aamir}}, \bibinfo {author} {\bibfnamefont {P.}~\bibnamefont {Jamet~Suria}},
		\bibinfo {author} {\bibfnamefont {J.~A.}\ \bibnamefont
			{Mar{\'i}n~Guzm{\'a}n}}, \bibinfo {author} {\bibfnamefont {C.}~\bibnamefont
			{Castillo-Moreno}}, \bibinfo {author} {\bibfnamefont {J.~M.}\ \bibnamefont
			{Epstein}}, \bibinfo {author} {\bibfnamefont {N.}~\bibnamefont
			{Yunger~Halpern}},\ and\ \bibinfo {author} {\bibfnamefont {S.}~\bibnamefont
			{Gasparinetti}},\ }\bibfield  {title} {\bibinfo {title} {Thermally driven
			quantum refrigerator autonomously resets a superconducting qubit},\ }\href
	{https://doi.org/10.1038/s41567-024-02708-5} {\bibfield  {journal} {\bibinfo
			{journal} {Nature Physics}\ }\textbf {\bibinfo {volume} {21}},\ \bibinfo
		{pages} {318} (\bibinfo {year} {2025})}\BibitemShut {NoStop}%
	\bibitem [{\citenamefont {Blok}\ and\ \citenamefont
		{Landi}(2025)}]{blok2025quantum}%
	\BibitemOpen
	\bibfield  {author} {\bibinfo {author} {\bibfnamefont {M.~S.}\ \bibnamefont
			{Blok}}\ and\ \bibinfo {author} {\bibfnamefont {G.~T.}\ \bibnamefont
			{Landi}},\ }\bibfield  {title} {\bibinfo {title} {Quantum thermodynamics for
			quantum computing},\ }\href {https://doi.org/10.1038/s41567-024-02764-x}
	{\bibfield  {journal} {\bibinfo  {journal} {Nature Physics}\ }\textbf
		{\bibinfo {volume} {21}},\ \bibinfo {pages} {187} (\bibinfo {year}
		{2025})}\BibitemShut {NoStop}%
	\bibitem [{\citenamefont {Karimi}\ \emph {et~al.}(2024)\citenamefont {Karimi},
		\citenamefont {Steffensen}, \citenamefont {Higginbotham}, \citenamefont
		{Marcus}, \citenamefont {Levy~Yeyati},\ and\ \citenamefont
		{Pekola}}]{Karimi2024bolometric}%
	\BibitemOpen
	\bibfield  {author} {\bibinfo {author} {\bibfnamefont {B.}~\bibnamefont
			{Karimi}}, \bibinfo {author} {\bibfnamefont {G.~O.}\ \bibnamefont
			{Steffensen}}, \bibinfo {author} {\bibfnamefont {A.~P.}\ \bibnamefont
			{Higginbotham}}, \bibinfo {author} {\bibfnamefont {C.~M.}\ \bibnamefont
			{Marcus}}, \bibinfo {author} {\bibfnamefont {A.}~\bibnamefont
			{Levy~Yeyati}},\ and\ \bibinfo {author} {\bibfnamefont {J.~P.}\ \bibnamefont
			{Pekola}},\ }\bibfield  {title} {\bibinfo {title} {Bolometric detection of
			josephson radiation},\ }\href {https://doi.org/10.1038/s41565-024-01770-7}
	{\bibfield  {journal} {\bibinfo  {journal} {Nature Nanotechnology}\ }\textbf
		{\bibinfo {volume} {19}},\ \bibinfo {pages} {1613} (\bibinfo {year}
		{2024})}\BibitemShut {NoStop}%
	\bibitem [{\citenamefont {Simbierowicz}\ \emph {et~al.}(2024)\citenamefont
		{Simbierowicz}, \citenamefont {Borrelli}, \citenamefont {Monarkha},
		\citenamefont {Nuutinen},\ and\ \citenamefont
		{Lake}}]{simbierowicz2024inherent}%
	\BibitemOpen
	\bibfield  {author} {\bibinfo {author} {\bibfnamefont {S.}~\bibnamefont
			{Simbierowicz}}, \bibinfo {author} {\bibfnamefont {M.}~\bibnamefont
			{Borrelli}}, \bibinfo {author} {\bibfnamefont {V.}~\bibnamefont {Monarkha}},
		\bibinfo {author} {\bibfnamefont {V.}~\bibnamefont {Nuutinen}},\ and\
		\bibinfo {author} {\bibfnamefont {R.~E.}\ \bibnamefont {Lake}},\ }\bibfield
	{title} {\bibinfo {title} {Inherent thermal-noise problem in addressing
			qubits},\ }\href {https://doi.org/10.1103/PRXQuantum.5.030302} {\bibfield
		{journal} {\bibinfo  {journal} {PRX Quantum}\ }\textbf {\bibinfo {volume}
			{5}},\ \bibinfo {pages} {030302} (\bibinfo {year} {2024})}\BibitemShut
	{NoStop}%
	\bibitem [{\citenamefont {Champain}\ \emph {et~al.}(2025)\citenamefont
		{Champain}, \citenamefont {Boschetto}, \citenamefont {Niebojewski},
		\citenamefont {Bertrand}, \citenamefont {Mauro}, \citenamefont {Bassi},
		\citenamefont {Schmitt}, \citenamefont {Jehl}, \citenamefont {Zihlmann},
		\citenamefont {Maurand}, \citenamefont {Niquet}, \citenamefont {Winkelmann},
		\citenamefont {Franceschi}, \citenamefont {Martinez},\ and\ \citenamefont
		{Brun}}]{champain2025heatresilientholespinqubit}%
	\BibitemOpen
	\bibfield  {author} {\bibinfo {author} {\bibfnamefont {V.}~\bibnamefont
			{Champain}}, \bibinfo {author} {\bibfnamefont {G.}~\bibnamefont {Boschetto}},
		\bibinfo {author} {\bibfnamefont {H.}~\bibnamefont {Niebojewski}}, \bibinfo
		{author} {\bibfnamefont {B.}~\bibnamefont {Bertrand}}, \bibinfo {author}
		{\bibfnamefont {L.}~\bibnamefont {Mauro}}, \bibinfo {author} {\bibfnamefont
			{M.}~\bibnamefont {Bassi}}, \bibinfo {author} {\bibfnamefont
			{V.}~\bibnamefont {Schmitt}}, \bibinfo {author} {\bibfnamefont
			{X.}~\bibnamefont {Jehl}}, \bibinfo {author} {\bibfnamefont {S.}~\bibnamefont
			{Zihlmann}}, \bibinfo {author} {\bibfnamefont {R.}~\bibnamefont {Maurand}},
		\bibinfo {author} {\bibfnamefont {Y.~M.}\ \bibnamefont {Niquet}}, \bibinfo
		{author} {\bibfnamefont {C.~B.}\ \bibnamefont {Winkelmann}}, \bibinfo
		{author} {\bibfnamefont {S.~D.}\ \bibnamefont {Franceschi}}, \bibinfo
		{author} {\bibfnamefont {B.}~\bibnamefont {Martinez}},\ and\ \bibinfo
		{author} {\bibfnamefont {B.}~\bibnamefont {Brun}},\ }\href
	{https://arxiv.org/abs/2509.15823} {\bibinfo {title} {A heat-resilient hole
			spin qubit in silicon}} (\bibinfo {year} {2025}),\ \Eprint
	{https://arxiv.org/abs/2509.15823} {arXiv:2509.15823} \BibitemShut {NoStop}%
	\bibitem [{\citenamefont {Ro{\ss}nagel}\ \emph {et~al.}(2016)\citenamefont
		{Ro{\ss}nagel}, \citenamefont {Dawkins}, \citenamefont {Tolazzi},
		\citenamefont {Abah}, \citenamefont {Lutz}, \citenamefont {Schmidt-Kaler},\
		and\ \citenamefont {Singer}}]{rossnagel2016single}%
	\BibitemOpen
	\bibfield  {author} {\bibinfo {author} {\bibfnamefont {J.}~\bibnamefont
			{Ro{\ss}nagel}}, \bibinfo {author} {\bibfnamefont {S.~T.}\ \bibnamefont
			{Dawkins}}, \bibinfo {author} {\bibfnamefont {K.~N.}\ \bibnamefont
			{Tolazzi}}, \bibinfo {author} {\bibfnamefont {O.}~\bibnamefont {Abah}},
		\bibinfo {author} {\bibfnamefont {E.}~\bibnamefont {Lutz}}, \bibinfo {author}
		{\bibfnamefont {F.}~\bibnamefont {Schmidt-Kaler}},\ and\ \bibinfo {author}
		{\bibfnamefont {K.}~\bibnamefont {Singer}},\ }\bibfield  {title} {\bibinfo
		{title} {A single-atom heat engine},\ }\href
	{https://doi.org/10.1126/science.aad6320} {\bibfield  {journal} {\bibinfo
			{journal} {Science}\ }\textbf {\bibinfo {volume} {352}},\ \bibinfo {pages}
		{325} (\bibinfo {year} {2016})}\BibitemShut {NoStop}%
	\bibitem [{\citenamefont {Thierschmann}\ \emph
		{et~al.}(2015{\natexlab{a}})\citenamefont {Thierschmann}, \citenamefont
		{S{\'a}nchez}, \citenamefont {Sothmann}, \citenamefont {Arnold},
		\citenamefont {Heyn}, \citenamefont {Hansen}, \citenamefont {Buhmann},\ and\
		\citenamefont {Molenkamp}}]{thierschmann2015three}%
	\BibitemOpen
	\bibfield  {author} {\bibinfo {author} {\bibfnamefont {H.}~\bibnamefont
			{Thierschmann}}, \bibinfo {author} {\bibfnamefont {R.}~\bibnamefont
			{S{\'a}nchez}}, \bibinfo {author} {\bibfnamefont {B.}~\bibnamefont
			{Sothmann}}, \bibinfo {author} {\bibfnamefont {F.}~\bibnamefont {Arnold}},
		\bibinfo {author} {\bibfnamefont {C.}~\bibnamefont {Heyn}}, \bibinfo {author}
		{\bibfnamefont {W.}~\bibnamefont {Hansen}}, \bibinfo {author} {\bibfnamefont
			{H.}~\bibnamefont {Buhmann}},\ and\ \bibinfo {author} {\bibfnamefont {L.~W.}\
			\bibnamefont {Molenkamp}},\ }\bibfield  {title} {\bibinfo {title}
		{Three-terminal energy harvester with coupled quantum dots},\ }\href
	{https://doi.org/10.1038/nnano.2015.176} {\bibfield  {journal} {\bibinfo
			{journal} {Nature Nanotechnology}\ }\textbf {\bibinfo {volume} {10}},\
		\bibinfo {pages} {854} (\bibinfo {year} {2015}{\natexlab{a}})}\BibitemShut
	{NoStop}%
	\bibitem [{\citenamefont {Jezouin}\ \emph {et~al.}(2013)\citenamefont
		{Jezouin}, \citenamefont {Parmentier}, \citenamefont {Anthore}, \citenamefont
		{Gennser}, \citenamefont {Cavanna}, \citenamefont {Jin},\ and\ \citenamefont
		{Pierre}}]{jezouin2013quantum}%
	\BibitemOpen
	\bibfield  {author} {\bibinfo {author} {\bibfnamefont {S.}~\bibnamefont
			{Jezouin}}, \bibinfo {author} {\bibfnamefont {F.~D.}\ \bibnamefont
			{Parmentier}}, \bibinfo {author} {\bibfnamefont {A.}~\bibnamefont {Anthore}},
		\bibinfo {author} {\bibfnamefont {U.}~\bibnamefont {Gennser}}, \bibinfo
		{author} {\bibfnamefont {A.}~\bibnamefont {Cavanna}}, \bibinfo {author}
		{\bibfnamefont {Y.}~\bibnamefont {Jin}},\ and\ \bibinfo {author}
		{\bibfnamefont {F.}~\bibnamefont {Pierre}},\ }\bibfield  {title} {\bibinfo
		{title} {Quantum limit of heat flow across a single electronic channel},\
	}\href {https://doi.org/10.1126/science.1241912} {\bibfield  {journal}
		{\bibinfo  {journal} {Science}\ }\textbf {\bibinfo {volume} {342}},\ \bibinfo
		{pages} {601} (\bibinfo {year} {2013})}\BibitemShut {NoStop}%
	\bibitem [{\citenamefont {Banerjee}\ \emph {et~al.}(2017)\citenamefont
		{Banerjee}, \citenamefont {Heiblum}, \citenamefont {Rosenblatt},
		\citenamefont {Oreg}, \citenamefont {Feldman}, \citenamefont {Stern},\ and\
		\citenamefont {Umansky}}]{banerjee2017observed}%
	\BibitemOpen
	\bibfield  {author} {\bibinfo {author} {\bibfnamefont {M.}~\bibnamefont
			{Banerjee}}, \bibinfo {author} {\bibfnamefont {M.}~\bibnamefont {Heiblum}},
		\bibinfo {author} {\bibfnamefont {A.}~\bibnamefont {Rosenblatt}}, \bibinfo
		{author} {\bibfnamefont {Y.}~\bibnamefont {Oreg}}, \bibinfo {author}
		{\bibfnamefont {D.~E.}\ \bibnamefont {Feldman}}, \bibinfo {author}
		{\bibfnamefont {A.}~\bibnamefont {Stern}},\ and\ \bibinfo {author}
		{\bibfnamefont {V.}~\bibnamefont {Umansky}},\ }\bibfield  {title} {\bibinfo
		{title} {Observed quantization of anyonic heat flow},\ }\href
	{https://doi.org/10.1038/nature22052} {\bibfield  {journal} {\bibinfo
			{journal} {Nature}\ }\textbf {\bibinfo {volume} {545}},\ \bibinfo {pages}
		{75} (\bibinfo {year} {2017})}\BibitemShut {NoStop}%
	\bibitem [{\citenamefont {Dutta}\ \emph {et~al.}(2022)\citenamefont {Dutta},
		\citenamefont {Umansky}, \citenamefont {Banerjee},\ and\ \citenamefont
		{Heiblum}}]{dutta2022isolated}%
	\BibitemOpen
	\bibfield  {author} {\bibinfo {author} {\bibfnamefont {B.}~\bibnamefont
			{Dutta}}, \bibinfo {author} {\bibfnamefont {V.}~\bibnamefont {Umansky}},
		\bibinfo {author} {\bibfnamefont {M.}~\bibnamefont {Banerjee}},\ and\
		\bibinfo {author} {\bibfnamefont {M.}~\bibnamefont {Heiblum}},\ }\bibfield
	{title} {\bibinfo {title} {Isolated ballistic non-abelian interface
			channel},\ }\href {https://doi.org/10.1126/science.abm6571} {\bibfield
		{journal} {\bibinfo  {journal} {Science}\ }\textbf {\bibinfo {volume}
			{377}},\ \bibinfo {pages} {1198} (\bibinfo {year} {2022})}\BibitemShut
	{NoStop}%
	\bibitem [{\citenamefont {Myers}\ \emph {et~al.}(2022)\citenamefont {Myers},
		\citenamefont {Abah},\ and\ \citenamefont {Deffner}}]{myers2022quantum}%
	\BibitemOpen
	\bibfield  {author} {\bibinfo {author} {\bibfnamefont {N.~M.}\ \bibnamefont
			{Myers}}, \bibinfo {author} {\bibfnamefont {O.}~\bibnamefont {Abah}},\ and\
		\bibinfo {author} {\bibfnamefont {S.}~\bibnamefont {Deffner}},\ }\bibfield
	{title} {\bibinfo {title} {Quantum thermodynamic devices: From theoretical
			proposals to experimental reality},\ }\href
	{https://doi.org/10.1116/5.0083192} {\bibfield  {journal} {\bibinfo
			{journal} {AVS Quantum Science}\ }\textbf {\bibinfo {volume} {4}},\ \bibinfo
		{pages} {027101} (\bibinfo {year} {2022})},\ \bibinfo {note} {and the
		references therein}\BibitemShut {NoStop}%
	\bibitem [{\citenamefont {Joulain}\ \emph {et~al.}(2016)\citenamefont
		{Joulain}, \citenamefont {Drevillon}, \citenamefont {Ezzahri},\ and\
		\citenamefont {Ordonez-Miranda}}]{joulain2016quantum}%
	\BibitemOpen
	\bibfield  {author} {\bibinfo {author} {\bibfnamefont {K.}~\bibnamefont
			{Joulain}}, \bibinfo {author} {\bibfnamefont {J.}~\bibnamefont {Drevillon}},
		\bibinfo {author} {\bibfnamefont {Y.}~\bibnamefont {Ezzahri}},\ and\ \bibinfo
		{author} {\bibfnamefont {J.}~\bibnamefont {Ordonez-Miranda}},\ }\bibfield
	{title} {\bibinfo {title} {Quantum thermal transistor},\ }\href
	{https://doi.org/10.1103/PhysRevLett.116.200601} {\bibfield  {journal}
		{\bibinfo  {journal} {Phys. Rev. Lett.}\ }\textbf {\bibinfo {volume} {116}},\
		\bibinfo {pages} {200601} (\bibinfo {year} {2016})}\BibitemShut {NoStop}%
	\bibitem [{\citenamefont {Dutta}\ \emph {et~al.}(2019)\citenamefont {Dutta},
		\citenamefont {Majidi}, \citenamefont {García~Corral}, \citenamefont
		{Erdman}, \citenamefont {Florens}, \citenamefont {Costi}, \citenamefont
		{Courtois},\ and\ \citenamefont {Winkelmann}}]{dutta2019direct}%
	\BibitemOpen
	\bibfield  {author} {\bibinfo {author} {\bibfnamefont {B.}~\bibnamefont
			{Dutta}}, \bibinfo {author} {\bibfnamefont {D.}~\bibnamefont {Majidi}},
		\bibinfo {author} {\bibfnamefont {A.}~\bibnamefont {García~Corral}},
		\bibinfo {author} {\bibfnamefont {P.~A.}\ \bibnamefont {Erdman}}, \bibinfo
		{author} {\bibfnamefont {S.}~\bibnamefont {Florens}}, \bibinfo {author}
		{\bibfnamefont {T.~A.}\ \bibnamefont {Costi}}, \bibinfo {author}
		{\bibfnamefont {H.}~\bibnamefont {Courtois}},\ and\ \bibinfo {author}
		{\bibfnamefont {C.~B.}\ \bibnamefont {Winkelmann}},\ }\bibfield  {title}
	{\bibinfo {title} {Direct probe of the seebeck coefficient in a
			kondo-correlated single-quantum-dot transistor},\ }\href
	{https://doi.org/10.1021/acs.nanolett.8b04398} {\bibfield  {journal}
		{\bibinfo  {journal} {Nano Letters}\ }\textbf {\bibinfo {volume} {19}},\
		\bibinfo {pages} {506} (\bibinfo {year} {2019})}\BibitemShut {NoStop}%
	\bibitem [{\citenamefont {Perrin}\ \emph {et~al.}(2015)\citenamefont {Perrin},
		\citenamefont {Burzurí},\ and\ \citenamefont {van~der
			Zant}}]{perrin2015single}%
	\BibitemOpen
	\bibfield  {author} {\bibinfo {author} {\bibfnamefont {M.~L.}\ \bibnamefont
			{Perrin}}, \bibinfo {author} {\bibfnamefont {E.}~\bibnamefont {Burzurí}},\
		and\ \bibinfo {author} {\bibfnamefont {H.~S.~J.}\ \bibnamefont {van~der
				Zant}},\ }\bibfield  {title} {\bibinfo {title} {Single-molecule
			transistors},\ }\href {https://doi.org/10.1039/C4CS00231H} {\bibfield
		{journal} {\bibinfo  {journal} {Chem. Soc. Rev.}\ }\textbf {\bibinfo {volume}
			{44}},\ \bibinfo {pages} {902} (\bibinfo {year} {2015})}\BibitemShut
	{NoStop}%
	\bibitem [{\citenamefont {Gupt}\ \emph {et~al.}(2022)\citenamefont {Gupt},
		\citenamefont {Bhattacharyya}, \citenamefont {Das}, \citenamefont {Datta},
		\citenamefont {Mukherjee},\ and\ \citenamefont {Ghosh}}]{gupt2022PRE}%
	\BibitemOpen
	\bibfield  {author} {\bibinfo {author} {\bibfnamefont {N.}~\bibnamefont
			{Gupt}}, \bibinfo {author} {\bibfnamefont {S.}~\bibnamefont {Bhattacharyya}},
		\bibinfo {author} {\bibfnamefont {B.}~\bibnamefont {Das}}, \bibinfo {author}
		{\bibfnamefont {S.}~\bibnamefont {Datta}}, \bibinfo {author} {\bibfnamefont
			{V.}~\bibnamefont {Mukherjee}},\ and\ \bibinfo {author} {\bibfnamefont
			{A.}~\bibnamefont {Ghosh}},\ }\bibfield  {title} {\bibinfo {title} {Floquet
			quantum thermal transistor},\ }\href
	{https://doi.org/10.1103/PhysRevE.106.024110} {\bibfield  {journal} {\bibinfo
			{journal} {Phys. Rev. E}\ }\textbf {\bibinfo {volume} {106}},\ \bibinfo
		{pages} {024110} (\bibinfo {year} {2022})}\BibitemShut {NoStop}%
	\bibitem [{\citenamefont {Wijesekara}\ \emph {et~al.}(2021)\citenamefont
		{Wijesekara}, \citenamefont {Gunapala},\ and\ \citenamefont
		{Premaratne}}]{wijesekara2021darlington}%
	\BibitemOpen
	\bibfield  {author} {\bibinfo {author} {\bibfnamefont {R.~T.}\ \bibnamefont
			{Wijesekara}}, \bibinfo {author} {\bibfnamefont {S.~D.}\ \bibnamefont
			{Gunapala}},\ and\ \bibinfo {author} {\bibfnamefont {M.}~\bibnamefont
			{Premaratne}},\ }\bibfield  {title} {\bibinfo {title} {Darlington pair of
			quantum thermal transistors},\ }\href
	{https://doi.org/10.1103/PhysRevB.104.045405} {\bibfield  {journal} {\bibinfo
			{journal} {Phys. Rev. B}\ }\textbf {\bibinfo {volume} {104}},\ \bibinfo
		{pages} {045405} (\bibinfo {year} {2021})}\BibitemShut {NoStop}%
	\bibitem [{\citenamefont {Perrin}\ \emph {et~al.}(2016)\citenamefont {Perrin},
		\citenamefont {Galán}, \citenamefont {Eelkema}, \citenamefont {Thijssen},
		\citenamefont {Grozema},\ and\ \citenamefont {van~der
			Zant}}]{perrin2016agate}%
	\BibitemOpen
	\bibfield  {author} {\bibinfo {author} {\bibfnamefont {M.~L.}\ \bibnamefont
			{Perrin}}, \bibinfo {author} {\bibfnamefont {E.}~\bibnamefont {Galán}},
		\bibinfo {author} {\bibfnamefont {R.}~\bibnamefont {Eelkema}}, \bibinfo
		{author} {\bibfnamefont {J.~M.}\ \bibnamefont {Thijssen}}, \bibinfo {author}
		{\bibfnamefont {F.}~\bibnamefont {Grozema}},\ and\ \bibinfo {author}
		{\bibfnamefont {H.~S.~J.}\ \bibnamefont {van~der Zant}},\ }\bibfield  {title}
	{\bibinfo {title} {A gate-tunable single-molecule diode},\ }\href
	{https://doi.org/10.1039/C6NR00735J} {\bibfield  {journal} {\bibinfo
			{journal} {Nanoscale}\ }\textbf {\bibinfo {volume} {8}},\ \bibinfo {pages}
		{8919} (\bibinfo {year} {2016})}\BibitemShut {NoStop}%
	\bibitem [{\citenamefont {Ghosh}\ \emph {et~al.}(2022)\citenamefont {Ghosh},
		\citenamefont {Gupt},\ and\ \citenamefont {Ghosh}}]{shuvadip2022univarsal}%
	\BibitemOpen
	\bibfield  {author} {\bibinfo {author} {\bibfnamefont {S.}~\bibnamefont
			{Ghosh}}, \bibinfo {author} {\bibfnamefont {N.}~\bibnamefont {Gupt}},\ and\
		\bibinfo {author} {\bibfnamefont {A.}~\bibnamefont {Ghosh}},\ }\bibfield
	{title} {\bibinfo {title} {Universal behavior of the coulomb-coupled
			fermionic thermal diode},\ }\href {https://www.mdpi.com/1099-4300/24/12/1810}
	{\bibfield  {journal} {\bibinfo  {journal} {Entropy}\ }\textbf {\bibinfo
			{volume} {24}} (\bibinfo {year} {2022})}\BibitemShut {NoStop}%
	\bibitem [{\citenamefont {Khan}\ \emph {et~al.}(2021)\citenamefont {Khan},
		\citenamefont {Potts}, \citenamefont {Lehmann}, \citenamefont {Thelander},
		\citenamefont {Dick}, \citenamefont {Samuelsson},\ and\ \citenamefont
		{Maisi}}]{khan2021efficient}%
	\BibitemOpen
	\bibfield  {author} {\bibinfo {author} {\bibfnamefont {W.}~\bibnamefont
			{Khan}}, \bibinfo {author} {\bibfnamefont {P.~P.}\ \bibnamefont {Potts}},
		\bibinfo {author} {\bibfnamefont {S.}~\bibnamefont {Lehmann}}, \bibinfo
		{author} {\bibfnamefont {C.}~\bibnamefont {Thelander}}, \bibinfo {author}
		{\bibfnamefont {K.~A.}\ \bibnamefont {Dick}}, \bibinfo {author}
		{\bibfnamefont {P.}~\bibnamefont {Samuelsson}},\ and\ \bibinfo {author}
		{\bibfnamefont {V.~F.}\ \bibnamefont {Maisi}},\ }\bibfield  {title} {\bibinfo
		{title} {Efficient and continuous microwave photoconversion in hybrid
			cavity-semiconductor nanowire double quantum dot diodes},\ }\href
	{https://doi.org/10.1038/s41467-021-25446-1} {\bibfield  {journal} {\bibinfo
			{journal} {Nature Communications}\ }\textbf {\bibinfo {volume} {12}},\
		\bibinfo {pages} {5130} (\bibinfo {year} {2021})}\BibitemShut {NoStop}%
	\bibitem [{\citenamefont {Upadhyay}\ \emph {et~al.}(2024)\citenamefont
		{Upadhyay}, \citenamefont {Golubev}, \citenamefont {Chang}, \citenamefont
		{Thomas}, \citenamefont {Guthrie}, \citenamefont {Peltonen},\ and\
		\citenamefont {Pekola}}]{upadhyay2024microwave}%
	\BibitemOpen
	\bibfield  {author} {\bibinfo {author} {\bibfnamefont {R.}~\bibnamefont
			{Upadhyay}}, \bibinfo {author} {\bibfnamefont {D.~S.}\ \bibnamefont
			{Golubev}}, \bibinfo {author} {\bibfnamefont {Y.-C.}\ \bibnamefont {Chang}},
		\bibinfo {author} {\bibfnamefont {G.}~\bibnamefont {Thomas}}, \bibinfo
		{author} {\bibfnamefont {A.}~\bibnamefont {Guthrie}}, \bibinfo {author}
		{\bibfnamefont {J.~T.}\ \bibnamefont {Peltonen}},\ and\ \bibinfo {author}
		{\bibfnamefont {J.~P.}\ \bibnamefont {Pekola}},\ }\bibfield  {title}
	{\bibinfo {title} {Microwave quantum diode},\ }\href
	{https://doi.org/10.1038/s41467-024-44908-w} {\bibfield  {journal} {\bibinfo
			{journal} {Nature Communications}\ }\textbf {\bibinfo {volume} {15}},\
		\bibinfo {pages} {Article number: 44908} (\bibinfo {year}
		{2024})}\BibitemShut {NoStop}%
	\bibitem [{\citenamefont {Moskalets}\ \emph {et~al.}(2008)\citenamefont
		{Moskalets}, \citenamefont {Samuelsson},\ and\ \citenamefont
		{B\"uttiker}}]{moskalets2008quantized}%
	\BibitemOpen
	\bibfield  {author} {\bibinfo {author} {\bibfnamefont {M.}~\bibnamefont
			{Moskalets}}, \bibinfo {author} {\bibfnamefont {P.}~\bibnamefont
			{Samuelsson}},\ and\ \bibinfo {author} {\bibfnamefont {M.}~\bibnamefont
			{B\"uttiker}},\ }\bibfield  {title} {\bibinfo {title} {Quantized dynamics of
			a coherent capacitor},\ }\href
	{https://doi.org/10.1103/PhysRevLett.100.086601} {\bibfield  {journal}
		{\bibinfo  {journal} {Phys. Rev. Lett.}\ }\textbf {\bibinfo {volume} {100}},\
		\bibinfo {pages} {086601} (\bibinfo {year} {2008})}\BibitemShut {NoStop}%
	\bibitem [{\citenamefont {Campaioli}\ \emph {et~al.}(2024)\citenamefont
		{Campaioli}, \citenamefont {Gherardini}, \citenamefont {Quach}, \citenamefont
		{Polini},\ and\ \citenamefont {Andolina}}]{Francesco2024colloquium}%
	\BibitemOpen
	\bibfield  {author} {\bibinfo {author} {\bibfnamefont {F.}~\bibnamefont
			{Campaioli}}, \bibinfo {author} {\bibfnamefont {S.}~\bibnamefont
			{Gherardini}}, \bibinfo {author} {\bibfnamefont {J.~Q.}\ \bibnamefont
			{Quach}}, \bibinfo {author} {\bibfnamefont {M.}~\bibnamefont {Polini}},\ and\
		\bibinfo {author} {\bibfnamefont {G.~M.}\ \bibnamefont {Andolina}},\
	}\bibfield  {title} {\bibinfo {title} {Colloquium: Quantum batteries},\
	}\href {https://doi.org/10.1103/RevModPhys.96.031001} {\bibfield  {journal}
		{\bibinfo  {journal} {Rev. Mod. Phys.}\ }\textbf {\bibinfo {volume} {96}},\
		\bibinfo {pages} {031001} (\bibinfo {year} {2024})}\BibitemShut {NoStop}%
	\bibitem [{\citenamefont {Ferraro}\ \emph {et~al.}(2026)\citenamefont
		{Ferraro}, \citenamefont {Cavaliere}, \citenamefont {Genoni}, \citenamefont
		{Benenti},\ and\ \citenamefont {Sassetti}}]{ferraro2026opportunities}%
	\BibitemOpen
	\bibfield  {author} {\bibinfo {author} {\bibfnamefont {D.}~\bibnamefont
			{Ferraro}}, \bibinfo {author} {\bibfnamefont {F.}~\bibnamefont {Cavaliere}},
		\bibinfo {author} {\bibfnamefont {M.~G.}\ \bibnamefont {Genoni}}, \bibinfo
		{author} {\bibfnamefont {G.}~\bibnamefont {Benenti}},\ and\ \bibinfo {author}
		{\bibfnamefont {M.}~\bibnamefont {Sassetti}},\ }\bibfield  {title} {\bibinfo
		{title} {Opportunities and challenges of quantum batteries},\ }\href
	{https://doi.org/10.1038/s42254-025-00906-5} {\bibfield  {journal} {\bibinfo
			{journal} {Nature Reviews Physics}\ }\textbf {\bibinfo {volume} {8}},\
		\bibinfo {pages} {115} (\bibinfo {year} {2026})}\BibitemShut {NoStop}%
	\bibitem [{\citenamefont {Kosloff}\ and\ \citenamefont
		{Levy}(2014)}]{kosloff2014quantum}%
	\BibitemOpen
	\bibfield  {author} {\bibinfo {author} {\bibfnamefont {R.}~\bibnamefont
			{Kosloff}}\ and\ \bibinfo {author} {\bibfnamefont {A.}~\bibnamefont {Levy}},\
	}\bibfield  {title} {\bibinfo {title} {Quantum heat engines and
			refrigerators: Continuous devices},\ }\href
	{https://doi.org/10.1146/annurev-physchem-040513-103724} {\bibfield
		{journal} {\bibinfo  {journal} {Annu. Rev. Phys. Chem.}\ }\textbf {\bibinfo
			{volume} {65}},\ \bibinfo {pages} {365} (\bibinfo {year} {2014})}\BibitemShut
	{NoStop}%
	\bibitem [{\citenamefont {Ghosh}\ \emph {et~al.}(2017)\citenamefont {Ghosh},
		\citenamefont {Latune}, \citenamefont {Davidovich},\ and\ \citenamefont
		{Kurizki}}]{arnab2017catalysis}%
	\BibitemOpen
	\bibfield  {author} {\bibinfo {author} {\bibfnamefont {A.}~\bibnamefont
			{Ghosh}}, \bibinfo {author} {\bibfnamefont {C.~L.}\ \bibnamefont {Latune}},
		\bibinfo {author} {\bibfnamefont {L.}~\bibnamefont {Davidovich}},\ and\
		\bibinfo {author} {\bibfnamefont {G.}~\bibnamefont {Kurizki}},\ }\bibfield
	{title} {\bibinfo {title} {Catalysis of heat-to-work conversion in quantum
			machines},\ }\href {https://doi.org/10.1073/pnas.1711381114} {\bibfield
		{journal} {\bibinfo  {journal} {Proceedings of the National Academy of
				Sciences}\ }\textbf {\bibinfo {volume} {114}},\ \bibinfo {pages} {12156}
		(\bibinfo {year} {2017})}\BibitemShut {NoStop}%
	\bibitem [{\citenamefont {Ghosh}\ \emph {et~al.}(2018)\citenamefont {Ghosh},
		\citenamefont {Gelbwaser-Klimovsky}, \citenamefont {Niedenzu}, \citenamefont
		{Lvovsky}, \citenamefont {Mazets}, \citenamefont {Scully},\ and\
		\citenamefont {Kurizki}}]{ghosh2018two-level}%
	\BibitemOpen
	\bibfield  {author} {\bibinfo {author} {\bibfnamefont {A.}~\bibnamefont
			{Ghosh}}, \bibinfo {author} {\bibfnamefont {D.}~\bibnamefont
			{Gelbwaser-Klimovsky}}, \bibinfo {author} {\bibfnamefont {W.}~\bibnamefont
			{Niedenzu}}, \bibinfo {author} {\bibfnamefont {A.~I.}\ \bibnamefont
			{Lvovsky}}, \bibinfo {author} {\bibfnamefont {I.}~\bibnamefont {Mazets}},
		\bibinfo {author} {\bibfnamefont {M.~O.}\ \bibnamefont {Scully}},\ and\
		\bibinfo {author} {\bibfnamefont {G.}~\bibnamefont {Kurizki}},\ }\bibfield
	{title} {\bibinfo {title} {Two-level masers as heat-to-work converters},\
	}\href {https://doi.org/10.1073/pnas.1805354115} {\bibfield  {journal}
		{\bibinfo  {journal} {Proceedings of the National Academy of Sciences}\
		}\textbf {\bibinfo {volume} {115}},\ \bibinfo {pages} {9941} (\bibinfo {year}
		{2018})}\BibitemShut {NoStop}%
	\bibitem [{\citenamefont {Binder}\ \emph {et~al.}(2019)\citenamefont {Binder},
		\citenamefont {Correa}, \citenamefont {Gogolin}, \citenamefont {Anders},\
		and\ \citenamefont {Adesso}}]{binder2019thermodynamics}%
	\BibitemOpen
	\bibfield  {author} {\bibinfo {author} {\bibfnamefont {F.}~\bibnamefont
			{Binder}}, \bibinfo {author} {\bibfnamefont {L.~A.}\ \bibnamefont {Correa}},
		\bibinfo {author} {\bibfnamefont {C.}~\bibnamefont {Gogolin}}, \bibinfo
		{author} {\bibfnamefont {J.}~\bibnamefont {Anders}},\ and\ \bibinfo {author}
		{\bibfnamefont {G.}~\bibnamefont {Adesso}},\ }\bibinfo {title}
	{Thermodynamics in the quantum regime: Fundamental aspects and new
		directions}\ (\bibinfo  {publisher} {Springer},\ \bibinfo {year} {2019})\
	\bibinfo {note} {and the references therein}\BibitemShut {NoStop}%
	\bibitem [{\citenamefont {Maity}\ \emph {et~al.}(2025)\citenamefont {Maity},
		\citenamefont {Chaki}, \citenamefont {Ghoshal},\ and\ \citenamefont
		{Sen}}]{maity2026quantum}%
	\BibitemOpen
	\bibfield  {author} {\bibinfo {author} {\bibfnamefont {A.}~\bibnamefont
			{Maity}}, \bibinfo {author} {\bibfnamefont {P.}~\bibnamefont {Chaki}},
		\bibinfo {author} {\bibfnamefont {A.}~\bibnamefont {Ghoshal}},\ and\ \bibinfo
		{author} {\bibfnamefont {U.}~\bibnamefont {Sen}},\ }\bibfield  {title}
	{\bibinfo {title} {Quantum heat transformers},\ }\href
	{https://doi.org/10.1088/1367-2630/ae2e34} {\bibfield  {journal} {\bibinfo
			{journal} {New Journal of Physics}\ }\textbf {\bibinfo {volume} {28}},\
		\bibinfo {pages} {014501} (\bibinfo {year} {2025})}\BibitemShut {NoStop}%
	\bibitem [{\citenamefont {Kurizki}\ and\ \citenamefont
		{Kofman}(2022)}]{kurizki2022thermodynamics}%
	\BibitemOpen
	\bibfield  {author} {\bibinfo {author} {\bibfnamefont {G.}~\bibnamefont
			{Kurizki}}\ and\ \bibinfo {author} {\bibfnamefont {A.}~\bibnamefont
			{Kofman}},\ }\href {https://books.google.co.in/books?id=WyZTEAAAQBAJ} {\emph
		{\bibinfo {title} {Thermodynamics and Control of Open Quantum Systems}}}\
	(\bibinfo  {publisher} {Cambridge University Press},\ \bibinfo {year}
	{2022})\BibitemShut {NoStop}%
	\bibitem [{\citenamefont {Giazotto}\ \emph {et~al.}(2006)\citenamefont
		{Giazotto}, \citenamefont {Heikkil\"a}, \citenamefont {Luukanen},
		\citenamefont {Savin},\ and\ \citenamefont
		{Pekola}}]{giazotto2006opportunities}%
	\BibitemOpen
	\bibfield  {author} {\bibinfo {author} {\bibfnamefont {F.}~\bibnamefont
			{Giazotto}}, \bibinfo {author} {\bibfnamefont {T.~T.}\ \bibnamefont
			{Heikkil\"a}}, \bibinfo {author} {\bibfnamefont {A.}~\bibnamefont
			{Luukanen}}, \bibinfo {author} {\bibfnamefont {A.~M.}\ \bibnamefont
			{Savin}},\ and\ \bibinfo {author} {\bibfnamefont {J.~P.}\ \bibnamefont
			{Pekola}},\ }\bibfield  {title} {\bibinfo {title} {Opportunities for
			mesoscopics in thermometry and refrigeration: Physics and applications},\
	}\href {https://doi.org/10.1103/RevModPhys.78.217} {\bibfield  {journal}
		{\bibinfo  {journal} {Rev. Mod. Phys.}\ }\textbf {\bibinfo {volume} {78}},\
		\bibinfo {pages} {217} (\bibinfo {year} {2006})}\BibitemShut {NoStop}%
	\bibitem [{\citenamefont {Pekola}\ \emph {et~al.}(1994)\citenamefont {Pekola},
		\citenamefont {Hirvi}, \citenamefont {Kauppinen},\ and\ \citenamefont
		{Paalanen}}]{pekola1994thermometry}%
	\BibitemOpen
	\bibfield  {author} {\bibinfo {author} {\bibfnamefont {J.~P.}\ \bibnamefont
			{Pekola}}, \bibinfo {author} {\bibfnamefont {K.~P.}\ \bibnamefont {Hirvi}},
		\bibinfo {author} {\bibfnamefont {J.~P.}\ \bibnamefont {Kauppinen}},\ and\
		\bibinfo {author} {\bibfnamefont {M.~A.}\ \bibnamefont {Paalanen}},\
	}\bibfield  {title} {\bibinfo {title} {Thermometry by arrays of tunnel
			junctions},\ }\href {https://doi.org/10.1103/PhysRevLett.73.2903} {\bibfield
		{journal} {\bibinfo  {journal} {Phys. Rev. Lett.}\ }\textbf {\bibinfo
			{volume} {73}},\ \bibinfo {pages} {2903} (\bibinfo {year}
		{1994})}\BibitemShut {NoStop}%
	\bibitem [{\citenamefont {Nahum}\ and\ \citenamefont
		{Martinis}(1993)}]{nahum1993ultrasensitive}%
	\BibitemOpen
	\bibfield  {author} {\bibinfo {author} {\bibfnamefont {M.}~\bibnamefont
			{Nahum}}\ and\ \bibinfo {author} {\bibfnamefont {J.~M.}\ \bibnamefont
			{Martinis}},\ }\bibfield  {title} {\bibinfo {title}
		{Ultrasensitive‐hot‐electron microbolometer},\ }\href
	{https://doi.org/10.1063/1.110237} {\bibfield  {journal} {\bibinfo  {journal}
			{Applied Physics Letters}\ }\textbf {\bibinfo {volume} {63}},\ \bibinfo
		{pages} {3075} (\bibinfo {year} {1993})}\BibitemShut {NoStop}%
	\bibitem [{\citenamefont {Meschke}\ \emph {et~al.}(2009)\citenamefont
		{Meschke}, \citenamefont {Peltonen}, \citenamefont {Courtois},\ and\
		\citenamefont {Pekola}}]{meschke2009calorimetric}%
	\BibitemOpen
	\bibfield  {author} {\bibinfo {author} {\bibfnamefont {M.}~\bibnamefont
			{Meschke}}, \bibinfo {author} {\bibfnamefont {J.~T.}\ \bibnamefont
			{Peltonen}}, \bibinfo {author} {\bibfnamefont {H.}~\bibnamefont {Courtois}},\
		and\ \bibinfo {author} {\bibfnamefont {J.~P.}\ \bibnamefont {Pekola}},\
	}\bibfield  {title} {\bibinfo {title} {Calorimetric readout of a
			superconducting proximity-effect thermometer},\ }\href
	{https://doi.org/10.1007/s10909-008-9854-y} {\bibfield  {journal} {\bibinfo
			{journal} {Journal of Low Temperature Physics}\ }\textbf {\bibinfo {volume}
			{154}},\ \bibinfo {pages} {190} (\bibinfo {year} {2009})}\BibitemShut
	{NoStop}%
	\bibitem [{\citenamefont {Halbertal}\ \emph {et~al.}(2016)\citenamefont
		{Halbertal}, \citenamefont {Cuppens}, \citenamefont {Ben~Shalom},
		\citenamefont {Embon}, \citenamefont {Shadmi}, \citenamefont {Anahory},
		\citenamefont {Naren}, \citenamefont {Sarkar}, \citenamefont {Uri},
		\citenamefont {Ronen}, \citenamefont {Myasoedov}, \citenamefont {Levitov},
		\citenamefont {Joselevich}, \citenamefont {Geim},\ and\ \citenamefont
		{Zeldov}}]{halbertal2016nanoscale}%
	\BibitemOpen
	\bibfield  {author} {\bibinfo {author} {\bibfnamefont {D.}~\bibnamefont
			{Halbertal}}, \bibinfo {author} {\bibfnamefont {J.}~\bibnamefont {Cuppens}},
		\bibinfo {author} {\bibfnamefont {M.}~\bibnamefont {Ben~Shalom}}, \bibinfo
		{author} {\bibfnamefont {L.}~\bibnamefont {Embon}}, \bibinfo {author}
		{\bibfnamefont {N.}~\bibnamefont {Shadmi}}, \bibinfo {author} {\bibfnamefont
			{Y.}~\bibnamefont {Anahory}}, \bibinfo {author} {\bibfnamefont {H.~R.}\
			\bibnamefont {Naren}}, \bibinfo {author} {\bibfnamefont {J.}~\bibnamefont
			{Sarkar}}, \bibinfo {author} {\bibfnamefont {A.}~\bibnamefont {Uri}},
		\bibinfo {author} {\bibfnamefont {Y.}~\bibnamefont {Ronen}}, \bibinfo
		{author} {\bibfnamefont {Y.}~\bibnamefont {Myasoedov}}, \bibinfo {author}
		{\bibfnamefont {L.~S.}\ \bibnamefont {Levitov}}, \bibinfo {author}
		{\bibfnamefont {E.}~\bibnamefont {Joselevich}}, \bibinfo {author}
		{\bibfnamefont {A.~K.}\ \bibnamefont {Geim}},\ and\ \bibinfo {author}
		{\bibfnamefont {E.}~\bibnamefont {Zeldov}},\ }\bibfield  {title} {\bibinfo
		{title} {Nanoscale thermal imaging of dissipation in quantum systems},\
	}\href {https://doi.org/10.1038/nature19843} {\bibfield  {journal} {\bibinfo
			{journal} {Nature}\ }\textbf {\bibinfo {volume} {539}},\ \bibinfo {pages}
		{407} (\bibinfo {year} {2016})}\BibitemShut {NoStop}%
	\bibitem [{\citenamefont {Karimi}\ \emph {et~al.}(2020)\citenamefont {Karimi},
		\citenamefont {Brange}, \citenamefont {Samuelsson},\ and\ \citenamefont
		{Pekola}}]{karimi2020reaching}%
	\BibitemOpen
	\bibfield  {author} {\bibinfo {author} {\bibfnamefont {B.}~\bibnamefont
			{Karimi}}, \bibinfo {author} {\bibfnamefont {F.}~\bibnamefont {Brange}},
		\bibinfo {author} {\bibfnamefont {P.}~\bibnamefont {Samuelsson}},\ and\
		\bibinfo {author} {\bibfnamefont {J.~P.}\ \bibnamefont {Pekola}},\ }\bibfield
	{title} {\bibinfo {title} {Reaching the ultimate energy resolution of a
			quantum detector},\ }\href {https://doi.org/10.1038/s41467-019-14247-2}
	{\bibfield  {journal} {\bibinfo  {journal} {Nature Communications}\ }\textbf
		{\bibinfo {volume} {11}},\ \bibinfo {pages} {367} (\bibinfo {year}
		{2020})}\BibitemShut {NoStop}%
	\bibitem [{\citenamefont {Pekola}\ and\ \citenamefont
		{Karimi}(2021)}]{pekola2021colloquium}%
	\BibitemOpen
	\bibfield  {author} {\bibinfo {author} {\bibfnamefont {J.~P.}\ \bibnamefont
			{Pekola}}\ and\ \bibinfo {author} {\bibfnamefont {B.}~\bibnamefont
			{Karimi}},\ }\bibfield  {title} {\bibinfo {title} {Colloquium: Quantum heat
			transport in condensed matter systems},\ }\href
	{https://doi.org/10.1103/RevModPhys.93.041001} {\bibfield  {journal}
		{\bibinfo  {journal} {Rev. Mod. Phys.}\ }\textbf {\bibinfo {volume} {93}},\
		\bibinfo {pages} {041001} (\bibinfo {year} {2021})}\BibitemShut {NoStop}%
	\bibitem [{\citenamefont {Majidi}\ \emph {et~al.}(2024)\citenamefont {Majidi},
		\citenamefont {Bergfield}, \citenamefont {Maisi}, \citenamefont {Höfer},
		\citenamefont {Courtois},\ and\ \citenamefont {Winkelmann}}]{Majidi2024Heat}%
	\BibitemOpen
	\bibfield  {author} {\bibinfo {author} {\bibfnamefont {D.}~\bibnamefont
			{Majidi}}, \bibinfo {author} {\bibfnamefont {J.~P.}\ \bibnamefont
			{Bergfield}}, \bibinfo {author} {\bibfnamefont {V.}~\bibnamefont {Maisi}},
		\bibinfo {author} {\bibfnamefont {J.}~\bibnamefont {Höfer}}, \bibinfo
		{author} {\bibfnamefont {H.}~\bibnamefont {Courtois}},\ and\ \bibinfo
		{author} {\bibfnamefont {C.~B.}\ \bibnamefont {Winkelmann}},\ }\bibfield
	{title} {\bibinfo {title} {Heat transport at the nanoscale and ultralow
			temperatures—implications for quantum technologies},\ }\href
	{https://doi.org/10.1063/5.0204207} {\bibfield  {journal} {\bibinfo
			{journal} {Applied Physics Letters}\ }\textbf {\bibinfo {volume} {124}},\
		\bibinfo {pages} {140504} (\bibinfo {year} {2024})}\BibitemShut {NoStop}%
	\bibitem [{\citenamefont {Dutta}\ \emph {et~al.}(2017)\citenamefont {Dutta},
		\citenamefont {Peltonen}, \citenamefont {Antonenko}, \citenamefont {Meschke},
		\citenamefont {Skvortsov}, \citenamefont {Kubala}, \citenamefont {K\"onig},
		\citenamefont {Winkelmann}, \citenamefont {Courtois},\ and\ \citenamefont
		{Pekola}}]{dutta2017thermal}%
	\BibitemOpen
	\bibfield  {author} {\bibinfo {author} {\bibfnamefont {B.}~\bibnamefont
			{Dutta}}, \bibinfo {author} {\bibfnamefont {J.~T.}\ \bibnamefont {Peltonen}},
		\bibinfo {author} {\bibfnamefont {D.~S.}\ \bibnamefont {Antonenko}}, \bibinfo
		{author} {\bibfnamefont {M.}~\bibnamefont {Meschke}}, \bibinfo {author}
		{\bibfnamefont {M.~A.}\ \bibnamefont {Skvortsov}}, \bibinfo {author}
		{\bibfnamefont {B.}~\bibnamefont {Kubala}}, \bibinfo {author} {\bibfnamefont
			{J.}~\bibnamefont {K\"onig}}, \bibinfo {author} {\bibfnamefont {C.~B.}\
			\bibnamefont {Winkelmann}}, \bibinfo {author} {\bibfnamefont
			{H.}~\bibnamefont {Courtois}},\ and\ \bibinfo {author} {\bibfnamefont
			{J.~P.}\ \bibnamefont {Pekola}},\ }\bibfield  {title} {\bibinfo {title}
		{Thermal conductance of a single-electron transistor},\ }\href
	{https://doi.org/10.1103/PhysRevLett.119.077701} {\bibfield  {journal}
		{\bibinfo  {journal} {Phys. Rev. Lett.}\ }\textbf {\bibinfo {volume} {119}},\
		\bibinfo {pages} {077701} (\bibinfo {year} {2017})}\BibitemShut {NoStop}%
	\bibitem [{\citenamefont {Cui}\ \emph {et~al.}(2017)\citenamefont {Cui},
		\citenamefont {Jeong}, \citenamefont {Hur}, \citenamefont {Matt},
		\citenamefont {Klöckner}, \citenamefont {Pauly}, \citenamefont {Nielaba},
		\citenamefont {Cuevas}, \citenamefont {Meyhofer},\ and\ \citenamefont
		{Reddy}}]{cui2017quantized}%
	\BibitemOpen
	\bibfield  {author} {\bibinfo {author} {\bibfnamefont {L.}~\bibnamefont
			{Cui}}, \bibinfo {author} {\bibfnamefont {W.}~\bibnamefont {Jeong}}, \bibinfo
		{author} {\bibfnamefont {S.}~\bibnamefont {Hur}}, \bibinfo {author}
		{\bibfnamefont {M.}~\bibnamefont {Matt}}, \bibinfo {author} {\bibfnamefont
			{J.~C.}\ \bibnamefont {Klöckner}}, \bibinfo {author} {\bibfnamefont
			{F.}~\bibnamefont {Pauly}}, \bibinfo {author} {\bibfnamefont
			{P.}~\bibnamefont {Nielaba}}, \bibinfo {author} {\bibfnamefont {J.~C.}\
			\bibnamefont {Cuevas}}, \bibinfo {author} {\bibfnamefont {E.}~\bibnamefont
			{Meyhofer}},\ and\ \bibinfo {author} {\bibfnamefont {P.}~\bibnamefont
			{Reddy}},\ }\bibfield  {title} {\bibinfo {title} {Quantized thermal transport
			in single-atom junctions},\ }\href {https://doi.org/10.1126/science.aam6622}
	{\bibfield  {journal} {\bibinfo  {journal} {Science}\ }\textbf {\bibinfo
			{volume} {355}},\ \bibinfo {pages} {1192} (\bibinfo {year}
		{2017})}\BibitemShut {NoStop}%
	\bibitem [{\citenamefont {Mosso}\ \emph {et~al.}(2017)\citenamefont {Mosso},
		\citenamefont {Drechsler}, \citenamefont {Menges}, \citenamefont {Nirmalraj},
		\citenamefont {Karg}, \citenamefont {Riel},\ and\ \citenamefont
		{Gotsmann}}]{mosso2017heat}%
	\BibitemOpen
	\bibfield  {author} {\bibinfo {author} {\bibfnamefont {N.}~\bibnamefont
			{Mosso}}, \bibinfo {author} {\bibfnamefont {U.}~\bibnamefont {Drechsler}},
		\bibinfo {author} {\bibfnamefont {F.}~\bibnamefont {Menges}}, \bibinfo
		{author} {\bibfnamefont {P.}~\bibnamefont {Nirmalraj}}, \bibinfo {author}
		{\bibfnamefont {S.}~\bibnamefont {Karg}}, \bibinfo {author} {\bibfnamefont
			{H.}~\bibnamefont {Riel}},\ and\ \bibinfo {author} {\bibfnamefont
			{B.}~\bibnamefont {Gotsmann}},\ }\bibfield  {title} {\bibinfo {title} {Heat
			transport through atomic contacts},\ }\href
	{https://doi.org/10.1038/nnano.2016.302} {\bibfield  {journal} {\bibinfo
			{journal} {Nature Nanotechnology}\ }\textbf {\bibinfo {volume} {12}},\
		\bibinfo {pages} {430} (\bibinfo {year} {2017})}\BibitemShut {NoStop}%
	\bibitem [{\citenamefont {Josefsson}\ \emph {et~al.}(2018)\citenamefont
		{Josefsson}, \citenamefont {Svilans}, \citenamefont {Burke}, \citenamefont
		{Hoffmann}, \citenamefont {Fahlvik~Svensson}, \citenamefont {Thelander},
		\citenamefont {Leijnse},\ and\ \citenamefont {Linke}}]{josefsson2018a}%
	\BibitemOpen
	\bibfield  {author} {\bibinfo {author} {\bibfnamefont {M.}~\bibnamefont
			{Josefsson}}, \bibinfo {author} {\bibfnamefont {A.}~\bibnamefont {Svilans}},
		\bibinfo {author} {\bibfnamefont {A.~M.}\ \bibnamefont {Burke}}, \bibinfo
		{author} {\bibfnamefont {E.~A.}\ \bibnamefont {Hoffmann}}, \bibinfo {author}
		{\bibfnamefont {S.}~\bibnamefont {Fahlvik~Svensson}}, \bibinfo {author}
		{\bibfnamefont {C.}~\bibnamefont {Thelander}}, \bibinfo {author}
		{\bibfnamefont {M.}~\bibnamefont {Leijnse}},\ and\ \bibinfo {author}
		{\bibfnamefont {H.}~\bibnamefont {Linke}},\ }\bibfield  {title} {\bibinfo
		{title} {A quantum-dot heat engine operating close to the thermodynamic
			efficiency limits},\ }\href {https://doi.org/10.1038/s41565-018-0200-5}
	{\bibfield  {journal} {\bibinfo  {journal} {Nature Nanotechnology}\ }\textbf
		{\bibinfo {volume} {13}},\ \bibinfo {pages} {920} (\bibinfo {year}
		{2018})}\BibitemShut {NoStop}%
	\bibitem [{\citenamefont {Volosheniuk}\ \emph {et~al.}(2026)\citenamefont
		{Volosheniuk}, \citenamefont {Conte}, \citenamefont {Pyurbeeva},
		\citenamefont {Baum}, \citenamefont {Vilas-Varela}, \citenamefont
		{Fern{\'a}ndez}, \citenamefont {Pe{\~{n}}a}, \citenamefont {van~der Zant},\
		and\ \citenamefont {Gehring}}]{volosheniuk2026asingle}%
	\BibitemOpen
	\bibfield  {author} {\bibinfo {author} {\bibfnamefont {S.}~\bibnamefont
			{Volosheniuk}}, \bibinfo {author} {\bibfnamefont {R.}~\bibnamefont {Conte}},
		\bibinfo {author} {\bibfnamefont {E.}~\bibnamefont {Pyurbeeva}}, \bibinfo
		{author} {\bibfnamefont {T.}~\bibnamefont {Baum}}, \bibinfo {author}
		{\bibfnamefont {M.}~\bibnamefont {Vilas-Varela}}, \bibinfo {author}
		{\bibfnamefont {S.}~\bibnamefont {Fern{\'a}ndez}}, \bibinfo {author}
		{\bibfnamefont {D.}~\bibnamefont {Pe{\~{n}}a}}, \bibinfo {author}
		{\bibfnamefont {H.~S.~J.}\ \bibnamefont {van~der Zant}},\ and\ \bibinfo
		{author} {\bibfnamefont {P.}~\bibnamefont {Gehring}},\ }\bibfield  {title}
	{\bibinfo {title} {A single-molecule quantum heat engine},\ }\href
	{https://doi.org/10.1021/acs.nanolett.5c04824} {\bibfield  {journal}
		{\bibinfo  {journal} {Nano Letters}\ }\textbf {\bibinfo {volume} {26}},\
		\bibinfo {pages} {984} (\bibinfo {year} {2026})}\BibitemShut {NoStop}%
	\bibitem [{\citenamefont {Dutta}\ \emph {et~al.}(2020)\citenamefont {Dutta},
		\citenamefont {Majidi}, \citenamefont {Talarico}, \citenamefont {Lo~Gullo},
		\citenamefont {Courtois},\ and\ \citenamefont
		{Winkelmann}}]{dutta2020single}%
	\BibitemOpen
	\bibfield  {author} {\bibinfo {author} {\bibfnamefont {B.}~\bibnamefont
			{Dutta}}, \bibinfo {author} {\bibfnamefont {D.}~\bibnamefont {Majidi}},
		\bibinfo {author} {\bibfnamefont {N.~W.}\ \bibnamefont {Talarico}}, \bibinfo
		{author} {\bibfnamefont {N.}~\bibnamefont {Lo~Gullo}}, \bibinfo {author}
		{\bibfnamefont {H.}~\bibnamefont {Courtois}},\ and\ \bibinfo {author}
		{\bibfnamefont {C.~B.}\ \bibnamefont {Winkelmann}},\ }\bibfield  {title}
	{\bibinfo {title} {Single-quantum-dot heat valve},\ }\href
	{https://doi.org/10.1103/PhysRevLett.125.237701} {\bibfield  {journal}
		{\bibinfo  {journal} {Phys. Rev. Lett.}\ }\textbf {\bibinfo {volume} {125}},\
		\bibinfo {pages} {237701} (\bibinfo {year} {2020})}\BibitemShut {NoStop}%
	\bibitem [{\citenamefont {Maillet}\ \emph {et~al.}(2020)\citenamefont
		{Maillet}, \citenamefont {Subero}, \citenamefont {Peltonen}, \citenamefont
		{Golubev},\ and\ \citenamefont {Pekola}}]{maillet2020electric}%
	\BibitemOpen
	\bibfield  {author} {\bibinfo {author} {\bibfnamefont {O.}~\bibnamefont
			{Maillet}}, \bibinfo {author} {\bibfnamefont {D.}~\bibnamefont {Subero}},
		\bibinfo {author} {\bibfnamefont {J.~T.}\ \bibnamefont {Peltonen}}, \bibinfo
		{author} {\bibfnamefont {D.~S.}\ \bibnamefont {Golubev}},\ and\ \bibinfo
		{author} {\bibfnamefont {J.~P.}\ \bibnamefont {Pekola}},\ }\bibfield  {title}
	{\bibinfo {title} {Electric field control of radiative heat transfer in a
			superconducting circuit},\ }\href
	{https://doi.org/10.1038/s41467-020-18163-8} {\bibfield  {journal} {\bibinfo
			{journal} {Nature Communications}\ }\textbf {\bibinfo {volume} {11}},\
		\bibinfo {pages} {4326} (\bibinfo {year} {2020})}\BibitemShut {NoStop}%
	\bibitem [{\citenamefont {Ronzani}\ \emph {et~al.}(2018)\citenamefont
		{Ronzani}, \citenamefont {Karimi}, \citenamefont {Senior}, \citenamefont
		{Chang}, \citenamefont {Peltonen}, \citenamefont {Chen},\ and\ \citenamefont
		{Pekola}}]{ronzani2018tunable}%
	\BibitemOpen
	\bibfield  {author} {\bibinfo {author} {\bibfnamefont {A.}~\bibnamefont
			{Ronzani}}, \bibinfo {author} {\bibfnamefont {B.}~\bibnamefont {Karimi}},
		\bibinfo {author} {\bibfnamefont {J.}~\bibnamefont {Senior}}, \bibinfo
		{author} {\bibfnamefont {Y.-C.}\ \bibnamefont {Chang}}, \bibinfo {author}
		{\bibfnamefont {J.~T.}\ \bibnamefont {Peltonen}}, \bibinfo {author}
		{\bibfnamefont {C.}~\bibnamefont {Chen}},\ and\ \bibinfo {author}
		{\bibfnamefont {J.~P.}\ \bibnamefont {Pekola}},\ }\bibfield  {title}
	{\bibinfo {title} {Tunable photonic heat transport in a quantum heat valve},\
	}\href {https://doi.org/10.1038/s41567-018-0199-4} {\bibfield  {journal}
		{\bibinfo  {journal} {Nature Physics}\ }\textbf {\bibinfo {volume} {14}},\
		\bibinfo {pages} {991} (\bibinfo {year} {2018})}\BibitemShut {NoStop}%
	\bibitem [{\citenamefont {Campbell~et. al.}(2026)}]{campbell2026roadmap}%
	\BibitemOpen
	\bibfield  {author} {\bibinfo {author} {\bibfnamefont {S.}~\bibnamefont
			{Campbell~et. al.}},\ }\bibfield  {title} {\bibinfo {title} {Roadmap on
			quantum thermodynamics},\ }\href {https://doi.org/10.1088/2058-9565/ae1e27}
	{\bibfield  {journal} {\bibinfo  {journal} {Quantum Science and Technology}\
		}\textbf {\bibinfo {volume} {11}},\ \bibinfo {pages} {012501} (\bibinfo
		{year} {2026})},\ \bibinfo {note} {and the references therein}\BibitemShut
	{NoStop}%
	\bibitem [{\citenamefont {Tiwari}\ \emph {et~al.}(2025)\citenamefont {Tiwari},
		\citenamefont {Bhattacharya},\ and\ \citenamefont
		{Banerjee}}]{devvrat2025quantum}%
	\BibitemOpen
	\bibfield  {author} {\bibinfo {author} {\bibfnamefont {D.}~\bibnamefont
			{Tiwari}}, \bibinfo {author} {\bibfnamefont {S.}~\bibnamefont
			{Bhattacharya}},\ and\ \bibinfo {author} {\bibfnamefont {S.}~\bibnamefont
			{Banerjee}},\ }\bibfield  {title} {\bibinfo {title} {Quantum thermal analogs
			of electric circuits: A universal approach},\ }\href
	{https://doi.org/10.1103/5x8m-bhgd} {\bibfield  {journal} {\bibinfo
			{journal} {Phys. Rev. Lett.}\ }\textbf {\bibinfo {volume} {135}},\ \bibinfo
		{pages} {020404} (\bibinfo {year} {2025})}\BibitemShut {NoStop}%
	\bibitem [{\citenamefont {Wang}\ \emph {et~al.}(2022)\citenamefont {Wang},
		\citenamefont {Wang}, \citenamefont {Wang},\ and\ \citenamefont
		{Ren}}]{wang2022cycleflux}%
	\BibitemOpen
	\bibfield  {author} {\bibinfo {author} {\bibfnamefont {L.}~\bibnamefont
			{Wang}}, \bibinfo {author} {\bibfnamefont {Z.}~\bibnamefont {Wang}}, \bibinfo
		{author} {\bibfnamefont {C.}~\bibnamefont {Wang}},\ and\ \bibinfo {author}
		{\bibfnamefont {J.}~\bibnamefont {Ren}},\ }\bibfield  {title} {\bibinfo
		{title} {Cycle flux ranking of network analysis in quantum thermal devices},\
	}\href {https://doi.org/10.1103/PhysRevLett.128.067701} {\bibfield  {journal}
		{\bibinfo  {journal} {Phys. Rev. Lett.}\ }\textbf {\bibinfo {volume} {128}},\
		\bibinfo {pages} {067701} (\bibinfo {year} {2022})}\BibitemShut {NoStop}%
	\bibitem [{\citenamefont {Tesser}\ \emph {et~al.}(2022)\citenamefont {Tesser},
		\citenamefont {Bhandari}, \citenamefont {Erdman}, \citenamefont {Paladino},
		\citenamefont {Fazio},\ and\ \citenamefont {Taddei}}]{tesser2022heat}%
	\BibitemOpen
	\bibfield  {author} {\bibinfo {author} {\bibfnamefont {L.}~\bibnamefont
			{Tesser}}, \bibinfo {author} {\bibfnamefont {B.}~\bibnamefont {Bhandari}},
		\bibinfo {author} {\bibfnamefont {P.~A.}\ \bibnamefont {Erdman}}, \bibinfo
		{author} {\bibfnamefont {E.}~\bibnamefont {Paladino}}, \bibinfo {author}
		{\bibfnamefont {R.}~\bibnamefont {Fazio}},\ and\ \bibinfo {author}
		{\bibfnamefont {F.}~\bibnamefont {Taddei}},\ }\bibfield  {title} {\bibinfo
		{title} {Heat rectification through single and coupled quantum dots},\ }\href
	{https://doi.org/10.1088/1367-2630/ac53b8} {\bibfield  {journal} {\bibinfo
			{journal} {New Journal of Physics}\ }\textbf {\bibinfo {volume} {24}},\
		\bibinfo {pages} {035001} (\bibinfo {year} {2022})}\BibitemShut {NoStop}%
	\bibitem [{\citenamefont {Gupt}\ \emph {et~al.}(2024)\citenamefont {Gupt},
		\citenamefont {Ghosh},\ and\ \citenamefont {Ghosh}}]{gupt2024graph}%
	\BibitemOpen
	\bibfield  {author} {\bibinfo {author} {\bibfnamefont {N.}~\bibnamefont
			{Gupt}}, \bibinfo {author} {\bibfnamefont {S.}~\bibnamefont {Ghosh}},\ and\
		\bibinfo {author} {\bibfnamefont {A.}~\bibnamefont {Ghosh}},\ }\bibfield
	{title} {\bibinfo {title} {Graph theoretic analysis of three-terminal quantum
			dot thermocouples: Onsager relations and spin-thermoelectric effects},\
	}\href {https://doi.org/10.1103/PhysRevB.109.125124} {\bibfield  {journal}
		{\bibinfo  {journal} {Phys. Rev. B}\ }\textbf {\bibinfo {volume} {109}},\
		\bibinfo {pages} {125124} (\bibinfo {year} {2024})}\BibitemShut {NoStop}%
	\bibitem [{\citenamefont {Ghosh}\ \emph {et~al.}(2026)\citenamefont {Ghosh},
		\citenamefont {Gupt},\ and\ \citenamefont {Ghosh}}]{ghosh2026inverse}%
	\BibitemOpen
	\bibfield  {author} {\bibinfo {author} {\bibfnamefont {S.}~\bibnamefont
			{Ghosh}}, \bibinfo {author} {\bibfnamefont {N.}~\bibnamefont {Gupt}},\ and\
		\bibinfo {author} {\bibfnamefont {A.}~\bibnamefont {Ghosh}},\ }\bibfield
	{title} {\bibinfo {title} {Inverse current in coupled transport: A quantum
			thermodynamic model},\ }\href {https://doi.org/10.1103/6r7j-j3n1} {\bibfield
		{journal} {\bibinfo  {journal} {Phys. Rev. Res.}\ }\textbf {\bibinfo {volume}
			{8}},\ \bibinfo {pages} {023166} (\bibinfo {year} {2026})}\BibitemShut
	{NoStop}%
	\bibitem [{\citenamefont {Wang}\ and\ \citenamefont
		{Li}(2007)}]{wang2007thermal}%
	\BibitemOpen
	\bibfield  {author} {\bibinfo {author} {\bibfnamefont {L.}~\bibnamefont
			{Wang}}\ and\ \bibinfo {author} {\bibfnamefont {B.}~\bibnamefont {Li}},\
	}\bibfield  {title} {\bibinfo {title} {Thermal logic gates: Computation with
			phonons},\ }\href {https://doi.org/10.1103/PhysRevLett.99.177208} {\bibfield
		{journal} {\bibinfo  {journal} {Phys. Rev. Lett.}\ }\textbf {\bibinfo
			{volume} {99}},\ \bibinfo {pages} {177208} (\bibinfo {year}
		{2007})}\BibitemShut {NoStop}%
	\bibitem [{\citenamefont {Paolucci}\ \emph {et~al.}(2018)\citenamefont
		{Paolucci}, \citenamefont {Marchegiani}, \citenamefont {Strambini},\ and\
		\citenamefont {Giazotto}}]{paolucci2018phasetunable}%
	\BibitemOpen
	\bibfield  {author} {\bibinfo {author} {\bibfnamefont {F.}~\bibnamefont
			{Paolucci}}, \bibinfo {author} {\bibfnamefont {G.}~\bibnamefont
			{Marchegiani}}, \bibinfo {author} {\bibfnamefont {E.}~\bibnamefont
			{Strambini}},\ and\ \bibinfo {author} {\bibfnamefont {F.}~\bibnamefont
			{Giazotto}},\ }\bibfield  {title} {\bibinfo {title} {Phase-tunable thermal
			logic: Computation with heat},\ }\href
	{https://doi.org/10.1103/PhysRevApplied.10.024003} {\bibfield  {journal}
		{\bibinfo  {journal} {Phys. Rev. Appl.}\ }\textbf {\bibinfo {volume} {10}},\
		\bibinfo {pages} {024003} (\bibinfo {year} {2018})}\BibitemShut {NoStop}%
	\bibitem [{\citenamefont {Ruokola}\ and\ \citenamefont
		{Ojanen}(2011)}]{ruokola2011single}%
	\BibitemOpen
	\bibfield  {author} {\bibinfo {author} {\bibfnamefont {T.}~\bibnamefont
			{Ruokola}}\ and\ \bibinfo {author} {\bibfnamefont {T.}~\bibnamefont
			{Ojanen}},\ }\bibfield  {title} {\bibinfo {title} {Single-electron heat
			diode: Asymmetric heat transport between electronic reservoirs through
			coulomb islands},\ }\href {https://doi.org/10.1103/PhysRevB.83.241404}
	{\bibfield  {journal} {\bibinfo  {journal} {Phys. Rev. B}\ }\textbf {\bibinfo
			{volume} {83}},\ \bibinfo {pages} {241404} (\bibinfo {year}
		{2011})}\BibitemShut {NoStop}%
	\bibitem [{\citenamefont {Koski}\ \emph {et~al.}(2015)\citenamefont {Koski},
		\citenamefont {Kutvonen}, \citenamefont {Khaymovich}, \citenamefont
		{Ala-Nissila},\ and\ \citenamefont {Pekola}}]{koski2015onchip}%
	\BibitemOpen
	\bibfield  {author} {\bibinfo {author} {\bibfnamefont {J.~V.}\ \bibnamefont
			{Koski}}, \bibinfo {author} {\bibfnamefont {A.}~\bibnamefont {Kutvonen}},
		\bibinfo {author} {\bibfnamefont {I.~M.}\ \bibnamefont {Khaymovich}},
		\bibinfo {author} {\bibfnamefont {T.}~\bibnamefont {Ala-Nissila}},\ and\
		\bibinfo {author} {\bibfnamefont {J.~P.}\ \bibnamefont {Pekola}},\ }\bibfield
	{title} {\bibinfo {title} {On-chip maxwell's demon as an information-powered
			refrigerator},\ }\href {https://doi.org/10.1103/PhysRevLett.115.260602}
	{\bibfield  {journal} {\bibinfo  {journal} {Phys. Rev. Lett.}\ }\textbf
		{\bibinfo {volume} {115}},\ \bibinfo {pages} {260602} (\bibinfo {year}
		{2015})}\BibitemShut {NoStop}%
	\bibitem [{\citenamefont {Thierschmann}\ \emph
		{et~al.}(2015{\natexlab{b}})\citenamefont {Thierschmann}, \citenamefont
		{Arnold}, \citenamefont {Mittermüller}, \citenamefont {Maier}, \citenamefont
		{Heyn}, \citenamefont {Hansen}, \citenamefont {Buhmann},\ and\ \citenamefont
		{Molenkamp}}]{thierschmann2015thermal}%
	\BibitemOpen
	\bibfield  {author} {\bibinfo {author} {\bibfnamefont {H.}~\bibnamefont
			{Thierschmann}}, \bibinfo {author} {\bibfnamefont {F.}~\bibnamefont
			{Arnold}}, \bibinfo {author} {\bibfnamefont {M.}~\bibnamefont
			{Mittermüller}}, \bibinfo {author} {\bibfnamefont {L.}~\bibnamefont
			{Maier}}, \bibinfo {author} {\bibfnamefont {C.}~\bibnamefont {Heyn}},
		\bibinfo {author} {\bibfnamefont {W.}~\bibnamefont {Hansen}}, \bibinfo
		{author} {\bibfnamefont {H.}~\bibnamefont {Buhmann}},\ and\ \bibinfo {author}
		{\bibfnamefont {L.~W.}\ \bibnamefont {Molenkamp}},\ }\bibfield  {title}
	{\bibinfo {title} {Thermal gating of charge currents with coulomb coupled
			quantum dots},\ }\href {https://doi.org/10.1088/1367-2630/17/11/113003}
	{\bibfield  {journal} {\bibinfo  {journal} {New Journal of Physics}\ }\textbf
		{\bibinfo {volume} {17}},\ \bibinfo {pages} {113003} (\bibinfo {year}
		{2015}{\natexlab{b}})}\BibitemShut {NoStop}%
	\bibitem [{\citenamefont {Zhang}\ \emph {et~al.}(2017)\citenamefont {Zhang},
		\citenamefont {Zhang}, \citenamefont {Ye}, \citenamefont {Lin},\ and\
		\citenamefont {Chen}}]{zhang2017three}%
	\BibitemOpen
	\bibfield  {author} {\bibinfo {author} {\bibfnamefont {Y.}~\bibnamefont
			{Zhang}}, \bibinfo {author} {\bibfnamefont {X.}~\bibnamefont {Zhang}},
		\bibinfo {author} {\bibfnamefont {Z.}~\bibnamefont {Ye}}, \bibinfo {author}
		{\bibfnamefont {G.}~\bibnamefont {Lin}},\ and\ \bibinfo {author}
		{\bibfnamefont {J.}~\bibnamefont {Chen}},\ }\bibfield  {title} {\bibinfo
		{title} {Three-terminal quantum-dot thermal management devices},\ }\href
	{https://doi.org/10.1063/1.4979977} {\bibfield  {journal} {\bibinfo
			{journal} {Applied Physics Letters}\ }\textbf {\bibinfo {volume} {110}},\
		\bibinfo {pages} {153501} (\bibinfo {year} {2017})}\BibitemShut {NoStop}%
	\bibitem [{\citenamefont {Whitney}\ \emph {et~al.}(2018)\citenamefont
		{Whitney}, \citenamefont {S{\'a}nchez},\ and\ \citenamefont
		{Splettstoesser}}]{whitney2018quantum}%
	\BibitemOpen
	\bibfield  {author} {\bibinfo {author} {\bibfnamefont {R.~S.}\ \bibnamefont
			{Whitney}}, \bibinfo {author} {\bibfnamefont {R.}~\bibnamefont
			{S{\'a}nchez}},\ and\ \bibinfo {author} {\bibfnamefont {J.}~\bibnamefont
			{Splettstoesser}},\ }\bibinfo {title} {Quantum thermodynamics of nanoscale
		thermoelectrics and electronic devices},\ in\ \href
	{https://doi.org/10.1007/978-3-319-99046-0_7} {\emph {\bibinfo {booktitle}
			{Thermodynamics in the Quantum Regime: Fundamental Aspects and New
				Directions}}},\ \bibinfo {editor} {edited by\ \bibinfo {editor}
		{\bibfnamefont {F.}~\bibnamefont {Binder}}, \bibinfo {editor} {\bibfnamefont
			{L.~A.}\ \bibnamefont {Correa}}, \bibinfo {editor} {\bibfnamefont
			{C.}~\bibnamefont {Gogolin}}, \bibinfo {editor} {\bibfnamefont
			{J.}~\bibnamefont {Anders}},\ and\ \bibinfo {editor} {\bibfnamefont
			{G.}~\bibnamefont {Adesso}}}\ (\bibinfo  {publisher} {Springer International
		Publishing},\ \bibinfo {address} {Cham},\ \bibinfo {year} {2018})\ pp.\
	\bibinfo {pages} {175--206}\BibitemShut {NoStop}%
	\bibitem [{\citenamefont {S\'anchez}\ and\ \citenamefont
		{B\"uttiker}(2011)}]{sanchez2011optimal}%
	\BibitemOpen
	\bibfield  {author} {\bibinfo {author} {\bibfnamefont {R.}~\bibnamefont
			{S\'anchez}}\ and\ \bibinfo {author} {\bibfnamefont {M.}~\bibnamefont
			{B\"uttiker}},\ }\bibfield  {title} {\bibinfo {title} {Optimal energy quanta
			to current conversion},\ }\href {https://doi.org/10.1103/PhysRevB.83.085428}
	{\bibfield  {journal} {\bibinfo  {journal} {Phys. Rev. B}\ }\textbf {\bibinfo
			{volume} {83}},\ \bibinfo {pages} {085428} (\bibinfo {year}
		{2011})}\BibitemShut {NoStop}%
	\bibitem [{\citenamefont {Pyurbeeva}\ and\ \citenamefont
		{Kosloff}(2026)}]{pyurbeeva2026quantum}%
	\BibitemOpen
	\bibfield  {author} {\bibinfo {author} {\bibfnamefont {E.}~\bibnamefont
			{Pyurbeeva}}\ and\ \bibinfo {author} {\bibfnamefont {R.}~\bibnamefont
			{Kosloff}},\ }\bibfield  {title} {\bibinfo {title} {Quantum dot thermal
			machines—a guide to engineering},\ }\bibfield  {journal} {\bibinfo
		{journal} {Entropy}\ }\textbf {\bibinfo {volume} {28}},\ \href
	{https://doi.org/10.3390/e28010002} {10.3390/e28010002} (\bibinfo {year}
	{2026})\BibitemShut {NoStop}%
	\bibitem [{\citenamefont {Ghosh}\ \emph {et~al.}(2012)\citenamefont {Ghosh},
		\citenamefont {Sinha},\ and\ \citenamefont {Ray}}]{ghosh2012fermionic}%
	\BibitemOpen
	\bibfield  {author} {\bibinfo {author} {\bibfnamefont {A.}~\bibnamefont
			{Ghosh}}, \bibinfo {author} {\bibfnamefont {S.~S.}\ \bibnamefont {Sinha}},\
		and\ \bibinfo {author} {\bibfnamefont {D.~S.}\ \bibnamefont {Ray}},\
	}\bibfield  {title} {\bibinfo {title} {Fermionic oscillator in a fermionic
			bath},\ }\href {https://doi.org/10.1103/PhysRevE.86.011138} {\bibfield
		{journal} {\bibinfo  {journal} {Phys. Rev. E}\ }\textbf {\bibinfo {volume}
			{86}},\ \bibinfo {pages} {011138} (\bibinfo {year} {2012})}\BibitemShut
	{NoStop}%
	\bibitem [{\citenamefont {Gupt}\ \emph {et~al.}(2021)\citenamefont {Gupt},
		\citenamefont {Bhattacharyya},\ and\ \citenamefont
		{Ghosh}}]{nikhil2021statistical}%
	\BibitemOpen
	\bibfield  {author} {\bibinfo {author} {\bibfnamefont {N.}~\bibnamefont
			{Gupt}}, \bibinfo {author} {\bibfnamefont {S.}~\bibnamefont
			{Bhattacharyya}},\ and\ \bibinfo {author} {\bibfnamefont {A.}~\bibnamefont
			{Ghosh}},\ }\bibfield  {title} {\bibinfo {title} {Statistical generalization
			of regenerative bosonic and fermionic stirling cycles},\ }\href
	{https://doi.org/10.1103/PhysRevE.104.054130} {\bibfield  {journal} {\bibinfo
			{journal} {Phys. Rev. E}\ }\textbf {\bibinfo {volume} {104}},\ \bibinfo
		{pages} {054130} (\bibinfo {year} {2021})}\BibitemShut {NoStop}%
	\bibitem [{\citenamefont {Damas}\ \emph {et~al.}(2023)\citenamefont {Damas},
		\citenamefont {de~Assis},\ and\ \citenamefont
		{de~Almeida}}]{damas2023cooling}%
	\BibitemOpen
	\bibfield  {author} {\bibinfo {author} {\bibfnamefont {G.~G.}\ \bibnamefont
			{Damas}}, \bibinfo {author} {\bibfnamefont {R.~J.}\ \bibnamefont
			{de~Assis}},\ and\ \bibinfo {author} {\bibfnamefont {N.~G.}\ \bibnamefont
			{de~Almeida}},\ }\bibfield  {title} {\bibinfo {title} {Cooling with fermionic
			thermal reservoirs},\ }\href {https://doi.org/10.1103/PhysRevE.107.034128}
	{\bibfield  {journal} {\bibinfo  {journal} {Phys. Rev. E}\ }\textbf {\bibinfo
			{volume} {107}},\ \bibinfo {pages} {034128} (\bibinfo {year}
		{2023})}\BibitemShut {NoStop}%
	\bibitem [{\citenamefont {Breuer}\ and\ \citenamefont
		{Petruccione}(2007)}]{breuer2002book}%
	\BibitemOpen
	\bibfield  {author} {\bibinfo {author} {\bibfnamefont {H.-P.}\ \bibnamefont
			{Breuer}}\ and\ \bibinfo {author} {\bibfnamefont {F.}~\bibnamefont
			{Petruccione}},\ }\href
	{https://doi.org/10.1093/acprof:oso/9780199213900.001.0001} {\emph {\bibinfo
			{title} {{The Theory of Open Quantum Systems}}}}\ (\bibinfo  {publisher}
	{Oxford University Press},\ \bibinfo {year} {2007})\BibitemShut {NoStop}%
	\bibitem [{\citenamefont {Strasberg}(2022)}]{strasberg2022quantum}%
	\BibitemOpen
	\bibfield  {author} {\bibinfo {author} {\bibfnamefont {P.}~\bibnamefont
			{Strasberg}},\ }\href@noop {} {\emph {\bibinfo {title} {Quantum Stochastic
				Thermodynamics: Foundations and Selected Applications}}}\ (\bibinfo
	{publisher} {Oxford University Press},\ \bibinfo {year} {2022})\BibitemShut
	{NoStop}%
	\bibitem [{\citenamefont {Millman}\ and\ \citenamefont
		{Halkias}(1972)}]{millman1972integrated}%
	\BibitemOpen
	\bibfield  {author} {\bibinfo {author} {\bibfnamefont {J.}~\bibnamefont
			{Millman}}\ and\ \bibinfo {author} {\bibfnamefont {C.}~\bibnamefont
			{Halkias}},\ }\href {https://books.google.co.in/books?id=biRTAAAAMAAJ} {\emph
		{\bibinfo {title} {Integrated Electronics: Analog and Digital Circuits and
				Systems}}},\ Electrical Engineering Series\ (\bibinfo  {publisher}
	{McGraw-Hill},\ \bibinfo {year} {1972})\BibitemShut {NoStop}%
	\bibitem [{\citenamefont {Dubos}\ \emph {et~al.}(2001)\citenamefont {Dubos},
		\citenamefont {Courtois}, \citenamefont {Pannetier}, \citenamefont {Wilhelm},
		\citenamefont {Zaikin},\ and\ \citenamefont {Sch\"on}}]{Dubos2001Josephson}%
	\BibitemOpen
	\bibfield  {author} {\bibinfo {author} {\bibfnamefont {P.}~\bibnamefont
			{Dubos}}, \bibinfo {author} {\bibfnamefont {H.}~\bibnamefont {Courtois}},
		\bibinfo {author} {\bibfnamefont {B.}~\bibnamefont {Pannetier}}, \bibinfo
		{author} {\bibfnamefont {F.~K.}\ \bibnamefont {Wilhelm}}, \bibinfo {author}
		{\bibfnamefont {A.~D.}\ \bibnamefont {Zaikin}},\ and\ \bibinfo {author}
		{\bibfnamefont {G.}~\bibnamefont {Sch\"on}},\ }\bibfield  {title} {\bibinfo
		{title} {Josephson critical current in a long mesoscopic s-n-s junction},\
	}\href {https://doi.org/10.1103/PhysRevB.63.064502} {\bibfield  {journal}
		{\bibinfo  {journal} {Phys. Rev. B}\ }\textbf {\bibinfo {volume} {63}},\
		\bibinfo {pages} {064502} (\bibinfo {year} {2001})}\BibitemShut {NoStop}%
	\bibitem [{\citenamefont {Esposito}\ \emph {et~al.}(2010)\citenamefont
		{Esposito}, \citenamefont {Lindenberg},\ and\ \citenamefont {den
			Broeck}}]{esposito2010entropy}%
	\BibitemOpen
	\bibfield  {author} {\bibinfo {author} {\bibfnamefont {M.}~\bibnamefont
			{Esposito}}, \bibinfo {author} {\bibfnamefont {K.}~\bibnamefont
			{Lindenberg}},\ and\ \bibinfo {author} {\bibfnamefont {C.~V.}\ \bibnamefont
			{den Broeck}},\ }\bibfield  {title} {\bibinfo {title} {Entropy production as
			correlation between system and reservoir},\ }\href
	{https://doi.org/10.1088/1367-2630/12/1/013013} {\bibfield  {journal}
		{\bibinfo  {journal} {New Journal of Physics}\ }\textbf {\bibinfo {volume}
			{12}},\ \bibinfo {pages} {013013} (\bibinfo {year} {2010})}\BibitemShut
	{NoStop}%
	\bibitem [{\citenamefont {Landi}\ and\ \citenamefont
		{Paternostro}(2021)}]{landi2021irreversible}%
	\BibitemOpen
	\bibfield  {author} {\bibinfo {author} {\bibfnamefont {G.~T.}\ \bibnamefont
			{Landi}}\ and\ \bibinfo {author} {\bibfnamefont {M.}~\bibnamefont
			{Paternostro}},\ }\bibfield  {title} {\bibinfo {title} {Irreversible entropy
			production: From classical to quantum},\ }\href
	{https://doi.org/10.1103/RevModPhys.93.035008} {\bibfield  {journal}
		{\bibinfo  {journal} {Rev. Mod. Phys.}\ }\textbf {\bibinfo {volume} {93}},\
		\bibinfo {pages} {035008} (\bibinfo {year} {2021})}\BibitemShut {NoStop}%
\end{thebibliography}
\end{document}